\documentclass[a4paper,11pt]{article}
\pdfoutput=1 
\RequirePackage{snapshot}  
\usepackage{jheppub} 
\usepackage{bm,amsmath,amssymb,slashed,graphicx,%
            enumerate,alltt,xspace,multirow,xcolor,mathrsfs}
\usepackage{fancyvrb}
\usepackage{booktabs,tabularx}
\usepackage{graphicx}
\usepackage{subcaption}
\usepackage{xspace}
\usepackage{cancel}
\usepackage[utf8]{inputenc} 
\usepackage[compat=1.0.0]{tikz-feynman}
\usepackage{scalerel}
\usepackage{layouts}
\usepackage{adjustbox}
\usepackage{pdflscape}
\usepackage[normalem]{ulem} 
\usepackage{lineno}
\usepackage{wasysym}

\usepackage{algorithm}
\usepackage[noEnd=true]{algpseudocodex}

\newcommand{\ps}{\text{\sc ps}}

\newcommand\BB{\bar{B}}

\newcommand\dBNLO{{dBNLO}\xspace}
\newcommand{\esme}{\text{ESME}\xspace}
\newcommand\ESME{\esme}

\newcommand\betaps{\beta_\text{\textsc{ps}}}

\newcommand\lambdaQCD{\ensuremath{\Lambda_{\scriptscriptstyle \rm QCD}}}
\newcommand{\mathd}{\mathrm{d}}

\newcommand{\PhiB}{\Phi_{\rm B}}
\newcommand{\PhiBR}{\Phi}
\newcommand{\Phirad}{\Phi_\text{rad}}

\newcommand\powhegbox{{\tt POWHEG-BOX}\xspace}
\newcommand\POWHEGBOX{\powhegbox}
\newcommand\powhegboxvt{{\tt POWHEG-BOX-V2}\xspace}
\newcommand\powhegboxres{{\tt POWHEG-BOX-RES}\xspace}

\newcommand\MiNLO{{\tt MiNLO}}

\newcommand\psm{\ensuremath{\mathrm{PS}}}
\newcommand\fks{\ensuremath{\mathrm{FKS}}}

\newcommand{\DeltaN}{N}

\usepackage[capitalise]{cleveref}

\makeatletter
\g@addto@macro\bfseries{\boldmath}
\makeatother

\definecolor{labelkey}{rgb}{0,0.5,0.0}
\definecolor{royalpurple}{rgb}{0.47, 0.32, 0.66}

\usepackage{listings}
\definecolor{darkred}{rgb}{0.6,0.0,0}
\definecolor{darkgreen}{rgb}{0,0.4,0}
\newcommand{\accept}[1]{\textcolor{darkgreen}{\it #1}}
\newcommand{\reject}[1]{\textcolor{darkred}{\it #1}}

\definecolor{grey}{rgb}{0.5,0.5,0.5}
\definecolor{rust}{rgb}{0.9,0.4,0.0}
\definecolor{lightblue}{rgb}{0.0,0.5,1.0}

\allowdisplaybreaks 

\definecolor{semiblue}{rgb}{0.3,0.3,0.8}
\newcommand{\logbook}[2]{}

\newcommand{\codefont}[1]{\texttt{#1}}

\newcommand{\MINT}{\codefont{MINT}\xspace}    
\newcommand\mint{\MINT}
\newcommand{\sherpa}{\codefont{Sherpa}\xspace}
\newcommand{\herwig}{\codefont{Herwig}\xspace}
\newcommand{\mgfive}{\codefont{MG5\_aMC}\xspace}

\newcommand{\pythia}{\codefont{Pythia}\xspace}

\newcommand{\hoppet}{\codefont{Hoppet}\xspace}
\newcommand{\disent}{\codefont{DISENT}\xspace}

\newcommand{\panscales}{PanScales\xspace}
\newcommand{\PanScales}{\panscales}
\newcommand{\panglobal}{\pg}
\newcommand{\PanGlobal}{\pg}
\newcommand{\PanLocal}{\pl}
\newcommand{\Pythia}{\pythia}

\newcommand{\pg}{{PanGlobal}\xspace}
\newcommand{\pl}{{PanLocal}\xspace}
\newcommand{\panlocal}{\pl}

\newcommand{\GeV}{\;\mathrm{GeV}}
\newcommand{\TeV}{\;\mathrm{TeV}}
\newcommand{\order}[1]{\mathcal{O}\left(#1\right)}
\newcommand{\as}{\alpha_s}

\newcommand{\Bbar}{{\bar B}}
\newcommand{\nB}{n_\text{\textsc{b}}}

\newcommand{\itilde}{{\tilde \imath}}
\newcommand{\jtilde}{{\tilde \jmath}}

\newcommand\sss{\mathchoice%
{\displaystyle}%
{\scriptstyle}%
{\scriptscriptstyle}%
{\scriptscriptstyle}%
}
\newcommand{\dis}{\text{\textsc{dis}}}

\newcommand{\rad}{\mathrm{rad}}

\newcommand{\rl}{r_{\rm \sss L}}

\PassOptionsToPackage{hypertexnames=false}{jheppub}

\tikzstyle{block} = [rectangle, minimum width=1.0cm, minimum height=0.75cm, thin, draw=black]
\tikzstyle{blob} = [circle, minimum width=0.5cm, thin, draw=black]
\tikzset{blackarrow/.style={-stealth, semithick, draw=black}}
\tikzset{connection/.style={inner sep=0,outer sep=0}}

\newcolumntype{C}{>{\centering\arraybackslash}X}

\title{Logarithmically-accurate and positive-definite NLO shower matching}

\preprint{\small CERN-TH-2025-004, OUTP-25-01P, Nikhef~2025-003}

\newcommand{\OXaff}{Rudolf Peierls Centre for Theoretical Physics, Clarendon Laboratory, Parks Road,
  University of Oxford, Oxford OX1 3PU, UK}
\newcommand{\ASCaff}{All Souls College, Oxford OX1 4AL, UK}
\newcommand{\CERNaff}{CERN, Theoretical Physics Department, CH-1211 Geneva 23, Switzerland}
\newcommand{\NIKHEFaff}{Nikhef, Theory Group, Science Park 105, 1098 XG, Amsterdam, The Netherlands}
\newcommand{\Monashaff}{School of Physics and Astronomy, Monash University, Wellington Rd, Clayton VIC-3800, Australia}
\newcommand{\CNRSaff}{Université Paris-Saclay, CNRS, CEA, Institut de physique théorique, 91191, Gif-sur-Yvette, France}
\newcommand{\UGRaff}{Departamento de F\'{i}sica Te\'{o}rica y del Cosmos, Universidad de Granada,
Campus de Fuentenueva, E-18071 Granada, Spain}

\author[a]{Melissa van Beekveld,}%
\author[b]{Silvia Ferrario Ravasio,}%
\author[c]{Jack Helliwell,}%
\author[b]{Alexander Karlberg,}%
\author[d,e]{Gavin~P.~Salam,}%
\author[c]{Ludovic Scyboz,}%
\author[f]{Alba Soto-Ontoso,}%
\author[g]{Gregory Soyez,}%
\author[d]{Silvia Zanoli}%

\affiliation[a]{\NIKHEFaff}
\affiliation[b]{\CERNaff}
\affiliation[c]{\Monashaff}
\affiliation[d]{\OXaff}
\affiliation[e]{\ASCaff}
\affiliation[f]{\UGRaff}
\affiliation[g]{\CNRSaff}

\date{Received: date / Accepted: \today}

\abstract{
  We present methods to achieve NLL+NLO accurate parton showering for
  processes with two coloured legs: neutral- and charged-current Drell--Yan, 
  and Higgs production in $pp$ collisions, as well as DIS and $e^+e^-$ to jets.
  The methods include adaptations of existing approaches, as well as a
  new NLO matching scheme, \ESME, that is positive-definite by
  construction.
  Our implementations of the methods within the \PanScales framework
  yield highly competitive NLO event generation speeds.
  We validate the fixed-order and combined resummation accuracy with
  tests in the limit of small QCD coupling and briefly touch on
  phenomenological comparisons to standard NLO results and to
  Drell--Yan data.
  The progress reported here is an essential step towards showers with
  logarithmic accuracy beyond NLL for processes with incoming hadrons.
}

\keywords{QCD, Parton Shower, NLO, Matching, Resummation, LHC, LEP
  \\[4em]
  \textit{For the purpose of Open Access, the authors have applied a CC BY
  public copyright licence to any Author Accepted Manuscript (AAM)
  version arising from this submission.}
}

\begin{document}

\maketitle

\section{Introduction}
\label{sec:introduction}
As the Large Hadron Collider (LHC) explores the electroweak scale and the energy frontier with
continually increasing luminosity, the demands on the accuracy of QCD
predictions grow ever more challenging.
The only predictive method that approaches physical realism while also
being accurate is to match parton showers with fixed order
calculations.
Next-to-leading order (NLO) matching is widely considered to be a
solved
problem~\cite{Frixione:2002ik,Nason:2004rx,Frixione:2007vw,Alioli:2010xd,Hoeche:2011fd,Alwall:2014hca},
and today most research on matching explores the question at
next-to-next-to-leading order (NNLO), see
e.g.\ \cite{Alioli:2013hqa,Hamilton:2013fea,Hoeche:2014aia,Monni:2019whf,Campbell:2021svd,Alioli:2023har,El-Menoufi:2024sys}.

With the advent of logarithmically-accurate parton
showers~\cite{Dasgupta:2020fwr,Hamilton:2020rcu,Karlberg:2021kwr,Hamilton:2021dyz,vanBeekveld:2022zhl,vanBeekveld:2022ukn,vanBeekveld:2023chs,Forshaw:2020wrq,Nagy:2020dvz,Nagy:2020rmk,Herren:2022jej,FerrarioRavasio:2023kyg,Assi:2023rbu,Preuss:2024vyu,Hoche:2024dee,vanBeekveld:2024wws},
it becomes necessary to revisit NLO matching and examine
how to combine fixed-order matching and
logarithmic accuracy in a consistent manner.
First steps at NLO were explored recently by some of us in
Ref.~\cite{Hamilton:2023dwb} for the simple case of unoriented
two-body decays. 
One of the main purposes of this paper is to implement logarithmically-consistent NLO matching for a wider variety of processes, specifically
$pp$ scattering producing a $Z/\gamma^*$, $W$, or a Higgs boson;
deep-inelastic scattering (DIS); and oriented two-body decays.
Our adaptations of standard methods for achieving this are discussed
in Section~\ref{sec:general_matching}.
That section also addresses our treatment of real radiation, using
multiplicative matching~\cite{Bengtsson:1986hr,Seymour:1994df} (cf.\
also POWHEG~\cite{Nason:2004rx}, KrkNLO~\cite{Jadach:2015mza} and
MAcNLOPS~\cite{Nason:2021xke}).
Furthermore it discusses how to generate the real radiation phase space efficiently
in the presence of substantial lepton asymmetries and highlights
subtleties in the treatment of flavour in projection-to-Born methods
in the presence of parity-violating contributions.

In revisiting NLO matching, the opportunity arises to ask whether it
is possible to resolve a long-standing issue, namely the presence of
negative-weight events.
These are a characteristic of all main modern NLO and NNLO matching
approaches, as well as NLO-merging
methods~\cite{Frederix:2012ps,Hoeche:2012yf}. 
Depending on the process, their fraction may range from about a
percent to tens of percent (see e.g.~\cite{Frederix:2020trv}).
This causes problems both for statistical convergence and for
machine-learning applications.
It also results in an effectively unphysical event sample.
There are, broadly speaking, two key sources of negative weights.
One is connected with the generation of real radiation, and is present
only in the MC@NLO method~\cite{Frixione:2002ik}, which adds and
subtracts contributions to a given shower's real radiation.
This source of negative weights is eliminated in the broad family of
multiplicative matching~\cite{Bengtsson:1986hr,Seymour:1994df}
methods, and is embodied in the POWHEG approach~\cite{Nason:2004rx}, as well as
KrkNLO~\cite{Jadach:2015mza} and MAcNLOPS~\cite{Nason:2021xke}.
Section~\ref{sec:slicing} presents a family of algorithms that resolve
the other part of the problem, namely guaranteeing a
positive-definite event-by-event NLO normalisation.
It also introduces a method to convert a slicing
calculation into the form of a subtraction.

An important part of the general \panscales approach is conclusive
numerical testing of the quoted accuracy.
It is to be kept in mind that NLO parton-shower matching brings
extra terms beyond strict NLO accuracy starting at relative order
$\alpha_s^2$, in particular, all-order logarithmic contributions.
One important exercise will therefore be, in
Section~\ref{sec:NLO-NNDL}, to compare our NLO matching directly to
NLO calculations with physical settings for the coupling.
We will
additionally verify NLO accuracy using the same type of $\as \to 0$
approach that we originally introduced for checking logarithmic
accuracy~\cite{Dasgupta:2020fwr}.
To the best of our knowledge, this is the first time such a test has
been carried out for NLO parton-shower matching.
We will also show next-to-next-to-double logarithmic (NNDL,
$\as^n L^{2n-2}$) accuracy tests for event-shape like observables.
As was the case for $e^+e^-$ matching~\cite{Hamilton:2023dwb}, the NNDL
tests are a crucial step on the way towards full
next-to-next-to-leading-logarithmic accuracy (NNLL,
$\as^n L^{n-1}$)~\cite{FerrarioRavasio:2023kyg,vanBeekveld:2024wws}.

Given that this is the first time that a parton shower demonstrably
reaches general NLL+NLO as well as event-shape NNDL accuracy, in Section~\ref{sec:pheno}
we will include a brief comparison to Drell--Yan data and then in
Section~\ref{sec:performance} discuss
event-generation speed, before concluding in
Section~\ref{sec:conclusions}.

Finally, some further technical details
are discussed in Appendices~\ref{app:summary-panscales}--\ref{app:swap}.

\section{Adaptations of existing NLO approaches}\label{sec:general_matching}

In this section we start with an overview of NLO matching and associated
notation.
We then briefly examine how we handle the matching of real
radiation, using a multiplicative method that all other parts of this
paper rely on.
We then consider two alternative ways for the generation of the Born
event with the NLO normalisation: one adapts the numerical approach of
the \powhegbox~\cite{Nason:2004rx,Frixione:2007vw,Alioli:2010xd}
framework to the \panscales{} showers, the other, specific to DIS,
uses an analytic Projection-to-Born (P2B)
methodology~\cite{Cacciari:2015jma,Hoche:2018gti}.

\subsection{Overview and notation}
\label{sec:nlo-matching-overview}
The strategies we follow in this work all belong to the multiplicative
matching category in which the cross section for the event, starting from 
a given Born phase-space point $\PhiB$, can be written as~\cite{Nason:2004rx}
\begin{multline}
  \label{eq:multiplicative}
  \mathd \sigma_\text{mult} =
  \BB(\PhiB)\,d \PhiB\,
  \left[
    S(v^\ps,\PhiB)
    \times \frac{R(\PhiB,\Phi_\text{rad})}{B_0(\PhiB)}  \,  d\Phirad
    \right]
    \times I_\ps(v^\ps, \PhiB,\Phirad)
    \,+\\+
    \Bbar(\PhiB) d\Phi_B S(v_{\min},\PhiB)
    \,.
\end{multline}
Let us focus for now on the first line.
Schematically, it depends on three main ingredients. 
The term in square brackets in Eq.~\eqref{eq:multiplicative} 
describes the generation of the first emission parameterised by $\Phi_\rad$, 
which is associated with a value of the parton shower (PS) ordering variable $v^{\ps}$. 
This depends on the Sudakov form factor, $S(v^{\ps},\PhiB)$, given by
\begin{equation}
  \label{eq:Sudakov}
  S(v^{\ps},\PhiB) = \exp\left[-\int_{v>v^\ps} \frac{R(\PhiB,\Phirad)}{B_0(\PhiB)} d \Phirad\right],
\end{equation}
which we note is computed using the full matrix element $R(\Phi)$
(with $\Phi \equiv \{\PhiB, \Phirad\}$), 
as opposed to the shower's approximate matrix element. 
We elaborate more on this term in Section~\ref{sec:real-radiation}.
Another ingredient is $I_\ps(v^\ps, \PhiBR)$ which denotes the 
subsequent iterations of the parton shower evolution, starting from $v^\ps$.
Lastly, the normalisation factor $\Bbar(\PhiB)$ is given at NLO by
\begin{equation}
  \label{eq:Bbar}
  \Bbar(\PhiB) = B_0(\PhiB) + \underbrace{V(\PhiB) +  \int R(\PhiB,\Phirad)
    \, d\Phirad}_{\text{relative order $\as$}} \, ,
\end{equation}
with $B_0(\PhiB)$ the Born matrix element, $V(\PhiB)$ the 1-loop
contribution and $R(\PhiB,\Phi_\text{rad})$ the real matrix element.
The correct NLO cross section is therefore obtained upon 
integration over the Born phase space
\begin{equation}
  \label{eq:dsigma-basic}
  \sigma =\int \Bbar(\PhiB) d\PhiB\,.
\end{equation}

Finally, we comment on the second line of
Eq.~(\ref{eq:multiplicative}).
It accounts for the no-radiation probability, with $v_{\min}$ the
transverse momentum scale of the non-perturbative shower cutoff.
Since it is a fairly standard term, in what follows we leave it out
from our explicit expressions, so as to lighten the notation.
However, in our practical implementations such a term is always
automatically included.

\subsection{Treatment of the real radiation}
\label{sec:real-radiation}

The shower branching kinematic variables are the dimensionful ordering
variable $v$ (e.g.\ a transverse momentum),
and two auxiliary angular variables $\bar \eta$ and
$\phi$ (cf.\ Appendix~\ref{app:summary-panscales})
and the core equation that we use for the
matched branching probability for any given partition of a dipole is
\begin{equation}
  \label{eq:2}
  \frac{dP}{d\ln v\, d\bar \eta\, d\phi} =
  \frac{1}{\mathd \PhiB}\frac{d\Phi}{d\ln v\, d\bar \eta\, d\phi}
  \frac{R_p(\Phi)}{B_0(\PhiB)}\,.
\end{equation}
Here $R_p(\Phi)$ is a positive \emph{partition} of the full matrix element for
the given final state, designed such that it has the full
singularities of the corresponding partition of the dipole, and only
those singularities.
This ensures that in the infrared the branching probability tends
to the shower branching probability, as is required notably for NNDL
accuracy~\cite{Hamilton:2023dwb}. 
The expressions for $\frac{d\Phi}{d\ln v d\bar \eta d\phi}$ 
together with our partitioning of the matrix elements are
outlined in Appendix~\ref{sec:regions}.
For Eq.~(\ref{eq:2}) to be sufficient for generating the full real
radiation, it is necessary that the shower branching map covers phase
space.
We have verified that this is the case for the \pg and \pl showers for
up to a total of 3 initial and final-state partons for all processes
that we consider here.
For the \pl showers, this required modifications outlined in
Appendix~\ref{app:panlocal}.

One practical consideration is the boundedness of
Eq.~(\ref{eq:2}).
In many of the cases that we examined, the bound from the parton
shower approximation $\frac{dP}{d\ln v\, d\bar \eta\, d\phi}$ was
sufficient also with the full matrix element.
One situation where it was not was in Drell--Yan production, somewhat
away from the $Z$ pole, e.g.\ around $m_{\ell\ell} =
130{-}140\GeV$ and more generally also for $W$ production and decay.
There, a substantial forward-backward asymmetry arises in the Born
matrix element, but not always in the corresponding real matrix
element, e.g.\ for transverse momenta of the order of $m_Z/2$.
In phase-space points where the asymmetry causes $B_0(\PhiB)$ to be
particularly small, this enhances the apparent branching probability.
Potential solutions include choosing a large overhead factor, which
would slow down event generation; or trying to adaptively determine a
grid of overhead factors, which might require a warm-up phase in the
event generation.

Instead we found it more convenient to simultaneously consider
the matrix elements for a given $\ell^+\ell^-$ final state and that
where the $\ell^+$ and $\ell^-$ momenta are swapped.
Starting from a given Born configuration (say $\ell^+\ell^-$), we
first make sure that we reproduce the correct total rate for
$\ell^+\ell^-$ or $\ell^-\ell^+$ in the real configuration. Then,
where necessary, we adjust the relative rates for $\ell^+\ell^-$ and
$\ell^-\ell^+$ by swapping the momenta of the $\ell^+$ and $\ell^-$
with an appropriate probability.
This approach, analogous to Eq.~(4) of
Ref.~\cite{FerrarioRavasio:2023kyg}, avoids a large overhead without
the use of an adaptive grid, allowing us to obtain efficient real
event generation without a substantial warm-up phase.
The concrete algorithm is described in
Appendix~\ref{app:swap}.\footnote{
  One open question is whether the swap algorithm could conceivably be
  generalised and made fully differential rather than discrete, e.g.\
  to sample massive vector boson decays after the NLO generation has
  been performed.
}

\subsection{\dBNLO: an adaptation of the \powhegbox  method
}
\label{sec:dbnlo}

For generic processes, the function $\bar B$ as 
defined by Eq.~\eqref{eq:Bbar} is not known analytically.
To numerically evaluate $\Bbar$, one approach that we adopt 
is similar to that used in the \powhegbox
framework~\cite{Frixione:2007vw}.
We dub this approach \dBNLO because, as we will see below, a correction to
the $\Bbar$ function is sometimes needed to reproduce NLO accuracy, owing to the fact
that the shower map may not agree with the parameterisation used in
the \powhegbox.

As in the original approach, the $\bar B$ function is trivially rewritten,
bringing all contributions in Eq.~\eqref{eq:Bbar} (see also
Eq.~\eqref{eq:Bbar-with-counterterm} below)
under the same integral sign:
\begin{align}
  \bar B(\PhiB) &= \int d X_1 d X_2 d X_3\tilde B(\PhiB, X_1, X_2,X_3)\,,   \label{eq:Bbar-MINT}
  \end{align}
where the equality between Eqs.~\eqref{eq:Bbar} and~\eqref{eq:Bbar-MINT}
effectively defines $\tilde B$. Here, the $X_i$ are the phase-space variables of
the real radiation, which are typically taken to be in a unit hypercube (through
simple transformation of the integration variables $\Phirad$). 

In order to evaluate Eq.~\eqref{eq:Bbar-MINT}, a specific map for the
real radiation phase space $\Phirad$ needs to be chosen. In
Ref.~\cite{Frixione:2007vw}, expressions were derived both within the
Frixione-Kunszt-Signer (FKS)~\cite{Frixione:1995ms,Frixione:1997np},
and Catani-Seymour (CS)~\cite{Catani:1996jh} subtraction schemes. 
On the other hand, in our approach the hardest emission is generated
using the \panscales{} showers. 
This is done so as to facilitate retaining the logarithmic accuracy,
as was discussed in Ref.~\cite{Hamilton:2023dwb}. 
The kinematic maps associated with the FKS (or CS) schemes are not
guaranteed to coincide with those implemented in the \panscales{}
showers in the hard region (which is the one relevant for matching).
A mismatch in the mapping from the Born to Born$+1$ phase space
between $\bar B$ and the parton shower will induce a spurious
$\mathcal{O}(\alpha_s)$, spoiling the NLO accuracy.

There are at least two possibilities to solve this issue. One could calculate
the counterterms associated with the \panscales{} showers so as to correctly sample
$\Bbar$ (these counterterms would need to be computed in $4-2\epsilon$
dimensions for each shower variant).
Alternatively, one can introduce a correction term to account for the
$\mathcal{O}(\alpha_s)$ difference, directly in four dimensions.
We opt for the latter and correct $\Bbar$ as
computed with the FKS parameterisation,\footnote{Using radiation variables  $\xi = \frac{2 E_k}{\sqrt{s}}, y = \cos \theta_{ik}, \phi$, where $k$ is the
emitted parton, $i$ is the emitter, and $\phi$ is an azimuthal angle.}
as this is the one that is implemented in the \powhegbox~\cite{Alioli:2010xd}. 
This solution has the advantage that, going forward, we also have the
option of using $\tilde B$ functions for different processes as
implemented in the \powhegbox.
The correction term, which we denote by $\Delta \tilde B$, 
can be calculated automatically. It is defined through
\begin{subequations}
  \label{eq:DeltaFKS}
  \begin{align}
    d\Phirad^\fks \tilde B^{\mathrm{corr.}}(\PhiB, X)
    &= d\Phirad^\fks\tilde
      B^\fks(\PhiB, X) + d\Phirad^\psm R(\Phi^\psm) -
      d\Phirad^\fks R(\Phi^\fks)
    \\
    &= d\Phirad^\fks \left[ \tilde B^\fks(\PhiB, X) + \Delta \tilde
      B(\PhiB, X) \right]\,,
  \end{align}
\end{subequations}
with $\Delta \tilde B$ defined as
\begin{equation}
  \label{eq:deltaB}
  \Delta \tilde B(\PhiB, X)
  \equiv
   \left|
     \frac{d\Phirad^\psm }{d\Phirad^\fks} \right|
   R(\Phi^\psm) - R(\Phi^\fks)\,.
\end{equation}
The phase-space measures
\begin{equation}
d\Phi^{\rm{PS/FKS}}=d\PhiB\ d\Phi_{\rm{rad}}^{\rm{PS/FKS}},
\end{equation}
implicitly define a map between the Born phase space, 
$\PhiB$, and the Born$+1$ phase space. The $\Delta \tilde
B(\PhiB, X)$ term is the main conceptual novelty of this matching
method, hence our choice of the \dBNLO name.

Regarding the technical implementation within the \PanScales framework,
we generate (unweighted) Born events according to $\bar{B}(\PhiB)$
following the \mint 
approach~\cite{Nason:2007vt}: in a first phase,
integration grids are generated using the adaptive importance-sampling {\tt VEGAS}
algorithm~\cite{Lepage:1977sw}. In a second phase, upper bounds are found for
the integrand $\tilde B$. In \mint~these upper bounds can be estimated with the
possibility of ``folding'' the radiation variables multiple times over the
integration range, so as to minimise the risk of the integrand being
negative-valued. We have implemented the folding procedure in our \panscales{}
framework, though in the following we typically show results without folding,
i.e.\ including negative weights.\footnote{In all processes we have investigated, 
the fraction of negative weights was below 5 per mille.}
 
Once upper bounds have been found,
the $\tilde B(\PhiB, X)$ function can be sampled randomly, 
and Born variables
are generated with an accept-reject algorithm, where we simply discard the
radiation coordinates $X$. This ensures that the Born phase space is sampled
according to the $\bar B(\PhiB)$ distribution.

The correction term $\Delta \tilde
B(\PhiB, X)$ is calculated at the same time as the $\tilde
B^\fks$ function, in a semi-automated numerical way.
When evaluating the contribution to the integrand for a given $(\PhiB, X)$, one
translates the FKS variables defining the real emission,
$\Phi_{\text{rad}}^\fks$, to the \panscales{} variables, $\Phi_{\text{rad}}^\psm$
(associated with the Jacobian $|d \Phi_\text{rad}^\psm / d \Phi_\text{rad}^\fks|$).
One can then perform that emission with full kinematics from the Born state
$\PhiB$ with any of the \panscales{} showers as well as our implementation of the
FKS map.
One then evaluates Eq.~(\ref{eq:deltaB}) at that phase space point.
Note that infrared and
collinear divergences cancel in $\Delta \tilde B$ and the correction is hence
finite.

We have implemented the \dBNLO method for
$e^+e^- \to \gamma^* \to q \bar q$, $pp \to Z$ for both the \pg
and \pl showers.
In these cases the $\tilde B^\text{FKS}$ function is known
analytically from Ref.~\cite{Frixione:2002ik}.
The $\Delta \tilde{B}$ is computed as explained above for
$e^+e^- \to q\bar q$ and is simply zero for $pp \to Z$, because the
FKS, \pg and \pl shower maps all act equivalently for the first
emission.

We have also considered $pp \to Z/\gamma^* \to \ell^+\ell^-$ for the \pg shower.
For this process, we interfaced the Fortran code from the \powhegbox
associated with Ref.~\cite{Alioli:2008gx}, in order to evaluate the
$\Bbar$.
Here too $\Delta \tilde B \equiv 0$ for \pg, whereas the \pl map
acts differently on $\gamma^*/Z$ decay products and so would require a
non-zero $\Delta \tilde B$, which we have yet to implement.
Note that when the lepton swaps of App.~\ref{app:swap}
are being used to optimise the generation of the Drell--Yan real matrix
element, we employ a suitably adapted version of the \powhegbox
Fortran code, to account for the fact that the real contribution in
the integral in Eq.~(\ref{eq:Bbar}) should involve the two lepton
permutations.
This modification is described in Appendix~\ref{app:swap}.

Thinking forward to future work, one potential advantage of the \dBNLO
method for generating the Born event is that it opens up the
possibility of reading in $\tilde{B}^{\rm{FKS}}$ from the \powhegbox
for matching generic processes.
The function $\Delta \tilde{B}$ would then be computed separately and
automatically.
Note however that for general processes the generation of the real
emission remains non-trivial and to maintain logarithmic accuracy
it is important for it to be generated in a way that is consistent
with the shower map and the shower's specific pattern of higher-order
corrections in various infrared limits. 

A final comment is that for \panglobal with $\betaps=0$, for the first
emission in any colour singlet process, the \powhegbox{} kinematic
map and ordering variable are identical to the
corresponding shower map and ordering variables.
This means that in principle it is also possible to shower events in the
Les Houches Event~\cite{Alioli:2013nda} (LHE) format produced from the
\powhegbox{} and retain NLL accuracy.
However, for such an interface, some practical aspects remain to be
implemented concerning the correct processing of the LHE files and the
setup of the corresponding \panscales event for subsequent showering.
Furthermore, with the information that is available in LHE files it
would not be possible to support the spin-correlation
component~\cite{Karlberg:2021kwr,Hamilton:2021dyz,vanBeekveld:2022zhl}
of NLL accuracy.

\subsection{Projection-to-Born}
\label{sec:p2b}

Generically, the P2B approach~\cite{Cacciari:2015jma} exploits the
fact that Eq.~\eqref{eq:Bbar} can be computed analytically for certain
processes, as a function of the Born kinematics, specifically where the
real branching leaves key Born invariants unchanged.
In a shower context, Ref.~\cite{Hoche:2018gti} used this at NNLO+PS
for DIS.\footnote{It is arguable whether a constant (N)NLO
  normalisation factor, as used e.g.\ in
  \cite{Bizon:2019tfo,Hamilton:2023dwb,Preuss:2024vyu}, also counts as
  P2B.
  In practice, our \panscales $H\to gg$ decay process has an inclusive
  NLO normalisation matching option that is classified as ``P2B'', but
  our nomenclature may evolve.}
In this paper, we apply it at NLO to DIS and Eq.~\eqref{eq:Bbar} is
known analytically in terms the proton structure
functions~\cite{ParticleDataGroup:2024cfk},
\begin{align}
  \label{eq:Bbar-dis}
 \bar{B}(\PhiB) =
 \frac{4\pi\alpha^2}{x_{\rm \sss DIS} Q^4_{\rm \sss
     DIS}}\left[\frac{1}{2}(1+(1- y_{\rm \sss DIS})^2)F_2
   -\frac{1}{2}y_{\rm \sss DIS}^2 F_L +x_{\rm \sss DIS}y_{\rm \sss
     DIS}(1-\frac12 y_{\rm \sss DIS})F_3\right] \,,
\end{align}
where the Born variables are $x_{\rm \sss DIS}$, 
$Q^2_{\rm \sss DIS}$ and 
$y_{\rm \sss DIS} = Q^2_{\rm \sss DIS}/(x_{\rm \sss DIS} s)$,
and $s$ is the collider centre-of-mass energy squared.
When writing Eq.~\eqref{eq:Bbar-dis} we have 
assumed a shower mapping that preserves the DIS invariants
$ Q^2_{\rm \dis}$ and $x_{\rm \sss DIS}$, 
as indeed happens for the \panscales showers. 
We use \hoppet~\cite{Salam:2008qg, HoppetInPrep} for the evaluation 
of the structure functions, $F_{2,3,L}$.

The expression in Eq.~\eqref{eq:Bbar-dis} contains an implicit sum
over all possible flavour channels.  In the context of parton shower
exclusive simulations, for a given event, we generate one specific
Born flavour channel and $\bar{B}$ needs to be known for that specific
flavour.
At LO, the Born flavour label is trivial, but this 
is not the case for NLO, where real diagrams might 
originate from several underlying Born configurations.
This is the case for gluon-initiated real corrections $g\ell \to
\ell' \bar{q} q'$, where the possible underlying Born channels are $q\ell \to
q'\ell'$ and $\bar q \ell \to \bar q' \ell'$. 
The \hoppet structure functions can be decomposed by flavour, but
for the gluon-induced axial ($F_3$) component there is an intrinsic
ambiguity in the assignment of a $g\ell \to X$ contribution to Born
flavour and anti-flavour structure functions, with only their
difference contributing to the cross section.
In practice that $F_3$ contribution to $g\ell \to X$ is effectively
set to zero.%
\footnote{
  In the neutral current $F_2$ and $F_L$ structure functions, quarks
  and anti-quarks contribute equally, while in the $F_3$ structure
  function, quark and anti-quark contributions appear with opposite
  signs, cf.\ Eq.~(18.18) of the 2024 edition of the Structure
  Functions review by the Particle Data
  Group~\cite{ParticleDataGroup:2024cfk}.
  A potential gluon-induced $F_3$ contribution would come from a
  convolution of a $C^{(3)}_{qg}$ coefficient function with the gluon
  distribution.
  However since the $C^{(3)}_{qg}$ convolution contributes equally to
  quarks and anti-quarks, its net contribution to the cross section
  differentially in $x$ and $Q^2$ will always be zero.
  Because of this, the $C^{(3)}_{qg}$ coefficient function term is
  conventionally simply set to be zero (note, for example, its absence
  in \cite{Moch:1999eb}).
  Consequently, in a standard structure-function based P2B approach,
  there is no NLO gluon-induced contribution that is attributed to
  $F_3$.
  \logbook{}{See logbook/2022-12-04-DIS-matching/DIS_matching.pdf for
    some further details}
} 
However, a given matched shower's backward evolution from quark or
anti-quark to gluons will in general not yield an ensemble of final
states with equivalent cross sections.
This implies that there can be a
mismatch between the NLO flavour versus anti-flavour assignment in
the structure functions and the true $\Bbar$ that is actually needed
for the given shower mapping.
To be able to use these structure functions, 
we can either partition the gluon-induced real correction consistently, which might not always be possible, or calculate the mismatch.

For photon-induced DIS, where $F_3$ is zero, any democratic partitioning of the gluon-induced matrix element,
such as the one we have implemented in
Eq.~(\ref{eq:DISisrq}), yields a result consistent with the structure function.
This is due to the fact that the matrix elements for 
$q\ell \to q \ell$ and $\bar{q}\ell \to \bar{q}\ell$ are identical.
This is no longer the case if we consider $Z$ or $W$ as mediators,
due to the axial component of the coupling.
This flavour mismatch will induce a correction to 
the flavour-decomposed $\bar{B}$. For the time being, therefore, with
the P2B method we focus only on the
photon-mediated DIS process and leave for future 
work the treatment of $Z$ and $W$ mediated process.
However, we note that for any infrared 
safe observable that does not depend on flavour,
we would still obtain NLO accuracy once we sum over all the flavours
if we use this partitioning for all the DIS processes.%
\footnote{Ref.~\cite{Hoche:2018gti} describes the implementation of a
NNLO+PS generator for DIS starting from the NNLO structure
functions.
The parton shower branching history is used to determine the
underlying Born flavour assignment for the configurations containing
one (and two) extra emissions, which does not match the partitioning
that is effectively present in the (N)NLO structure functions on the
final integrated result.
We stress that the mismatch cannot be seen for any flavour-summed
observable, hence the NNLO accuracy of the generator presented in
Ref.~\cite{Hoche:2018gti} is not impacted for such observables.
}

\section{Positive-definite NLO event generation}
\label{sec:slicing}
When NLO matching methods for parton showers were first
developed~\cite{Frixione:2002ik,Nason:2004rx}, the advance was
sufficiently revolutionary that a small fraction of negative weights
was considered a price well worth paying.
However, as NLO parton-shower matching has evolved to become the
default accuracy for essentially all studies, and a foundation for
first NNLO shower matching methods, the question of negative-weight
events is taking on greater importance.
Firstly, for a fraction $f$ of negative-weight events, the statistics
required for a given accuracy scale as $1/(1-2f)^2$.
For example, already for $f=0.15$ this doubles the required
statistics.
There are key LHC studies where this is a limiting
factor~\cite{CMS:2022psv} and nowadays this is widely considered to be
a problem~\cite{Mandrik:2017vly,HSFPhysicsEventGeneratorWG:2020gxw,ATLAS:2021yza,Campbell:2022qmc}.
The issue of negative weights turns out to be challenging also with
modern machine-learning (ML) approaches (see
e.g.~\cite{%
  ATLAS:2022rws,
  ATLAS:2023ajo,
  CMS:2024jdl,
  ATLAS:2024nab,
  ATLAS:2024jry,
  ATLAS:2024ynn
}), which typically assume a
physical, i.e.\ positive-definite event stream.
Indeed, one could argue that the core goal of Monte Carlo event
simulation, which is to provide a physically realistic simulation of
high-energy collisions, is in some way not being met if there is even
a single negative-weight event.

There are several strategies in the literature to address the question
of negative weights.
Some are intended to be used as an intrinsic part of the NLO
generation code, for example
folding~\cite{Nason:2007vt,Frederix:2020trv} and Born
spreading~\cite{Frederix:2023hom} and related
methods~\cite{Shyamsundar:2025nzn,Shyamsundar:2025mfw} (see below for
further discussion). Methods to reduce the fraction of negative
weights have also been explored within the \sherpa
framework~\cite{Danziger:2021xvr}.
Other methods effectively modify a sample after it has been generated,
notably cell
resampling~\cite{Andersen:2021mvw,Andersen:2023cku,Andersen:2024mqh}
and machine-learning based neural resampler
methods~\cite{Nachman:2020fff}.
In general these methods reduce the fraction of negative weights, but
do not completely eliminate them.\footnote{The neural resampler method
  promises to eliminate negative event weights, as long as the cross
  section is positive in a given phase space region. As discussed
  below, this is not always the case.}
In almost all cases, the reduction comes at the cost of a speed penalty, a
potentially hard-to-quantify NLO bias, a sample that no longer has uniform
weights and/or an after-burner stage that complicates the overall
event-generation workflow.

The purpose of this section is to introduce a new method that ensures
the absence of negative weights, intrinsically as part of the event
generation, while maintaining speed and guaranteed NLO
accuracy.
In Section~\ref{sec:neg-event-sources} we discuss the various
potential sources of negative weights (see also the discussion
of Ref.~\cite{Frederix:2020trv}).
One main source is addressed by treatments of real radiation that
involve just multiplicative or (positive-definite) additive
matching~\cite{Bengtsson:1986hr,Seymour:1994df,Nason:2004rx,Jadach:2015mza,Nason:2021xke},
cf.\ our choices in Section~\ref{sec:real-radiation}.
Section~\ref{sec:Bbar-integer-alg} then introduces a generic method to
address the other non-trivial source, connected with the Monte Carlo
evaluation of the NLO $\Bbar$ normalisation.
It exploits a Sudakov exponentiation, and we will refer to the
resulting generic class of algorithms as ``Exponentiated Subtraction
for Matching Events,'' ESME.
It can be seen as a generalisation of a Poisson process, and we note
that Poisson processes have been explored for matching also in past
work~\cite{Lavesson:2008ah,Lonnblad:2012ix}.
Section~\ref{sec:ESME-core} then provides a specific implementation that
combines real and NLO normalisation into a single algorithm.  
Finally, Section~\ref{sec:counterterm-from-slicing} highlights a
translation that we have used between slicing and subtraction that
facilitates the use of our algorithm with the \PanScales parton showers.

\subsection{The origins of negative weights in standard matching approaches}
\label{sec:neg-event-sources}
As discussed in Section~\ref{sec:nlo-matching-overview}, the weight of
a Born event, at NLO accuracy, should be generated according to
\begin{equation}
  \label{eq:dsigma-basic-in-esme-section}
  d\sigma = \Bbar(\PhiB) d\PhiB\,,
\end{equation}
where $\Bbar(\PhiB)$ is given in Eq.~\eqref{eq:Bbar}. 
In the most common NLO matching approaches, MC@NLO and POWHEG,
equations like Eq.~(\ref{eq:Bbar})\footnote{Note that in MC@NLO $R(\Phi)$ 
is the shower approximation to the real matrix element.} are evaluated
with the help of FKS~\cite{Frixione:1995ms} or
dipole~\cite{Catani:1996jh,Catani:1996vz} subtraction counterterms 
\begin{equation}
  \label{eq:Bbar-with-counterterm}
  \Bbar(\PhiB) = B_0(\PhiB) + 
  \underbrace{
    V(\PhiB) +    
    C_\text{int}(\PhiB)
    + 
    \int \left[R(\Phi) - C(\Phi) \right] \, d\Phi_\text{rad}
  }_{\text{relative order $\as$}}\,.
\end{equation}
Generically, $C(\Phi)$ is a counterterm that satisfies $R-C \to 0$ in the
soft and/or collinear limits for $\Phi_\text{rad}$, and that
is sufficiently simple to be integrated analytically
\begin{equation}
  \label{eq:Cint}
  C_\text{int}(\PhiB)  = \int C(\Phi)\, d\Phi_\text{rad}\,.
\end{equation}
If we assume that we have positive-definite parton distribution
functions (PDFs), as in
recent work from the NNPDF group~\cite{Cruz-Martinez:2024cbz}, 
there are three sources of negative weights in common matching procedures.

The \emph{first source of negative weights} lies in the fact that the
contents of the underbrace in Eq.~(\ref{eq:Bbar-with-counterterm}) may
genuinely be large and negative.
For example if considering a process such as $Z+\text{jet}$
production (as the Born process), then in the limit of small-$p_t$ for
the jet the underbrace will go as $ - B_0(\PhiB) \times 2\as C_F/\pi \ln^2
M_Z/p_t$,
and the overall $\Bbar$ as $B_0(\PhiB)(1 - 2\as C_F/\pi \ln^2 M_Z/p_t)$.
For sufficiently small $p_t$, this will go negative.
In this case the physical origin is clear.\footnote{And as a result
  there is an obvious physically-motivated solution in the MiNLO
  approach~\cite{Hamilton:2012np}, which generates the Born event with
  a Sudakov, whose expansion cancels the negative $\as \ln^2 M_Z/p_t$
  term.
  Alternatively, if one nests NLO $Z$ and NLO $Z+\text{jet}$ showering
  then one may use formulas such as those present in
  Refs.~\cite{Hartgring:2013jma,Li:2016yez,Campbell:2021svd,vanBeekveld:2024qxs}
  which cancel the negative $\as \ln^2 M_Z/p_t$ through the structure of
  the nested NLO terms.  }
However in general there may be a range of situations where the NLO
coefficient is large and negative and the physical origin will not
always be obvious.

A \emph{second source of negative weights} is connected with the
way the integral in Eq.~(\ref{eq:Bbar-with-counterterm}) is
evaluated.
In general, it requires a Monte Carlo evaluation, and this is often
done with just a single $\Phi_\text{rad}$ sample for a given $\PhiB$.
Even if the underbrace is positive when carrying out the full
integration, in a Monte Carlo evaluation with a limited number of
$\Phi_\text{rad}$ points, for a given $\PhiB$ one may
end up sampling a set of $\Phi_\text{rad}$ phase space points such
that the underbrace appears large and negative.
The main mitigation measure that is used for this is
folding~\cite{Nason:2007vt}, which splits the real phase space into
distinct regions and samples each of them for any given $\PhiB$.
This can improve the situation quite substantially, albeit at a speed
cost.
Other
techniques~\cite{Frederix:2023hom,Shyamsundar:2025nzn,Shyamsundar:2025mfw}
seek to reorganise the integrand.
This can reduce the fraction of negative weights without any impact on
speed, but it arguably adds complexity to the formulation of the
method.
As they stand, none of these method provide a guarantee of positivity.

In purely additive matching schemes, notably the MC@NLO approach, one
has a \textit{third source of negative weights}.\footnote{Called
  $\mathbb{H}$ in e.g.\ Ref.~\cite{Frederix:2020trv}; there it is
  further split into \textbf{N.1} and \textbf{N.2}.
  The first and second sources that we discussed above correspond,
  together, to $\mathbb{S}$ in Ref.~\cite{Frederix:2020trv}, or
  equivalently \textbf{N.3}.
}
In such an approach the $\Bbar_s(\PhiB)$ function reads 
\begin{equation}
  \label{eq:Bbar_s}
  \Bbar_s(\PhiB) = B_0(\PhiB) + \underbrace{V(\PhiB) +  \int R_s(\Phi)
    \, d\Phi_\text{rad}}_{\text{relative order $\as$}}\,,
\end{equation}
where $R_s(\Phi)$ is the shower's approximation of the real matrix
element.
The Born event generation (with its subsequent showering) is then to
be supplemented with an additional stream of events, which generates
\begin{equation}
  \label{eq:R-minus-Rs}
  d\Phi (R - R_s),
\end{equation}
leading to negative weights when $R < R_s$.

In the rest of this section, we will show how to eliminate all sources of
negative weights and so guarantee positive-weight events.
In the simple cases that we have implemented, this is achieved without
any speed penalty relative to the public NLO matching codes that we
have tried.

\subsection{Exponentiated subtraction for $\Bbar$}
\label{sec:Bbar-integer-alg}

\subsubsection{The general algorithm}
Here we present an algorithm that converts any subtraction integral of
the form Eq.~(\ref{eq:Bbar-with-counterterm}) into an event-by-event
integer, with the option to bound the integer and to control higher-order
terms in the Monte Carlo average to some given order.
The underlying principles of this algorithm can serve as a basis for a
wide range of variants.

As a starting point, we assume a phase-space generation in which one can factorise
the radiation phase space $d\Phi_\text{rad}$ into an ordering variable
$v$ and a 2-dimensional remainder,
\begin{equation}
  \label{eq:dPhirad-vPhi2}
  d\Phi_\text{rad} = d\ln v\, d\Phi_2\, J,
\end{equation}
where $J$ is a Jacobian.
Standard FKS~\cite{Frixione:1995ms} and
Catani-Seymour~\cite{Catani:1996vz} phase-space generation lend
themselves to this organisation, as reflected in their use for
parton-shower style real-emission generation in
\powhegbox~\cite{Frixione:2007vw} and \sherpa~\cite{Schumann:2007mg}.
For the purposes of the discussion below, it may be useful to think of
$v$ as being equivalent to a transverse momentum.
As with a standard shower, we define a Sudakov factor
\begin{multline}
  \label{eq:Bbar-Sudakov}
  \Delta(v) = \exp\left[- \int_v^{v_\text{max}}
    \frac{dv'}{v'} \rho(v')  \right]
  \quad\text{with}\quad
  \\
  \rho(v) = \int d\Phi_2\, J\,
  \frac{M(\Phi)}{B_0(\PhiB)},
  \quad
  M(\Phi) \ge \max[R(\Phi),C(\Phi)]
\end{multline}
where $M(\Phi)$ is a generic overestimate function that is always at
least as large as the maximum of $R(\Phi)$ and $C(\Phi)$ (this
requirement could be loosened here, but will be needed later in
Section~\ref{sec:ESME-core}). 
In the discussion below we take it to always be of the same order in
$\as$ as the real matrix element, and to share the same infrared and
collinear scaling properties.
By definition, $R(\Phi)$ is positive definite and we assume a
subtraction scheme in which $C(\Phi)$ is also positive definite,
possibly after a suitable sum over partitions.\footnote{The
  $C(\Phi)\ge 0$ restriction can, we believe, be lifted simply by
  replacing $\max[R(\Phi),C(\Phi)]$, below, with
  $\max[R(\Phi),C(\Phi), R(\Phi)-C(\Phi)]$.
  $M(\Phi)$ is generally trivial to find in the infrared.
  It may be more complicated in the hard region if $R(\Phi)/B_0(\PhiB)$
  grows large, however in that case it is conceptually straightforward
  to add a separate stream of events that accounts for any regions
  where $M(\Phi)$ is not sufficiently large, using standard
  unweighting methods.
  Typically we would expect $C(\Phi)/B_0(\PhiB)$ to remain under good
  control insofar as the counterterm is constructed from the Born
  multiplied by a factorised emission.
}
Furthermore, the integration Eq.~(\ref{eq:Bbar-Sudakov})
covers the full real phase space, and we have $\Delta(v_{\max})=1$
and $\Delta(0)=0$.
With this we can introduce our core procedure,
Algorithm~\ref{alg:sub-to-int}. 

\begin{algorithm}[h!]
  \caption{General algorithm to convert NLO subtraction integral to integer}
  \label{alg:sub-to-int}
\begin{algorithmic}[1]
  \State{Set $\nB = 1$ and $v=v_{\max}$\label{alg:sub-to-int:start}} 
  \While{$v > v_\text{min}$}
  \State generate next $v$ and $\Phi_2$ according to Sudakov
  with density $\rho(v) d\ln v$, Eq.~(\ref{eq:Bbar-Sudakov})%
  \label{alg:sub-to-int:lnv}
  \State generate random number $0<r<1$
  \If{$r < |R(\Phi)-C(\Phi)|/M(\Phi)$}
    \State \textbf{if} $R(\Phi) > C(\Phi)$:  $\nB \to \nB + 1$   \label{alg:sub-to-int:plus}
    \State \textbf{else}:\qquad\qquad\;\;\quad $\nB \to \nB - 1$ \label{alg:sub-to-int:minus}
  \EndIf
  \EndWhile
  \State \textbf{return} $\nB$
\end{algorithmic}
\end{algorithm}
Algorithm~\ref{alg:sub-to-int} calculates an event-by-event normalisation factor
$\nB$ that multiplies $B_0(\PhiB)$ and whose average across many
events with the same $\PhiB$ is intended to satisfy
\begin{equation}
  \label{eq:nB-result}
  \langle \nB \rangle  =  1 +  r\,,\qquad r \equiv \int \frac{R(\Phi)
    - C(\Phi)}{B_0(\PhiB)} \, d\Phi_\text{rad}\,. 
\end{equation}
This means that Algorithm~\ref{alg:sub-to-int} can be used in the
evaluation of $\Bbar$ in
Eq.~\eqref{eq:Bbar-with-counterterm}, apart from the $V+C_\text{int}$
contribution, which we will discuss explicitly below in Section~\ref{sec:ESME-core}.
It achieves this by using the standard Sudakov veto algorithm
to choose a phase space point to sample.
However, rather than generating an emission as in the normal Sudakov
veto algorithm (with probability $R/M$), it increments or decrements
$\nB$ (according to the value of $(R-C)/M$).
No real emission is ever generated.
A second difference relative to the normal Sudakov veto algorithm is
that after a change in $\nB$, the algorithm continues down in $v$,
allowing subsequent further changes in $\nB$.
It only stops when it reaches $v_{\min}$.
If using a fixed coupling inside Algorithm~\ref{alg:sub-to-int},
$v_{\min}$ does not necessarily have to be connected with the shower
cutoff, rather it is akin to a technical
cutoff in standard subtraction procedures.

We can demonstrate Eq.~\eqref{eq:nB-result} as follows.
The probability that the algorithm will have triggered
step~\ref{alg:sub-to-int:lnv} in a specific $d\ln v$ window is given by
$\rho(v) d\ln v$. 
Given the $\ln v$ value and $\Phi_2$ phase-space point, the algorithm
will increment or decrement $\nB$ with conditional probabilities
$P_{+}$ or $P_{-}$ respectively
\begin{subequations}
  \begin{align}
    &\text{if $R(\Phi)>C(\Phi)$, increment $\nB$ with probability}\; P_{+} = \frac{R(\Phi)-C(\Phi)}{M(\Phi)}\,,
    \\
    &\text{if $R(\Phi)<C(\Phi)$, decrement $\nB$ with probability}\; P_{-} = \frac{C(\Phi)-R(\Phi)}{M(\Phi)}\,,
  \end{align}
\end{subequations}
or otherwise leave $\nB$ unchanged.
Writing out the integrals for $\rho(v)$, this then gives the following
result for the average of $\nB$,
\begin{align}
  \label{eq:nB-proof}
  \langle \nB \rangle
  &= 1 + 
    \int \frac{dv}{v} d\Phi_2   J\,
    \frac{M(\Phi)}{B_0(\PhiB)}\left( P_{+}\Theta(P_{+}) \,-\, P_{-}\Theta(P_{-}) \right)\,,
\end{align}
which simplifies exactly to Eq.~(\ref{eq:nB-result}).

Algorithm~\ref{alg:sub-to-int} always gives an integer as its
output.
It is particularly simple to analyse if $R-C$ always has the same
sign.
For example if we always have $R-C>0$, then the probability distribution
for $\nB$ is exactly given by a Poisson distribution, i.e.\
$P(\nB) = e^{-r} r^{\nB-1}/(\nB-1)!$ for $\nB\ge 1$, from which it is
clear to see again that $\langle \nB\rangle = 1 + r$.
We note that the use of a Poisson process to obtain a matching weight
has been discussed before, notably in the context of NLO
merging~\cite{Lavesson:2008ah,Lonnblad:2012ix}.\footnote{We thank a
  referee for bringing this to our attention.}

In the general case, $R-C$ may sometimes be positive, sometimes
negative.
Then, all but a fraction $\order{\as}$ of the time, the integer
that is returned is $\nB=1$, a consequence of the fact that $r$ 
in Eq.~(\ref{eq:nB-result}) is of order $\as$.
%
%
A fraction $\order{\as}$ of the time, the integer will be $\nB=0$ or
$\nB=2$.
A fraction $\order{\as^2}$ of the time, the integer will be $\nB=-1$ or
$\nB=3$, and so forth.
Thus, if we are interested just in NLO accuracy, we can discard any
events with $\nB < 0$.
Similarly, we are free to replace $\nB \to \min(\nB,p)$, where $p$ is
some integer $p \ge 2$.
Assuming we know how to generate unweighted Born events,
(positive-definite) unweighted NLO events can then simply be obtained
by enhancing the Born event generation cross section by a factor $p$
and then accepting any given Born event with probability $\nB/p$.

There is considerable freedom in adapting
Algorithm~\ref{alg:sub-to-int} according to one's needs.
Below, in Section~\ref{sec:ESME-core}, we will present a variant that
incorporates the real event generation into the same loop, and is NLO
accurate and relatively fast.
Here we comment briefly on the scope for designing an algorithm that
is positive-definite, bounded and that reproduces
Eq.~(\ref{eq:nB-result}) up to and including relative order $\as^m$
for any choice of positive integer $m$.
The adaptation is remarkably simple: one simply multiplies the density
$\rho(v)$ in the Sudakov by $m$, and increments or
decrements $\nB$ in steps \ref{alg:sub-to-int:plus} and
\ref{alg:sub-to-int:minus} by $1/m$ rather than $1$.
Without any bounds on $\nB$, one still reproduces
Eq.~(\ref{eq:nB-result}) exactly.
With a positivity bound $\nB\ge 0$, at least $m+1$ decrement steps are
needed to trigger the bound, i.e.\ the bound affects the results
starting only at order $\as^{m+1}$.
Analogously with an upper bound $p$ (with $p$ at least $2$).
A detailed mathematical analysis confirming these points is given
below in Section~\ref{sec:ESME-general-anl}. 

There are also various potential adaptations concerning the speed of the
algorithm, i.e.\ essentially the number of times one must evaluate
$R(\Phi)$.
In particular, to reproduce Eq.~(\ref{eq:nB-result}) up to and
including order $\as^m$, one must allow the algorithm to go through
steps \ref{alg:sub-to-int:lnv}--\ref{alg:sub-to-int:minus} at least
$m$ times.
However, after $m$ steps have taken place, there is freedom to simply
exit the algorithm even if $v > v_{\min}$.
In general we expect a (modest) speed penalty in going to higher $m$,
due to the higher Sudakov density and the larger number of steps that
must be carried out before one is allowed to exit the algorithm.

A final comment concerns $v_{\min}$.
One option is to take $v_{\min}$ to be the same as in
Eq.~(\ref{eq:multiplicative}), i.e.\ the scale of the non-perturbative
shower cutoff.
Another possibility is take the limit $v_{\min} \to 0$, which is an
option as long as the $R(\Phi)$ and $C(\Phi)$ use a fixed-order
expansion of the coupling (and PDF), not an all-order form with its Landau pole.
If one wishes to obtain exact agreement with a standard NLO
calculation, it is the $v_{\min} \to 0$ option that must be used, with both the
coupling and any PDF factors being evaluated at the renormalisation
and factorisation scales of the hard process.
The difference between the two options will generally involve a
non-perturbative power correction associated with the scale
$v_{\min} \sim \lambdaQCD$, as well as beyond-NLO contributions
associated with the different possible choices for the running of the
coupling and PDF.

We refer to procedures in the family of Algorithm~\ref{alg:sub-to-int}
as ``Exponentiated Subtraction'' and their use for matching showers
with fixed-order calculations as ``Exponentiated Subtraction for
Matching Events'' (\ESME). 

\subsubsection{Mathematical analysis of exponentiated subtraction}
\label{sec:ESME-general-anl}
Let us define
\begin{subequations}
  \label{eq:r-plus-minus}
  \begin{align}
    r_+
    &\equiv
      +\int \frac{R(\Phi) - C(\Phi)}{B_0(\PhiB)}\Theta(R(\Phi)>C(\Phi)) \, d\Phi_\text{rad}\,,
    \\
    r_-
    &\equiv
      -\int \frac{R(\Phi) - C(\Phi)}{B_0(\PhiB)}\Theta(R(\Phi)<C(\Phi)) \, d\Phi_\text{rad}\,, 
  \end{align}
\end{subequations}
such that $r$ in Eq.~(\ref{eq:nB-result}) is given by $r = r_+ - r_-$,
with both $r_+$ and $r_-$ of order
$\as$, by construction.
Let us separately count the number of increments that occur, $n_+$ and
the number of decrements, $n_-$.
The final $\nB$ will be given by $\nB = 1 + n_+ - n_-$.
The probability of a given number of increments is given by a Poisson
distribution
\begin{equation}
  \label{eq:nplus}
  P_+(n_+) = e^{-r_+} \frac{(r_+)^{n_+}}{n_+!}\,,
\end{equation}
and similarly for the number of decrements,
\begin{equation}
  \label{eq:nminu}
  P_-(n_-) = e^{-r_-} \frac{(r_-)^{n_-}}{n_-!}\,.
\end{equation}
With no bounds on the number of increments or decrements, we have
\begin{equation}
  \label{eq:nBavg}
  \langle n_\pm\rangle = r_\pm\,,
  \qquad
  \langle \nB\rangle = 1 + \langle n_+\rangle \,-\, \langle n_-\rangle
  = 1 + r\,,
\end{equation}
as expected.
Next let us consider the probability $P(\DeltaN)$ to obtain a given value of
$\DeltaN \equiv n_+ - n_-$ (with $\nB = 1 + \DeltaN$),
\begin{subequations}
  \begin{align}
    P(\DeltaN)
    &= \sum_{n_+ = +\DeltaN}^\infty P_+(n_+) P_{-}(n_+ -
    \DeltaN)\,, \qquad  \DeltaN \ge 0\,,
    \\
    P(\DeltaN)
    &= \sum_{n_- = -\DeltaN}^\infty P_-(n_-) P_{+}(n_- +
    \DeltaN)\,, \qquad  \DeltaN \le 0\,.
  \end{align}
\end{subequations}
The sums can be evaluated to yield
\logbook{}{maths/ESME-foundation-maths.nb}
\begin{equation}
  \label{eq:P-deltan-result}
    P(\DeltaN)
    =
    e^{-(r_+ + r_-)} \left(\frac{r_+}{r_-}\right)^{\DeltaN/2} I_{|\DeltaN|}(2\sqrt{r_+r_-})\,,
\end{equation}
where $I_n(x)$ is the modified Bessel function of the first kind,
whose series expansion is $I_n(2x) = \sum_{k=0}
\frac{x^{n+2k}}{k!(n+k)!}$. 
%
%
 From the first term in that series it is evident that
Eq.~(\ref{eq:P-deltan-result}) is consistent with the expectation that
$P(\DeltaN) \sim \as^{|\DeltaN|}$,
recalling that the $r_\pm$ are of order $\as$.
Furthermore, one sees that
$P(\DeltaN) = P_\pm(\DeltaN) + \order{\as^{|\DeltaN+1|}}$ for
$\DeltaN \gtrless 0$, which is as expected given that the lowest order
route to achieve a given value of $\DeltaN$ involves only increments
($\DeltaN > 0$), or only decrements ($\DeltaN < 0$).

Now let us suppose that we truncate $\DeltaN$ in such a way
that when $\DeltaN > \DeltaN_{\max}$, we reset it to
$\DeltaN_{\max}$ and when $\DeltaN < -\DeltaN_{\max}$ we reset it to
$-\DeltaN_{\max}$.
Thus if $\DeltaN_{\max}=1$, this is like bounding $0 \le \nB \le 2$,
i.e.\ whenever $\nB>2$ after completing
Algorithm~\ref{alg:sub-to-int}, $\nB$ is reset to $2$, and whenever
$\nB<0$ after completing Algorithm~\ref{alg:sub-to-int} $\nB$ is reset
to $0$. 
The average bounded $\nB$ is given by
\begin{equation}
  \label{eq:average-bounded-nB-def}
  \langle \nB\rangle_{\DeltaN_{\max}} \equiv
  1
  + \sum_{\DeltaN = 0}^{ \DeltaN_{\max}} \DeltaN [P(\DeltaN) - P(-\DeltaN)]
  \;+\!\!\! \sum_{\DeltaN = \DeltaN_{\max}+1}^{\infty} \!\!\!\!\!\DeltaN_{\max} [P(\DeltaN) - P(-\DeltaN)]\,.
\end{equation}
We have not been able to evaluate this in general, however it is
possible to establish the structure of the first non-trivial terms,
\logbook{}{%
  We see this explicitly in the mathematica file up to
  $\DeltaN_{\max}=5$.
  In terms of an analytic derivation, the first deviation from the
  exact result is when we get $|\DeltaN| = \DeltaN_{\max}+1$, where the
  mismatch $\DeltaN - \DeltaN_{\max} = \mp 1$ and the probability of
  that $\DeltaN$ is $P_{\pm}(|\DeltaN_{\max}|+1)$, which gives the
  result that is shown.
}
\begin{equation}
  \label{eq:average-bounded-nB-expanded}
  \langle \nB\rangle_{\DeltaN_{\max}}
  = 1 + (r_+ - r_-)
  - \frac{1}{(\DeltaN_{\max} + 1)!} \left(r_+^{\DeltaN_{\max}+1} - r_-^{\DeltaN_{\max}+1}\right)
  + \order{r_{\pm}^{\DeltaN_{\max}+2}}\,.
\end{equation}
Thus if $r_+$ and $r_-$ are both of order $\as$, free of any
logarithmic enhancements, the difference between
$\langle \nB\rangle_{\DeltaN_{\max}}$ and the unbounded
$\langle \nB\rangle$ becomes of order $\as^{\DeltaN_{\max}+1}$, as
expected, and the coefficient multiplying that difference is
factorially suppressed.

Finally, let us consider the variant of the algorithm that is designed
to guarantee positive-definite results with control of terms up to and
including $\as^m$.
As discussed above, one multiplies $\rho(v)$ by a positive integer
$m$, which introduces an extra factor of $m$ in front of every
occurrence of $r_\pm$ above.
Additionally, one defines 
\begin{equation}
  \label{eq:alg1-multm}
  \nB^{(m)} = 1 + \frac{\DeltaN}{m}\,,
\end{equation}
and takes $\DeltaN_{\max} = m$.
That yields
\begin{equation}
  \label{eq:average-bounded-nB-def-mvariant}
  \langle \nB^{(m)}\rangle_{\DeltaN_{\max} = m}
  = 1 + (r_+ - r_-)
  - \frac{m^m}{(m + 1)!} \left(r_+^{m+1} - r_-^{m+1}\right)
  + \order{r_{\pm}^{m+2}}\,.
\end{equation}
If we consider $m=3$, which would be sufficient to eliminate spurious
terms up to and including N$^3$LO, the coefficient multiplying
$\left(r_+^{m+1} - r_-^{m+1}\right)$ is $27/24$, which is comfortably
close to $1$.

A final comment is that it is important that $r_+$ and $r_-$ are
separately of order $\as$, free of logarithmic enhancements, i.e.\ one
should not rely on large cancellations between $r_+$ and $r_-$, or
between their difference and $C_\text{int}$ in
Eq.~(\ref{eq:Bbar-with-counterterm}).
In general we expect that this condition will be satisfied for
single-scale problems and local subtraction schemes that avoid
restrictive phase space cuts in the counterterm.
For multi-scale problems, instead, the dependence on $\DeltaN_{\max}$
or $m$ would provide a measure of potential higher order terms that is
distinct from, and thus complementary to, scale variation.
It would be interesting to further explore this in future work,
keeping in mind that multi-scale problems anyway need a dedicated
treatment, for example as in the \MiNLO\ approach~\cite{Hamilton:2012np,Hamilton:2013fea}.

\subsection{An \ESME algorithm with joint reals and subtractions}
\label{sec:ESME-core}

Here, we adapt the algorithm of Section~\ref{sec:Bbar-integer-alg} not
only to ensure NLO accuracy with positive-definite weights, but to
organise it such that the effective $\Bbar$ evaluation and the
real-emission generation share evaluations of the real matrix element.
This helps reduce the total number of real matrix-element and
associated PDF evaluations, and so can contribute to faster NLO event
generation.
It should be seen as just one among many possible algorithms founded
on the principles of Section~\ref{sec:Bbar-integer-alg}.

One consideration is that in Section~\ref{sec:Bbar-integer-alg} we
left free the details of how to incorporate the $V(\PhiB) +
C_\text{int}(\PhiB)$ contributions in
Eqs.~(\ref{eq:Bbar-with-counterterm})~and~\eqref{eq:Cint}, while here
we will give specific prescriptions.
The starting point of our method is that we will generate Born events
with a weight $\Bbar_C(\PhiB)$, defined as
\begin{equation}
  \label{eq:Bbar-C}
  \Bbar_C(\PhiB) = B_0(\PhiB) + 
  V(\PhiB) +    
    C_\text{int}(\PhiB)\,.
\end{equation}
Standard approaches instead generate the Born events with weight
$\Bbar(\PhiB)$.
The key difference is that $\Bbar_C(\PhiB)$ does not involve the
Monte Carlo integral over $d\Phi_\text{rad}$ in
Eq.~(\ref{eq:Bbar-with-counterterm}).
This has a potential practical advantage, namely that to obtain the
weight for any given Born configuration, one does not need a separate
explicit integration over the real phase space. 
However, it obviously misses part of the overall $\Bbar$
normalisation.
We will recover the normalisation through the use of two non-unitary
streams of events, which will account for the
$\int d\Phi_\text{rad} [R(\Phi) - C(\Phi)]$ contribution to $\Bbar$.
Physically, if $R(\Phi) < C(\Phi)$, then $\Bbar_C(\PhiB)$ is too large
and we need to eliminate some of the events generated with weight
$\Bbar_C(\PhiB)$ (stream 1).
If instead $R(\Phi) > C(\Phi)$, this implies that $\Bbar_C(\PhiB)$
is too small and we need an extra source of events (stream 2).

Each of the two streams will effectively account for specific parts of
Algorithm~\ref{alg:sub-to-int}, which, we recall, precisely evaluates
$\int d\Phi_\text{rad} [R(\Phi) - C(\Phi)]$. 
The algorithm for \ref{alg:born+nlo-rej} will address the situations
in Algorithm~\ref{alg:sub-to-int} where $\nB=1$ or $\nB=0$, i.e.\ 
$\nB$ is unchanged or decremented.
Conversely, the algorithm for \ref{alg:nlo-add} will provide an
additional source of events to account for the situations in
Algorithm~\ref{alg:sub-to-int} where $\nB$ is incremented.
In other words, stream 1 will discard a fraction $\order{\as}$ of
events relative to the Born rate, while stream 2 will add a fraction
$\order{\as}$ of events.
The sum of the two streams will also generate the hardest emission in such a way
as to produce the correct real matrix element.

\renewcommand{\thealgorithm}{Stream~1}
\begin{algorithm}
  \caption{(ESME) Born + NLO rejection}
  \label{alg:born+nlo-rej}
\begin{algorithmic}[1]
  \State{Generate Born event according to $\Bbar_C$ distribution and
    set $v=v_{\max}$}
  \While{$v > v_\text{min}$}
  \State generate next $v$ and
  $\Phi_2$ according to Sudakov with density $\rho(v) d\ln v$,
  Eq.~(\ref{eq:Bbar-Sudakov})%
  \State generate random number $0<r<1$
  \If{$C(\Phi) > R(\Phi)$}
    \State \textbf{if} $r > C(\Phi)/M(\Phi)$:  veto emission
     \State \textbf{else if} $r > R(\Phi)/M(\Phi)$:  \textbf{return}  \reject{reject event}
    \label{alg:born+nlo-rej:reject}
    \State \textbf{else}: accept emission and \textbf{return} \accept{continue shower, accept event}
    \label{alg:born+nlo-rej:accept-emsn}
  \Else
    \State \textbf{if} $r > C(\Phi)/M(\Phi)$: veto emission
    \State \textbf{else}: accept emission and \textbf{return} \accept{continue shower, accept event}
    \label{alg:born+nlo-rej:accept-emsn-C}
    \EndIf
    \EndWhile
 \State \textbf{return}  \accept{accept event}
\end{algorithmic}
\end{algorithm}

Let us first look at the
algorithm for~\ref{alg:born+nlo-rej}. 
Step~\ref{alg:born+nlo-rej:reject} is a critical part of the algorithm,
because it is the only step that is non-unitary.
Specifically, it rejects the event with probability
$[C(\Phi)-R(\Phi)]/M(\Phi)$.
It is the direct analogue of step~\ref{alg:sub-to-int:minus} of
Algorithm~\ref{alg:sub-to-int}, which decrements $\nB$.
In standard NLO approaches for evaluating
Eq.~(\ref{eq:Bbar-with-counterterm}), such regions with $C>R$ would be
associated with a risk of negative-weight events.
Because we account for that region through a rejection mechanism,
that danger does not arise here.
Aside from that, the $C>R$ branch is very much the standard Sudakov
veto algorithm, accepting the emission with probability
$R(\Phi)/M(\Phi)$ in step \ref{alg:born+nlo-rej:accept-emsn}r.
In the other branch, $C\le R$, the stream 1 algorithm deviates from
the standard Sudakov algorithm, because the emission is accepted with
probability $C(\Phi)/M(\Phi)$ rather than $R(\Phi)/M(\Phi)$.
The missing difference $R(\Phi) - C(\Phi)$, which connects with the
increment of $\nB$ in step~\ref{alg:sub-to-int:plus} of
Algorithm~\ref{alg:sub-to-int}, will be accounted for in the algorithm
for \ref{alg:nlo-add}. 
\renewcommand{\thealgorithm}{Stream~2}
\begin{algorithm}[!h]
  \caption{(ESME) NLO addition}
  \label{alg:nlo-add}
\begin{algorithmic}[1]
  \State{Generate Born event according to $B_0$ (or $\Bbar_C$) distribution and
    set $v=v_{\max}$}
  \While{$ v >  v_\text{min}$}
  \State generate next $v$ and
  $\Phi_2$ according to Sudakov with density $\rho(v) d\ln v$,
  Eq.~(\ref{eq:Bbar-Sudakov})%
  \State generate random number $0<r<1$
  \If{$C(\Phi) > R(\Phi)$}
    \State \textbf{if} $r > R(\Phi)/M(\Phi)$:  veto emission
    \State \textbf{else}:  \textbf{return}  \reject{reject event}
      \label{alg:nlo-add:rejCgtR}
  \Else
    \State \textbf{if} $r > R(\Phi)/M(\Phi)$: veto emission
    \State \textbf{else if} $r > C(\Phi)/M(\Phi)$: accept emsn,
    \textbf{return} \accept{continue shower, accept event}
    \label{alg:nlo-add:accept}
    \State \textbf{else}:  \textbf{return}  \reject{reject event}
      \label{alg:nlo-add:rejRgtC}
    \EndIf
  \EndWhile
  \State \textbf{return}  \reject{reject event}
\end{algorithmic}
\end{algorithm}

\begin{figure}
  \centering
  \includegraphics{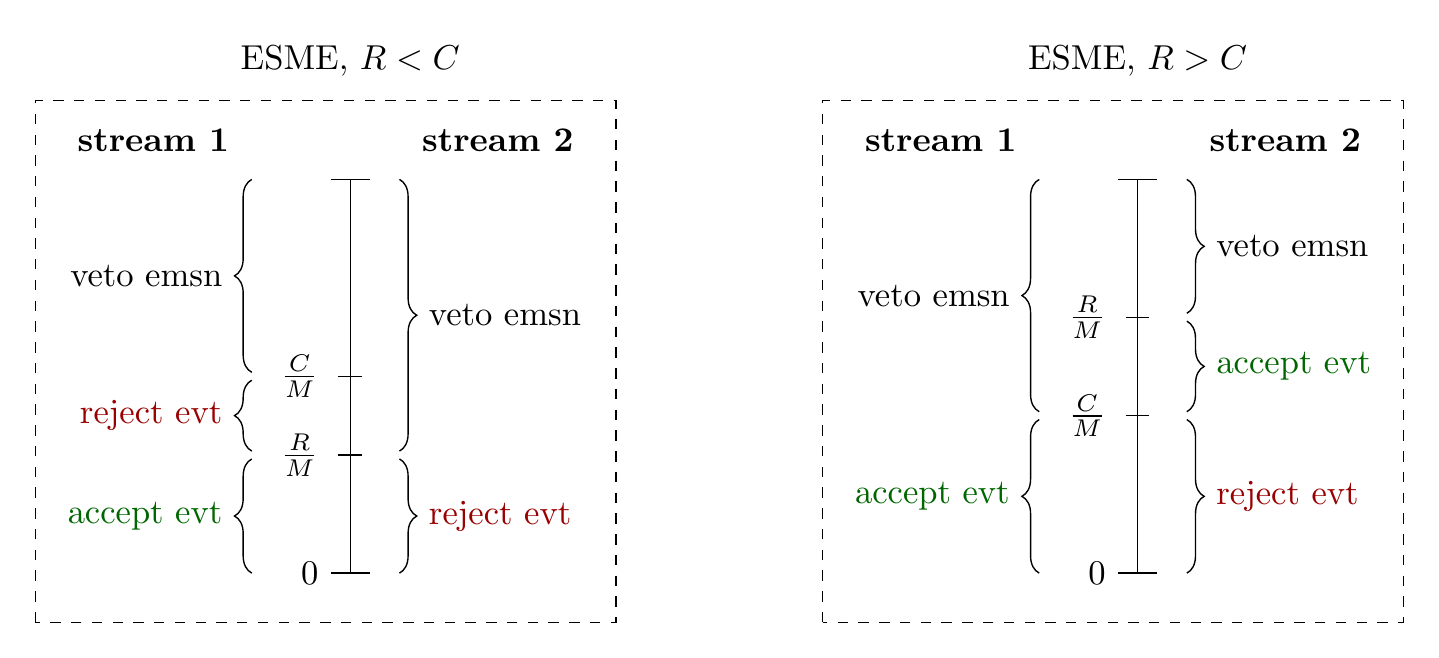}
  \caption{Simple illustration of the different possible actions in
    the two streams of the \ESME algorithm with joint reals and
    subtractions.
    The actions are shown separately for the cases $R(\Phi)<C(\Phi)$ (left)
    and $R(\Phi)>C(\Phi)$ (right).
    In each case, when summing the two streams, one sees that the
    ``accept evt'' action occurs with total weight $R/M$.
    One can also verify that the contribution to the total event rate
    change relative to the $\Bbar_C$ normalisation is $(R-C)/M$.
    Recall that the default action in stream 1 (2) is to accept
    (reject) the event if the shower scale reaches $v_{\min}$ --- only
    when the action is different from the stream's default is the
    total event rate affected.
  }
  \label{fig:stream-cartoon}
\end{figure}

Specifically, stream~2's step~\ref{alg:nlo-add:accept} occurs
with probability proportional to $R(\Phi)-C(\Phi)$.
It also generates a real emission, compensating the missing
contribution for real emissions in stream~1's step
\ref{alg:born+nlo-rej:accept-emsn-C}.
Stream~2's step~\ref{alg:nlo-add:accept} is also the only
step that leads to an \emph{event} being accepted in that stream.
To order $\as$, the corresponding probability,
$[R(\Phi)-C(\Phi)]/M(\Phi)$ (when $R>C$) exactly matches the
probability for incrementing $\nB$ in step~\ref{alg:sub-to-int:plus}
of Algorithm~\ref{alg:sub-to-int}.
The behaviour of the overall algorithm is illustrated also in
Fig.~\ref{fig:stream-cartoon}. 

The combination of the two streams in reminiscent in some ways of the
MAcNLOPS method~\cite{Nason:2021xke}, while stream 1 alone is
similarly reminiscent of KrKNLO~\cite{Jadach:2015mza,Delorme:2025teo}.
But, where those references aimed to eliminate negative-weight events
when trying to obtain the correct real part of the showering, here our
intention is to also address difficulties that arise with the overall
normalisation.

A further comment is that each stream exits the main matching loop as
soon as a non-trivial action has taken place (i.e.\ reject event, or accept
the emission and continue normal showering of the accepted event).
If $M(\Phi)$ is chosen carefully enough, i.e.\ to be of order $\as$,
then there is an $\order{1}$ probability of exiting the loop at each
stage, leading to an $\order{1}$ total number iterations around the
loop.
This is to be contrasted with the default formulation of Algorithm
\ref{alg:sub-to-int}, which would typically require a number of steps
proportional to $\ln^2 v_{\max}/v_{\min}$, most of which would bring
no action because $|R-C| \ll M$ for small $v$.

Overall the above approach addresses the \textit{second} source of
negative events discussed in Section~\ref{sec:neg-event-sources},
i.e.\ the one associated with the NLO normalisation, as well as
potentially speeding up the generation by not requiring evaluation of
the $d\Phirad (R-C)$ integral directly in the Born.
The \textit{third} source of negative events (associated with the
generation of real radiation) was already implicitly addressed by our
use of multiplicative matching, cf.~Section~\ref{sec:real-radiation}.
There remains a potential for the \textit{first} source of negative events,
i.e.\ there could be Born phase space regions where
$\Bbar_C(\PhiB) < 0$ because of large negative NLO contributions.
Contrary to the other two sources, in this case there is often a clear
physical reason, notably due to the breakdown of perturbative
convergence for standard fixed-order calculations in the presence of
disparate scales.
In any region where this occurs, plain NLO cross sections are
anyway devoid of physical meaning, and we are therefore free to
reorganise the perturbative series so as to make it positive definite.
In particular, defining $B_0$ and $\as B_{C,1}$ as the LO and NLO
contributions to $\Bbar_C$, we have the freedom to use
\begin{equation}
  \label{eq:BbarC-posdef}
  \Bbar_C(\PhiB) = B_0 [1 + f(\as B_{C,1}/B_0)],
\end{equation}
where $f(x)$ is any function that satisfies $f(x) = x + \order{x^2}$.
If we additionally choose $f(x)$ such that $f(x) > -1$ for all $x$,
then we will guarantee the total absence of negative-weight events.
In practice we take 
\begin{equation}
f(x) =
\begin{cases}
x & \quad \mbox{ for } x \geq 0 \\
\tanh x & \quad \mbox{ for } x<0\, ,
\end{cases}
\end{equation}
so that spurious higher-order contributions start only
at relative order $\as^3$ and large positive $K$ factors 
are not modified.\footnote{
  It is natural to ask whether one could use such an approach at the
  level of the \emph{integrand} in standard NLO matching.
  While we cannot rule out that it might work, it is potentially more
  delicate.
  For example taking $f(x) = \max(-1,x)$ and integrating $\int_0^1 dv
  [1 +f(- \as v^{-2/3})]$ yields $1-3\as + 2 \as^{3/2}$.
  \logbook{4d23e1f71}{see maths/higher-order-reduction-in-integrand.nb}%
  This is \emph{not} correct, because of the $\as^{3/2}$ term,
  which is parametrically larger than a NNLO contribution.
  To what extent such an issue would arise in practice depends
  critically on the adaptive phase space generation for $d\Phirad$.  }

It is important to be aware that some aspects of the algorithm as
formulated in this section still induce spurious $\as^2$
contributions.
Some of these can be mitigated: for example cross terms between the
$\Bbar_C$ normalisation and the stream 2 additions can be eliminated by
generating stream 2 with a weight $B_0$.
In stream 1, there are $\as^2$ effects from the product of
$\as B_{C,1}$ and the order $\as$ event-rejection probability, which
can be eliminated by renormalising that probability with a suitable
factor.\footnote{%
  Specifically, if $\Bbar_C/B_0 > 1$, then the event-rejection
  probability, $[R(\Phi)-C(\Phi)]/M(\Phi)$ that is used in step~\ref{alg:born+nlo-rej:reject} of
  \ref{alg:born+nlo-rej} is divided by $\Bbar_C/B_0$.
  Otherwise it is multiplied by $2 - \Bbar_C/B_0 = 1 - \as B_{C,1}/B_0$.
}
These mitigation strategies are included by default in our
  implementation, and in the phenomenological results
  presented in Section~\ref{sec:pheno+perf}. Yet other spurious
second-order contributions are intrinsically associated with the
structure of the algorithm.
Their elimination would require adaptation of some of the techniques
discussed at the end of Section~\ref{sec:Bbar-integer-alg}.
We will briefly discuss the size of these terms below, in
Section~\ref{sec:NLO-tests}.

The algorithms for each of the two streams both contain a $v_{\min}$
cutoff, as in Eq.~(\ref{eq:dsigma-basic-in-esme-section}) and
Algorithm~\ref{alg:sub-to-int}.
Given the interplay with real radiation, here $v_{\min}$ is always
taken to be the shower non-perturbative cutoff.
Technically that could introduce non-perturbative power corrections,
however we have not found any evidence of these being numerically
significant.

A final comment is that there is freedom also to replace stream 2 with
a direct generation of real radiation events in proportion to
$d\Phi (R-C) \Theta(R-C)$.
For now we have not explicitly explored this option because of the
need for an additional warm-up phase in order to efficiently sample
the corresponding phase space.

\subsection{Counterterm from slicing}
\label{sec:counterterm-from-slicing}

When combining multiplicative matching of
Section~\ref{sec:real-radiation} and the \ESME treatment of
Section~\ref{sec:ESME-core} for NLO normalisation, it is convenient to
have a counterterm $C(\Phi)$ that can be easily represented in the
same $v$, $\bar \eta$ and $\phi$ variables as used for the shower.
As in the preceding sections, there is considerable freedom in how to
approach this.
Here we outline the specific route we have taken, highlighting the
ability to relate a slicing calculation to a shower-based subtraction
approach, which we dub a slice-to-subtraction approach.
In this subsection, to illustrate the approach, we focus on
$e^+e^- \to \text{2 jets}$ and the \PanGlobal shower, with other 2-leg
processes discussed in Appendix~\ref{sec:S2S-incoming-hadrons}.

The starting point will be a slicing calculation of the NLO rate for
producing a given Born configuration with a tight constraint on any
additional radiation.
At NLO, conversion between calculations for different slicing
variables is relatively straightforward and we will choose a slicing
variable that coincides with the shower ordering variable $v$
everywhere in phase space in the limit of small $v$.
Specifically, for $e^+e^- \to q\bar q$ and the \PanGlobal shower
with $\betaps=0$, the NLO normalisation with a slicing
constraint $v < e^{-|L|}$, is given by
\begin{equation}
  \label{eq:Sigma-v}
  \Bbar_\text{PG}(v < e^{-|L|}) = B_0(\PhiB) \left[ 1 -
    \frac{\alpha_s C_F}{2\pi}\left(4 L^2 + 6 L  +
      \frac{\pi^2}{3} + 2
    \right)\right]\,,
\end{equation}
valid for large and negative $L$. This result was obtained by adapting 
the calculation presented in \cite{Medves:2022ccw}.

In general, it is possible to convert a slicing calculation into a
subtraction calculation by constructing a counterterm $C(\Phi)$ that
has the full QCD behaviour in the soft and/or collinear regions and can in
practice be integrated over the full phase space above some arbitrary
small $v$.
Normally this is done with a counterterm that lives in the actual
real-radiation phase space.
However, one is free to choose a counterterm that is non-zero even for
values of shower generation 
variables that do not map to valid phase space regions, e.g.\ if 
this facilitates the integration of the counterterm.\footnote{For the
  specific case of the \PanGlobal final-state shower that we use as an
  illustration here, the counterterm phase space will actually
  coincide exactly with the genuine real phase space.}
Expressed in terms of the shower phase-space generation variables $\ln
v$ and $\bar\eta$ (cf.\ Appendix~\ref{app:summary-panscales}), we
take the counterterm to be
\begin{align}
   \frac{C(\Phi)}{B_0(\PhiB)} d\Phirad \to \frac{dv}{v} d\bar \eta
  \frac{d\phi}{2\pi} 
  \frac{\as}{\pi}  z P_{gq}(z),
  \quad
  \ln z = \bar\eta - \bar\eta_{\max},
  \quad
  0 < \bar\eta < \bar\eta_{\max} = \ln Q/v,
\end{align}
{where $Q$ is the total centre-of-mass energy,}
for the quark and similarly for the anti-quark.
Here $P_{gq}(z)=C_F(1+(1-z)^2)/z$ is the usual LO splitting function.
The integration above ${Q}e^{-|L|}$ is simple and, after a sum over
(half) dipoles, gives 
\begin{equation}
  \label{eq:Cint-PG}
  \frac{C_\text{int}(v > {Q}e^{-|L|})}{B_0(\PhiB)}
  =
  2 \int^{{Q}}_{{Q} e^{-|L|}} \frac{dv}{v}
  \int_{0}^{\ln {Q}/v} d\bar \eta
  \int_0^{2\pi} \frac{d\phi}{2\pi} \frac{\as}{\pi}  z P_{gq}(z)
  = 
  \frac{\alpha_s C_F}{2\pi}\left( 4 L^2 + 6L + 7\right),
\end{equation}
for large $|L|$. 
From this, we can work out $\Bbar_C$ as in Eq.~(\ref{eq:Bbar-C}) with
this specific counterterm,
\begin{align}
  \Bbar_C
  =
  \Sigma_\text{PG}^\text{NLO}(v <{Q} e^{-|L|}) + C_\text{int}(v > {Q} e^{-|L|})
  \label{eq:BbarC-res}
  =
  B_0(\PhiB) \left[ 1
    + \frac{\alpha_s C_F}{2\pi}\left(5 - \frac{\pi^2}{3}\right)\right]\,.
\end{align}

\section{NLO and NNDL tests}
\label{sec:NLO-NNDL}
In this section we validate the fixed-order and the logarithmic accuracy of our matched predictions. In particular, in Sec.~\ref{sec:NLO-tests} we show that the $\mathcal{O}(\alpha_s)$ expansion of such predictions reproduces NLO calculations.
In Sec.~\ref{sec:NNDL-tests} we instead demonstrate that we achieve NNDL accuracy for a series of continuously global observables.
\subsection{NLO tests}
\label{sec:NLO-tests}

We start by testing the relative $\mathcal{O}(\alpha_s)$ accuracy of the matched shower algorithms.
For each process and observable, we first carry out a comparison to a
standard NLO calculation with phenomenological settings for the
coupling and shower parameters.
For generic observables, NLO-matched showers give predictions that differ from the pure NLO result, because of higher-order differences that can come from e.g. the Sudakov form factor and its intrinsic connection to the
momentum mapping during showering. 
In the \esme method, an additional source of higher-order differences comes from the elimination of negative event weights.
Given that, in general, matched showers will not exactly agree with a pure NLO calculation, it is important to check that any differences are
genuinely of higher order and not a small mistake in the NLO
coefficient.
Accordingly, we also explore the $\as \to 0$ limit of the NLO
matched shower to isolate the pure NLO coefficient in the shower and
thus conclusively establish the shower's NLO accuracy in any given
matching scheme.
This procedure is inspired by the standard \PanScales approach for
testing logarithmic accuracy, but we believe that it is the first time
that it has been applied to tests of NLO shower matching.

\subsubsection{$e^+e^-$ collisions}

A non-trivial observable with which to test the NLO accuracy for the
$e^+e^- \to \gamma^*\to q \bar q$ process is 
the polar angle of the thrust axis, $\cos \theta_T$.
At NLO its differential distribution is given analytically
by~\cite{Lampe:1992au,Mateu:2013gya}
\begin{align}
  \label{eq:ee-thrust-angle}
	\frac{d\sigma_\text{an}}{d\cos \theta_T} &= \sigma_0 \left[ \frac{3}{8}\left(
	1+\cos^2 \theta_T \right) R_U + \left( 1-3\cos^2 \theta_T \right) R_L
	\right]\,,  \\
	R_U &= 1 + \frac{\alpha_s}{\pi} \frac{3 C_F}{4} +
	\mathcal{O}(\alpha_s^2)\,,\qquad
	R_L = \frac{\alpha_s}{\pi} \frac{3 C_F}{8} \left( 8 \ln \frac{3}{2} - 3
	\right) + \mathcal{O}(\alpha_s^2)\,,
\end{align}
with $\sigma_0$ the inclusive Born cross section.
Note that showers (including the \panscales showers) typically do not
preserve the thrust axis, even in the presence of just one emission.
Therefore the tests will verify both the effective $\Bbar$ for the
Born configuration, as well as the structure of the real radiation.

\begin{figure}
	\centering
	\includegraphics[width=0.48\textwidth,page=1]{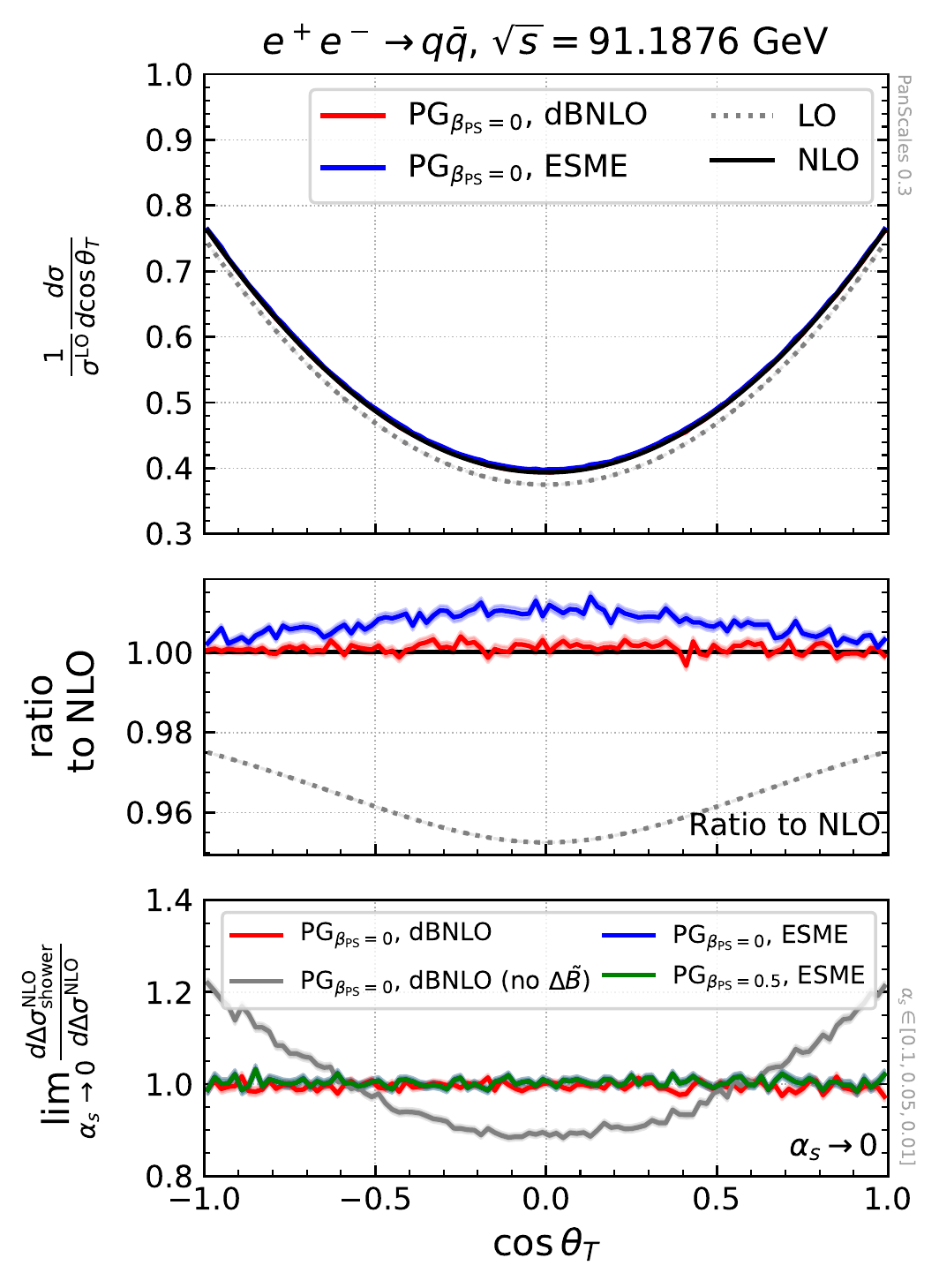}
	\caption{Tests of NLO-matched showers, showing the oriented thrust axis
          distribution in $e^+e^- \to \gamma^* \to q\bar q$ collisions.
          The top panel shows the ratio to the total Born cross
          section for a phenomenological setup with $\sqrt{s} =
          91.1876\GeV$, $\as(\sqrt{s}) = 0.118$ and a showering cutoff
          of $0.5\GeV$. 
          The middle panel shows the ratio to the differential NLO
          cross section with the same settings for $\sqrt{s}$ and $\as$.
          The bottom panel shows the ratio of the pure NLO coefficient
          in the matched shower to the known exact NLO coefficient,
          i.e.\ the ratio in Eq.~(\ref{eq:alphas-nlo-coeff}). 
        }
	\label{fig:ee-NLO-tests-thrust-axis}
\end{figure}

In Fig.~\ref{fig:ee-NLO-tests-thrust-axis}, we show fixed-order tests of the
$\cos \theta_T$ distribution.
In the top panel, we show NLO results with the \PanGlobal
$\betaps = 0$ shower, for a phenomenological setup with
$\sqrt{s} = 91.1876\GeV$, with the \dBNLO and \esme methods, using
only $\gamma^*$ exchange.
We compare it to the analytical LO and NLO results from
Eq.~(\ref{eq:ee-thrust-angle}).
The middle panel shows the ratio to the total
NLO result.
Qualitatively the shower is similar to the NLO result, both in
normalisation and shape, though there is about a $1\%$ offset in the
\esme method.
The \dBNLO method shows essentially no statistically significant
offset. 
The difference between them is one measure of the size of the higher-order
corrections associated with the elimination of negative weights
and the observed $1\%$ effect is consistent with the expected order of
magnitude of an order $\as^2$ term.

To verify the correctness of the pure NLO coefficient we examine
\begin{align}
	\label{eq:alphas-nlo-coeff}
\lim_{\alpha_s \to 0} \frac{d\Delta \sigma^{\rm NLO}_{\rm shower}}{d\Delta
	\sigma^{\rm NLO}_{\rm an.}}\,, \quad d\Delta \sigma^{\rm NLO} \equiv
	d\sigma^{\rm NLO} - d\sigma^{\rm LO}\,,
\end{align} 
where $d\sigma^{\rm LO}$ ($d\sigma^{\rm NLO}$) is the leading-order
(next-to-leading-order) differential cross section for a 
specific observable.
Note that for the analytic result, 
$\Delta \sigma^{\rm NLO}_{\rm an.}$ is a pure (relative) 
$\mathcal{O}(\alpha_s)$ correction.
In the shower case, 
any higher-order corrections will be eliminated
by taking the $\alpha_s\to 0$ limit.
The $\alpha_s\to 0$ extrapolation is performed from runs at three
values of $\alpha_s \in [0.1,0.05,0.01]$.
The bottom panel of Fig.~\ref{fig:ee-NLO-tests-thrust-axis} shows
Eq.~(\ref{eq:alphas-nlo-coeff}).
The result is consistent with $1$, to within statistical errors,
confirming the NLO correctness of the \esme implementation.
That same panel shows the NLO test for other NLO matching choices:
\esme with the \PanGlobal shower and $\betaps=0.5$ and \dBNLO with
PanGlobal $\betaps=0.0$, confirming their NLO correctness as well.
\logbook{e453bc3b}{logbook/2024-01-17-slice-NLO/2024-12-18-costh-maxemsn-checks
  for checks of origin, which indicate that the issue is not coming
  from emissions beyond the first.}

The bottom panel of Fig.~\ref{fig:ee-NLO-tests-thrust-axis} also shows the
$\as \to 0$ result in the \dBNLO method with the $\Delta \tilde B$
term artificially set to zero (grey curve).
As expected, this does not agree with the true NLO correction,
confirming the necessity of the $\Delta \tilde B$ term.
Note that distributions that are insensitive to --- or averaged over
--- the orientation of the event would be unaffected by the absence of
$\Delta \tilde B$ correction.
Equivalently, for the case of oriented $e^+e^-$ events, the effect of
the $\Delta \tilde B$ correction is visible only in the longitudinal
component of the NLO coefficient in Eq.~(\ref{eq:ee-thrust-angle}).
This is because for one single final-state emission, the \panglobal
and the FKS maps are identical up to an overall event rotation, and
the \panlocal map is formally identical to the FKS map up to the
partitioning of singular regions.
 
\subsubsection{$pp$ tests}
\label{sec:NLO-pp-tests}
In this section we examine colour-singlet production in $pp$ collisions.
We consider three processes, namely (i) neutral-current Drell --Yan ($pp \to Z/\gamma^* \to e^+ e^-$), (ii) charged-current Drell--Yan ($pp \to W^+ \to e^+ \nu_e $), and (iii) Higgs production in gluon fusion $pp \to H$. 
For our tests, we use
the positive-definite \texttt{NNPDF40MC\_nlo\_as\_01180} PDF
set~\cite{Cruz-Martinez:2024cbz}, with $\as(m_Z) = 0.118$.
We use a centre-of-mass energy of $13.6\TeV$. We work in the $G_\mu$ electroweak scheme~\cite{Denner:2000bj} and take the following input parameters~\cite{ParticleDataGroup:2022pth}
\begin{align}
& m_Z =91.1876\GeV\,,\quad m_W = 80.377\GeV\,, \quad  G_F = 1.16639 \cdot 10^{-5} \GeV^{-2}\,, \quad \label{eq:EWparams}\\
&\notag m_H =
125\GeV \,,\quad \Gamma_Z = 2.4952\GeV\,, \quad \Gamma_W = 2.085\GeV\,. 
\end{align} 
In the Higgs case, we also use the infinite top mass limit~\cite{Dawson:1990zj,Djouadi:1991tka}.
To obtain the LO and NLO baselines, we use
MCFM~v10.3~\cite{Campbell:1999ah,Campbell:2011bn,Campbell:2015qma,Boughezal:2016wmq,Campbell:2019dru}.
For phenomenological results we take the event-by-event di-lepton (or Higgs) invariant mass as our central renormalisation scale and carry out 7-point scale variation to show uncertainty bands.
For the extraction of the pure NLO coefficient, we instead use a fixed renormalisation and factorisation scale equal to the on-shell mass of the produced boson, as given in Eq.~\eqref{eq:EWparams}. 

\begin{figure}
	\centering
	\includegraphics[width=0.48\textwidth,page=6]{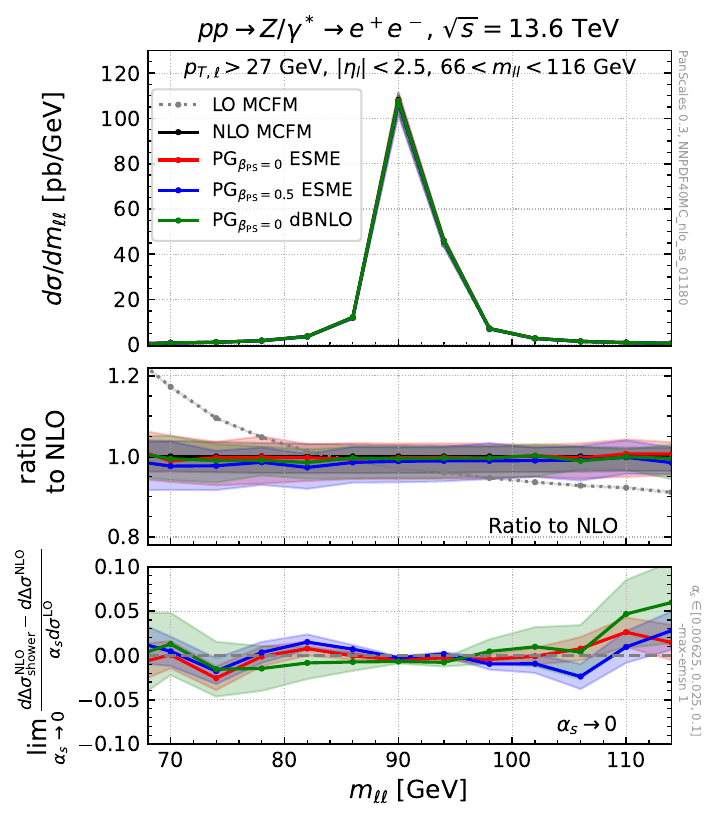}
	\includegraphics[width=0.48\textwidth,page=7]{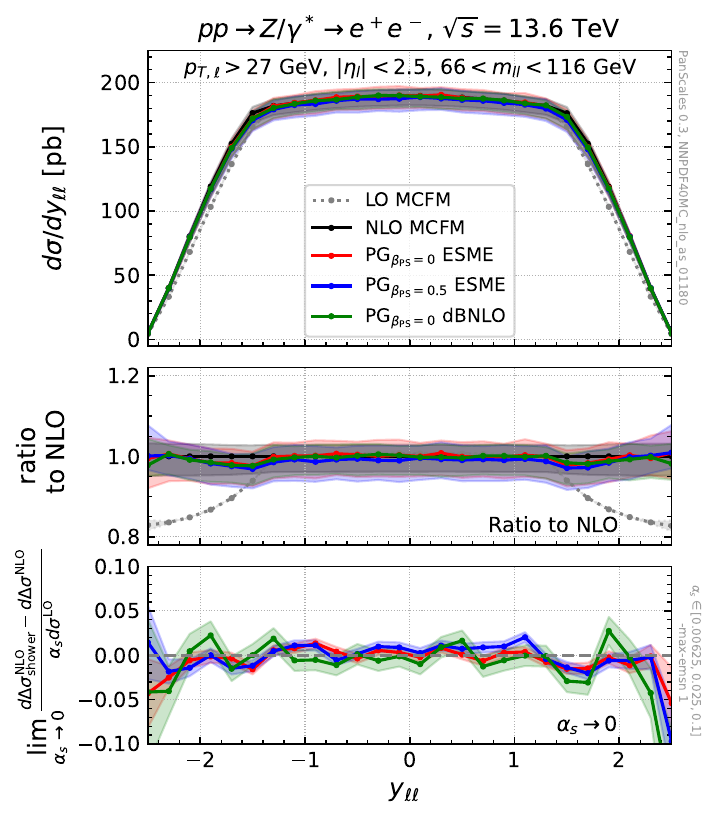}
	\caption{NLO tests for the $pp\to Z/\gamma^* \to e^+ e^-$
          process with cuts on the lepton transverse momentum and
          rapidity.
          Left (right): the invariant mass (rapidity) of the
          colour singlet.
          The top and middle panels show results with phenomenological
          settings, compared to NLO predictions from MCFM.
          Bands correspond to 7-scale uncertainty, $m_{\ell\ell}/2\le
          \mu_R,\mu_F \le 2m_{\ell\ell}$ with $1/2\le \mu_R/\mu_F \le2$.
          The bottom panel shows the ratio of the shower NLO
          coefficient (extracted in an $\as \to 0$ limit) to the NLO
          coefficient from MCFM.
          The bands represent the combined statistical
          uncertainty on the ratio.
        }
	\label{fig:pp2Zdec-NLO}
 \end{figure}

\begin{figure}
	\centering
	\includegraphics[width=0.48\textwidth,page=12]{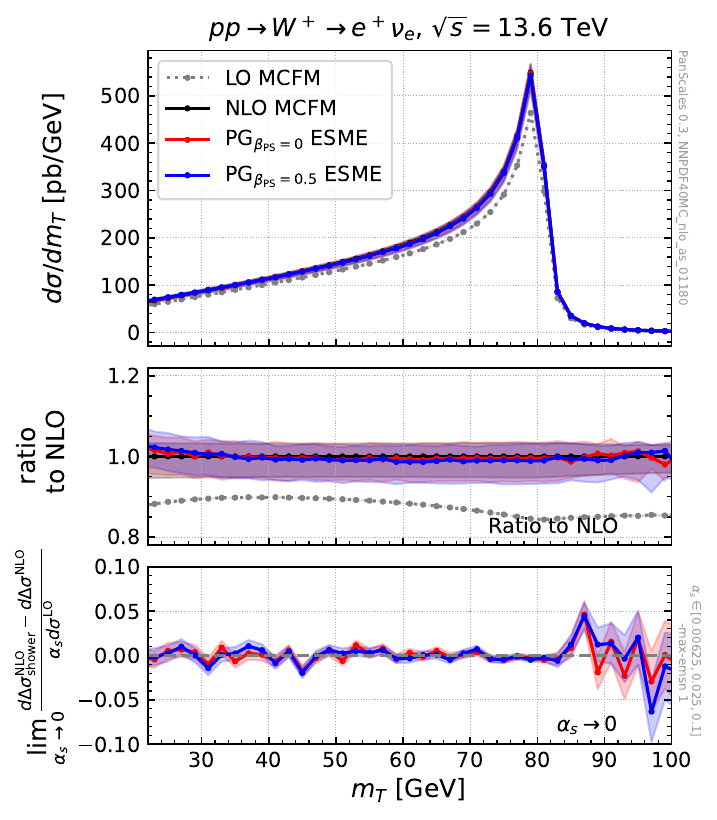}
	\includegraphics[width=0.48\textwidth,page=11]{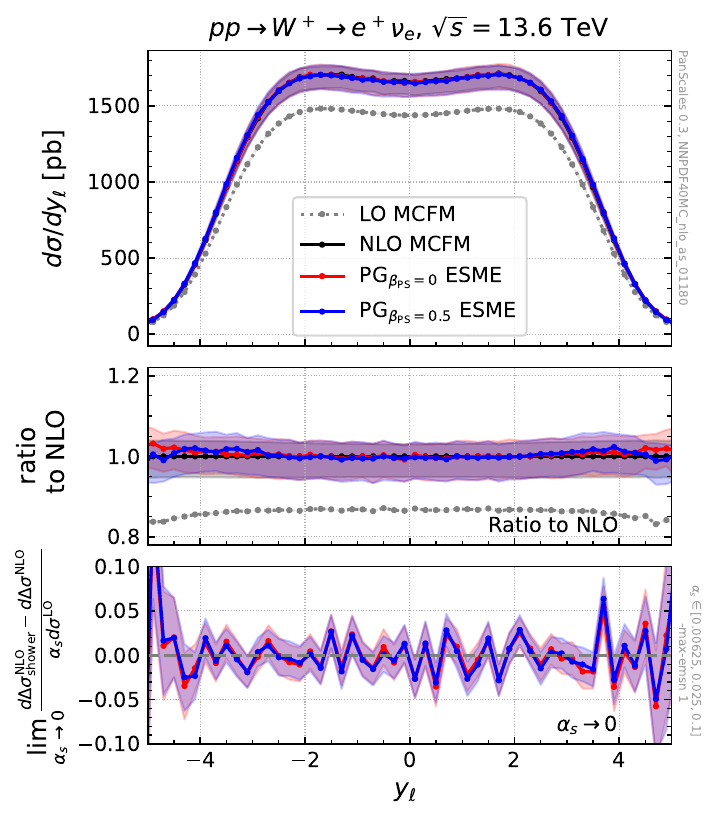}
	\caption{
          Analogue of Fig.~\ref{fig:pp2Zdec-NLO}, showing the $W$
          transverse mass and the charged lepton rapidity, without
          lepton or missing 
          momentum cuts.
        }
	\label{fig:pp2Wdec-NLO}
\end{figure}
      
Fig.~\ref{fig:pp2Zdec-NLO} shows results for $pp\to Z/\gamma^* \to
e^+e^-$ with the following lepton cuts: $p_{t\ell}>27\GeV$ and
$|\eta_\ell| < 2.5$, $66 < m_{\ell\ell} < 116 \GeV$.
The left-hand panel is for the distribution of $m_{\ell\ell}$, the
lepton-pair invariant mass; the right-hand panel is for the rapidity
of the boson (or equivalently, the lepton pair).
As in Fig.~\ref{fig:ee-NLO-tests-thrust-axis}, the upper panels show
the differential cross section, while the middle panels show the ratio to
NLO.
One observes agreement to within about a percent for both ESME and dBNLO.
Note that with our specific lepton cuts, the NLO $K$-factor is quite
close to $1$ near the $Z$ mass and for central rapidities.
This comes from an interplay between a positive NLO effect in the
total cross section and negative NLO effect due to the cuts.
That interplay is also responsible for much of the kinematic
dependence of the NLO $K$-factor.
The lower panels show the $\as \to 0$ test of NLO accuracy.
Given that the NLO coefficient is close to zero in parts of the phase
space, we show a ratio to $\as$ times the LO result,
\begin{equation}
  \label{eq:limitNLOdiff}
  \lim_{\as\to0} \frac{%
    d\Delta \sigma^{\rm NLO}_{\rm shower}
    - d\Delta \sigma^{\rm NLO}}{\as\, d\sigma^{\rm LO}}\,.
\end{equation}
Note that for these NLO accuracy tests (i.e.\ the bottom panel of each
plot), we limit the shower to the
first emission, since higher numbers of emissions can only modify
$\as^2$ terms and beyond.
We also freeze the PDF at the factorisation scale used in the
fixed-order calculation (the $Z$ on-shell mass in this case) and we
use a fixed renormalisation scale, independent of $m_{\ell\ell}$
(again, $m_Z$).
We see agreement to within statistical uncertainties, shown as a band.
Fig.~\ref{fig:pp2Wdec-NLO} shows analogous tests for $W^+$ production,
without lepton cuts, and the conclusions are similar.
Since the lepton rapidity distribution is sensitive to the $V-A$
structure of the $W$ interaction, it provides a direct check of the
lepton swap procedure discussed in Section~\ref{sec:real-radiation}.

\begin{figure}
	\centering
	\includegraphics[width=0.48\textwidth,page=1]{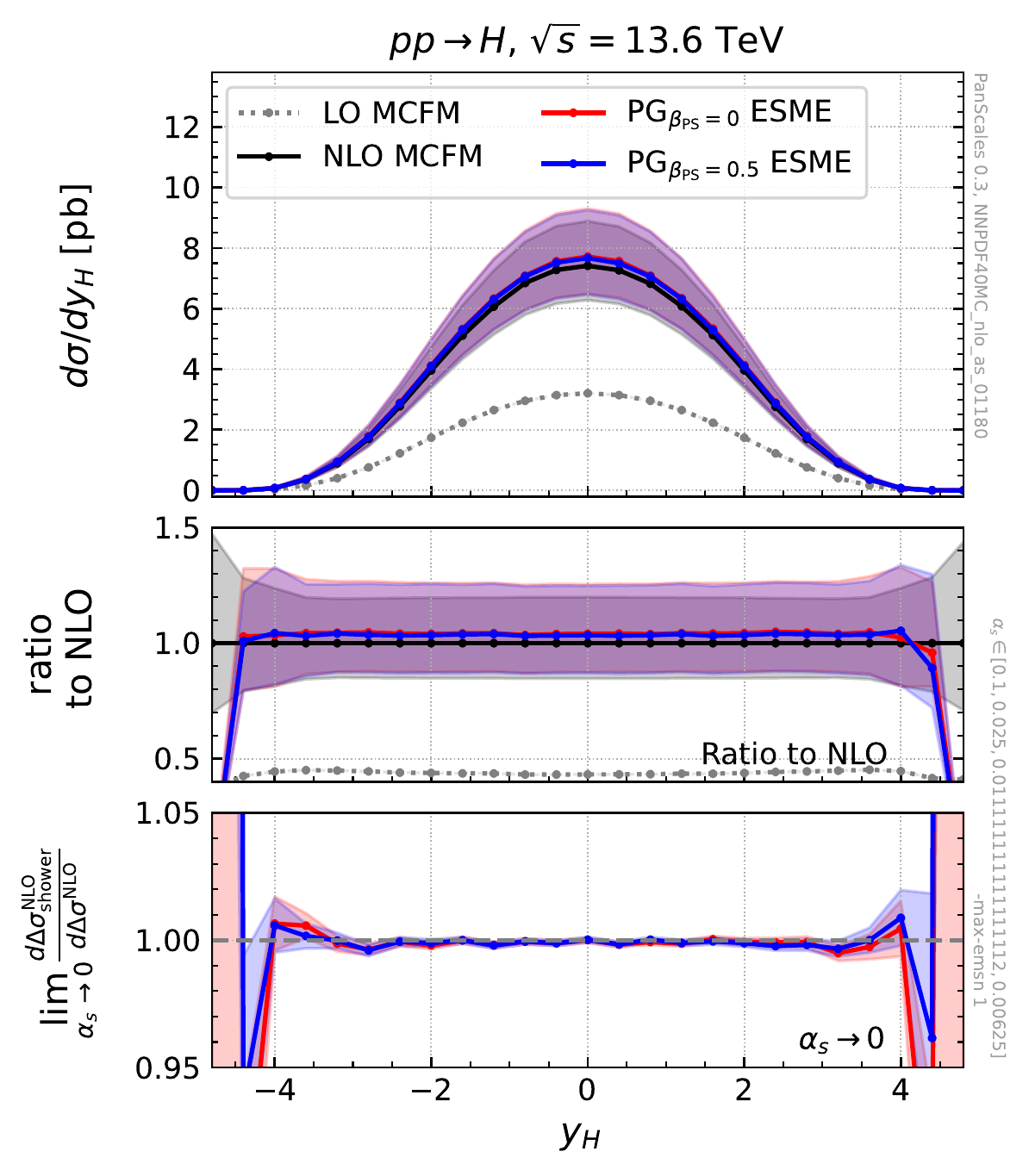}
	\includegraphics[width=0.48\textwidth,page=2]{final-figs-paper/fo-checks-pp-dis.pdf}
	\caption{
          Analogue of Fig.~\ref{fig:pp2Zdec-NLO} for Higgs production,
          showing the Higgs rapidity distribution (left) and the
          transverse momentum distribution (right).
        }
	\label{fig:pp2H-NLO}
\end{figure}

In Fig.~\ref{fig:pp2H-NLO} we provide results for Higgs production.
In the left panel, we illustrate the $H$ rapidity: here too the conclusions are similar, though in the middle panel the
ratio to NLO (middle panel) deviates from $1$ by about $3{-}4\%$
rather than the $1\%$ seen for $Z$ and $W$ production.
Note, however, that this deviation is numerically small compared to the
size of scale uncertainties.
The NLO coefficient itself is in excellent agreement with the MCFM
result. 
The right-hand panel illustrates the $H$ transverse momentum.
As this observable is highly sensitive to all-order corrections, it is
no surprise that a parton shower differs significantly from a NLO
result, and the $\alpha_s \to 0$ limit is necessary to verify that
differences are indeed purely due to higher-order effects.

\subsubsection{DIS tests}
\label{sec:NLO-DIS-tests}

\begin{figure}
	\centering
	\includegraphics[width=0.48\textwidth,page=17]{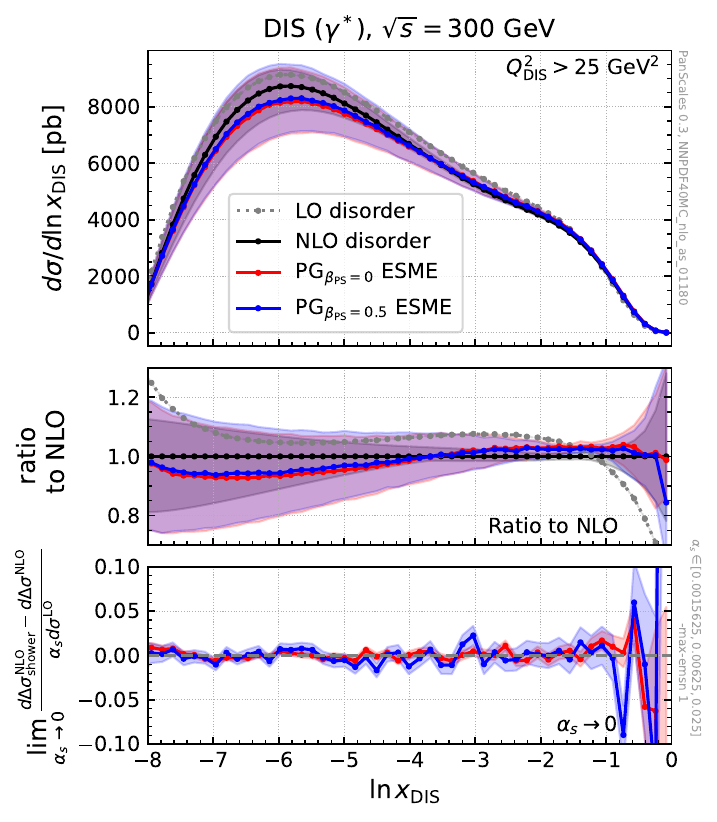}
	\includegraphics[width=0.48\textwidth,page=18]{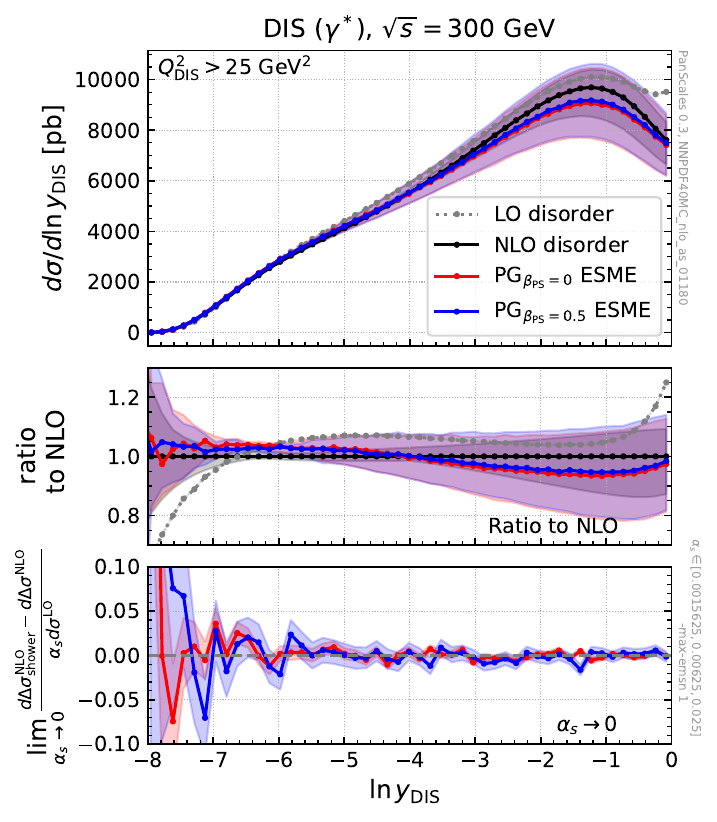}
	\caption{
          Analogue of Fig.~\ref{fig:pp2Zdec-NLO} for DIS.
          The left-hand plot shows $\ln x_{\rm DIS}$, the right-hand
          one $\ln y_{\rm DIS}$.
          The reference LO and NLO results have been obtained with the
          \texttt{disorder} code.
        }
	\label{fig:DIS-NLO}
\end{figure}

For our first set of DIS tests, in Fig.~\ref{fig:DIS-NLO}, we consider photon-mediated $e^-p$ collisions at $\sqrt s = 300$~GeV.
The left and right-hand plots show the distributions in the $x_\dis = \frac{Q_{\dis}^2}{2p.q}$ and $y_\dis=\frac{p.q}{p.k}$
variables respectively.
Here $p$ and $k$ are the incoming proton and electron momenta and $q$ is the photon momentum.
The plots include a
constraint $Q_{\dis}^2 > 25\,\GeV^2$.
We obtain our reference NLO results from the \texttt{disorder}
code~\cite{Karlberg:2024hnl} which itself relies on
\hoppet~\cite{Salam:2008qg} and \disent~\cite{Catani:1996vz}.
For the phenomenological predictions, we use as central renormalisation and factorisation scale $Q_{\dis}$, while for the extraction of the pure $\alpha_s$ correction we use a fixed scale equal to the on-shell $Z$ mass.

The two NLO matching methods that we have explored in DIS are \esme,
which we have implemented with the \PanGlobal shower, and P2B, which
works with both \PanGlobal and \PanLocal showers.
For the P2B method, since the \PanScales showers conserve $x_\dis$ and
$y_\dis$, we expect the NLO shower results to be identical to pure NLO
predictions.
We have verified that this is the case.
In Fig.~\ref{fig:DIS-NLO} we therefore focus on the \esme method.
The phenomenological predictions from the \esme method, in the upper
two panels, are in agreement with the exact NLO to within about $5\%$,
and well within the scale uncertainty bands.\footnote{We also explored
  a modification of \ref{alg:nlo-add} without steps
  \ref{alg:nlo-add:rejCgtR} and \ref{alg:nlo-add:rejRgtC} and it
  brought the \esme NLO results closer to pure NLO.
  We note however (not shown) that the \esme NLO result is remarkably
  close to the actual NNLO result.\logbook{}{See additional plots in
    the main plot file, with green NNLO curve}
  Indeed, in several instances in our investigations, we have observed
  that higher-order freedom in the formulation of \esme brings shape
  differences that are similar to those seen in NLO$\to$NNLO, and
  not hinted at by normal NLO scale variation.
  We believe that further study on this question may be of interest in any
  future work that addresses higher orders and their uncertainties
  more comprehensively.  }
The $\as \to 0$ NLO accuracy test, in the bottom panel, demonstrates
the correctness of the shower's pure NLO contribution, to within
statistical uncertainties.

\begin{figure}
	\centering
	\includegraphics[width=0.48\textwidth,page=1]{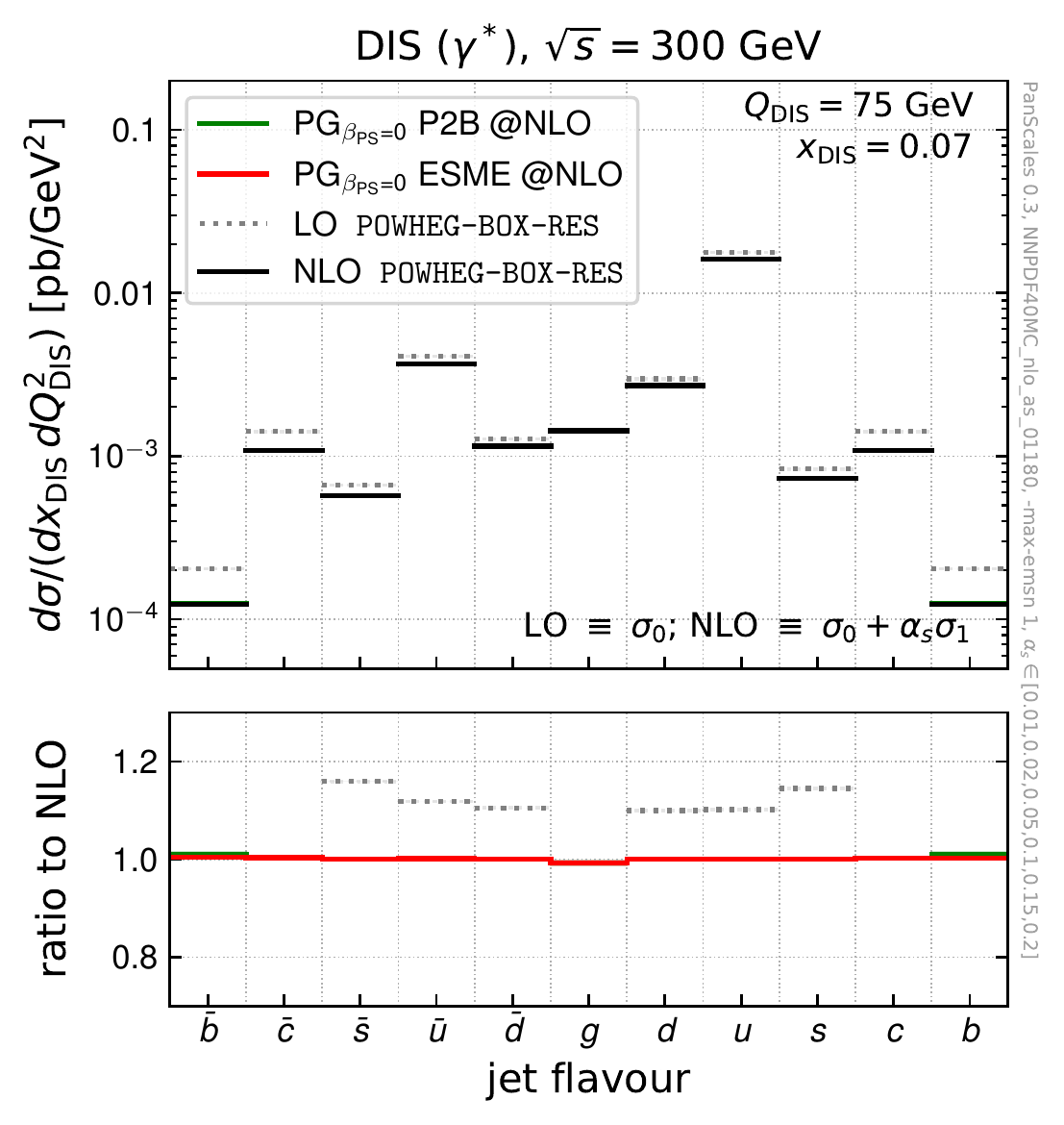}
	\includegraphics[width=0.48\textwidth,page=2]{final-figs-paper/DIS_flavour_paper.pdf}
	\caption{NLO flavour tests for the DIS process, showing the
          distribution of net flavour of the jet with largest
          light-cone component along the current direction.
          The left-hand plot is for photon exchange where both the P2B and
          \esme methods are expected (and observed) to agree with NLO.
          The right-hand plot is for the full $\gamma^*/Z$
          contribution where one sees that the P2B approach differs
          from NLO, due to the missing gluon-induced axial contribution.
          The reference LO and NLO results have been produced 
          running the \powhegboxres \texttt{DIS} generator.
          The \ESME and P2B predictions are 
          shown at pure NLO, i.e.\ the LO
          and NLO coefficients have been extracted from an $\as \to 0$
          extrapolation and then the physical pure NLO result has been
          obtained by adding them with the actual $\as(Q)$ multiplying
          the NLO coefficient. 
          This choice provides a phenomenologically relevant indication of the 
          size of NLO effects, while avoiding any potential confusion
          associated with differing treatments of orders beyond NLO.
        }
	\label{fig:DIS-flavour}
\end{figure}

Our second set of tests in DIS considers the net flavour of the
leading jet, in order to study the effects discussed in
Section~\ref{sec:p2b}, specifically the impact of the gluon-induced
axial ($F_3$) component in the P2B method.
To probe the issue, we identify jets with the DIS
version~\cite{vanBeekveld:2023chs} of the Cambridge/Aachen
algorithm~\cite{Dokshitzer:1997in,Wobisch:1998wt}, implemented as a
FastJet~\cite{Cacciari:2011ma} plugin, used in the Breit frame, and consider the
jet with the largest light-cone component in the current direction.
As has been extensively discussed in the
literature~\cite{Caletti:2022hnc,Czakon:2022wam,Gauld:2022lem,Caola:2023wpj}, the jet flavour is
not an infrared safe quantity for standard jet algorithms, because of
configurations associated with a pair of soft quarks, starting at
$\order{\as^2}$.\footnote{The algorithms proposed in those articles
  have yet to be extended to DIS.}
We therefore limit our study to events where we generate just the
first shower emission, which ensures that the infrared unsafe
configuration is not present.
As a reference, we use the NLO+PS event generator of
Ref.~\cite{Banfi:2023mhz}, which 
can perform fixed-order (``stage 2'') differential calculations 
with explicit flavour dependence thanks to the \texttt{detailedNLO} 
feature of the \powhegboxres framework~\cite{Jezo:2015aia}.
We still consider $e^-p$ collisions at $\sqrt{s}=300\GeV$, but we fix
$x_\dis = 0.07$ and $Q_{\dis} = 75\GeV$, a combination that helps 
enhance the relative size of the gluon-induced axial contribution.
For this test, we also use $\mu_F=\mu_R=Q_{\dis}$.
Fig.~\ref{fig:DIS-flavour} shows the cross section in bins of the
leading jet's net flavour, with just photon exchange (left) and full
$\gamma^*/Z$ exchange (right).
At LO, the results are driven entirely by the flavour distribution of
the proton PDF and the associated quark charges.
At NLO, with just photon exchange, the P2B and \esme methods both
agree with the predictions from \POWHEGBOX.
With $\gamma^*/Z$ exchange, while the \esme method is correct at
NLO, one sees that the P2B method is not.
The differences relative to NLO are generally larger for anti-quark
flavours and for flavours where the ratio of gluon-induced to Born
contributions is enhanced.
One can imagine various resolutions of this issue, for example
supplementing P2B with a flavour-related $\Delta\Bbar$ contribution
or, perhaps, through an adaptation of the swap technique of
Appendix~\ref{app:swap}, applied between outgoing quark and
anti-quark flavours in the gluon-induced channel.
We leave their investigation to future work.

\subsection{NNDL tests}
\label{sec:NNDL-tests}
In this section we provide a numerical demonstration that the matching
algorithms that we have introduced bring NNDL accuracy for event-shape
like observables, i.e.\ control
of terms $\as^n L^{2n-p}$ with $p\le 2$.
This is specifically in the cumulative cross section,
$\Sigma_\text{PS}(\as,L)$, for a given dimensionless observable to
have a value $v$ less than $e^{L}$ (with $L$ large and negative).
If done properly, NLO matching together with NLL-accurate parton
showers should automatically provide NNDL event-shape accuracy in
$\Sigma_\text{PS}(\as,L)$.
Testing the NNDL accuracy provides a key validation of one of the
necessary ingredients towards general NNLL accuracy in
$\ln \Sigma_\text{PS}(\as,L)$.
For a matched shower to achieve NNDL precision, the cumulative cross
section must satisfy~\cite{Hamilton:2023dwb}
\begin{align}
	\label{eq:nndl-test}
	\raisebox{2.25mm}{\(\lim\limits_{\substack{\alpha_s \to 0 \\  \xi \, {\rm fixed}}}\)} \;
	\frac{\Sigma_{\rm PS}(\alpha_s, L) - \Sigma_{\rm NNDL}(\alpha_s, L)}{\alpha_s \Sigma_{\rm DL}(\alpha_s, L)} = 0\,.
\end{align}
Note that $\alpha_s=\alpha_s(Q)$ and that $\xi = \alpha_s L^2$ 
is kept fixed when evaluating Eq.~\eqref{eq:nndl-test}, 
so as to isolate the pure NNDL contribution to $v$. 
In practice, we numerically evaluate Eq.~\eqref{eq:nndl-test} 
by first running the showers with fixed values of $\alpha_s =
0.1/N^2$ with $N \in \left\{6,7,12,24\right\}$ and then 
performing a polynomial extrapolation in powers of $\sqrt{\alpha_s}$ 
so as to obtain the $\alpha_s\to 0$ limit.
Typically the extrapolation uses a subset of the points.
We estimate a systematic uncertainty associated with this 
procedure by considering a second extrapolation with different
$\alpha_s$ values.  
For all processes, we use $\xi=1.296$. 

To test the showers, we use observables based on the
Lund-plane~\cite{Andersson:1988gp,Dreyer:2018nbf} picture, as
introduced for \PanScales NLL testing in $e^+e^-$~\cite{Dasgupta:2020fwr},
$pp$~\cite{vanBeekveld:2022ukn,vanBeekveld:2022zhl} and
DIS~\cite{vanBeekveld:2023chs} collisions.
These Lund-based event shapes are defined as
\begin{equation}
	\label{eq:evth-shapes}
	M_{\beta} = \max_{j\in {\rm decl.}}\left\{\frac{|p_{t,j}|}{Q}e^{-\beta y_j}\right\}\,
  ,\quad S_{\beta} =
	\sum_{j\in {\rm decl.}} \frac{|p_{t,j}|}{Q}e^{-\beta |\eta_j|}\,.
\end{equation}
where $\beta$ is a free parameter and we explore three values, $\beta \in \lbrace 0,0.5,1 \rbrace$. 
In Eq.~(\ref{eq:evth-shapes}), the max and sum run over primary
declusterings, which are suitably defined depending on the
hard-scattering process.

In $e^+ e^-$ collisions the whole
event is clustered into two jets using the
Cambridge~\cite{Dokshitzer:1997in} algorithm.
For each of these jets we undo the last clustering, and
define $p_{t,j} = E_j |\sin \theta_{ij}|$ and $\eta_j = -\ln \tan
\frac{\theta_{ij}}{2}$, with $E_j$ the energy of the softer particle in the
branching, and $\theta_{ij}$ the angle between the two particles. 
We repeat the declustering following the harder subjet, 
such that only the set of primary declusterings is
considered~\cite{Dreyer:2018nbf,Dasgupta:2018nvj,Dasgupta:2020fwr}.
In proton-proton collisions we cluster 
the full event with the Cambridge/Aachen (C/A)
algorithm~\cite{Dokshitzer:1997in,Wobisch:1998wt} into jets with 
radius $R =1$. We then calculate the transverse momentum and 
rapidity of each of the jets with respect to the beam, 
which defines $p_{t,j}$ and $\eta_j$~\cite{vanBeekveld:2022ukn}.
For deep-inelastic scattering we use the algorithm 
defined in Appendix~C of Ref.~\cite{vanBeekveld:2023chs}, and 
specifically Eqs.~(C.4), (C.5) of that reference to define the 
transverse momentum and rapidity that enter into Eq.~\eqref{eq:evth-shapes}. 

Resummed predictions for $M_{\beta}$ and $S_\beta$ at NNLL accuracy
will be presented in Ref.~\cite{2024FutureResummations} for all
processes considered in this work. We have used the NNDL expansion of
those results to test the showers.

For processes with incoming protons, we use the toy PDF set described
in Appendix A.3 of Ref.~\cite{vanBeekveld:2022ukn}, designed
specifically for logarithmic accuracy tests.
For quark-initiated processes, we assume an initial $d$ quark.
The runs are carried out using the NODS colour scheme~\cite{Hamilton:2020rcu,vanBeekveld:2022zhl}, 
and turning off spin correlations~\cite{Hamilton:2021dyz,Karlberg:2021kwr,vanBeekveld:2022zhl} as
they do not affect the NNDL accuracy. 
In all plots, the $\alpha_s$ values used to obtain the central value 
of the extrapolation and its error are quoted on the side of each figure in grey. 
Finally, all NNDL tests are performed for fixed Born kinematics and flavour.  
\begin{figure}[bt!]
	\centering
	\includegraphics[width=\textwidth,page=1]{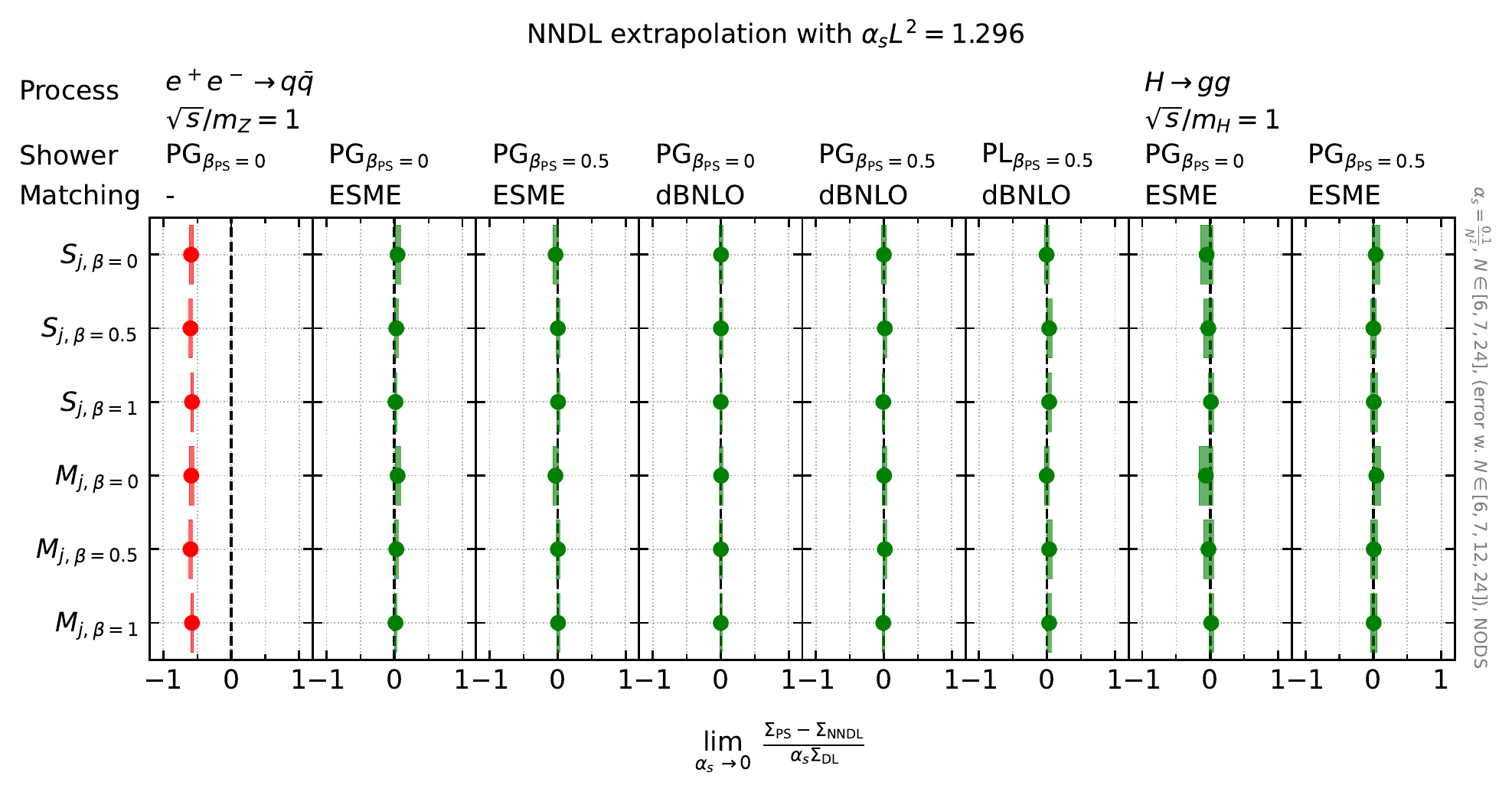}
	\caption{Results of the NNDL accuracy tests,
          Eq.~(\ref{eq:nndl-test}), at fixed $\xi = \alpha_s L^2$ for
          $e^+e^-\to q \bar q$ and $H \to gg$ with
          $ \sqrt{s}/m_X = 1$.
          The tests are carried out for a fixed Born configuration (for
          $e^+e^-\to q \bar q$ we use $\cos\theta_{q,\text{beam}} =
          0.5$ and for $H \to gg$ we align the gluons along the
          $z$-axis).
        }
	\label{fig:ee-NNDL-tests}
\end{figure}
\begin{figure}[tb!]
  \centering
  \includegraphics[width=\textwidth,page=1]{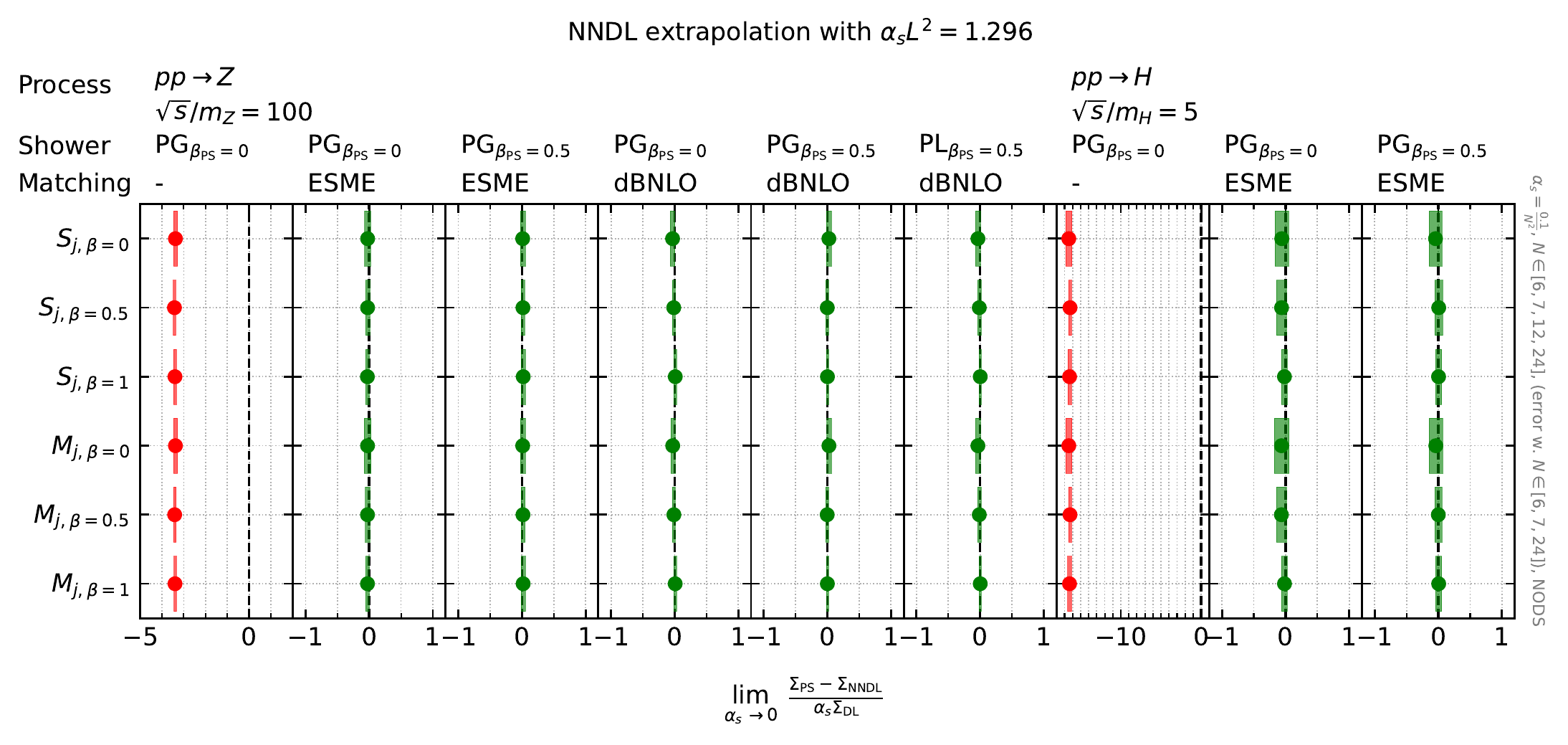}
  \caption{
    Results of the NNDL accuracy tests at fixed $\xi = \alpha_s L^2$
    for $pp\to Z$ with $\sqrt{s}/m_Z = 100$ and $pp\to H$ with
    $\sqrt{s}/m_H = 5$.
    The rapidity of the colour singlet is set to 0.  For the case of
    $Z$ production, we consider $d\bar{d}\to Z$ as the Born flavour
    configuration.
  }
\label{fig:pp-NNDL-tests}
\end{figure}

We begin by examining the $e^+e^-$ results.
NNDL tests have already been carried out for event shape observables
in Ref.~\cite{Hamilton:2023dwb}.
There, the tests used a generator without (oriented) NLO
normalisation, by comparing to an NNDL calculation that is divided by the total NLO cross section.
In Fig.~\ref{fig:ee-NNDL-tests}, instead, we have the explicit NLO normalisation in the generator, with both \esme and \dBNLO methods and the NNDL calculation is correspondingly normalised to the total LO cross section.
With NLO matching, all combinations of observable, matching scheme and
process are consistent with NNDL accuracy.  

We test for NNDL accuracy in $pp$ collisions in
Fig.~\ref{fig:pp-NNDL-tests}, which shows \esme and \dBNLO results
for $Z$ production and \esme results for Higgs production, including one
case of matching with the \PanLocal shower. 
For these tests, we fix the colour-singlet rapidity to be 0.
The left-most column of the plot shows that without matching there is
a large discrepancy, illustrating the power of the test to diagnose
potential issues.
All the matched results are in agreement with NNDL prediction.
%
\begin{figure}[t!]
	\centering
	\includegraphics[width=0.7\textwidth,page=1]{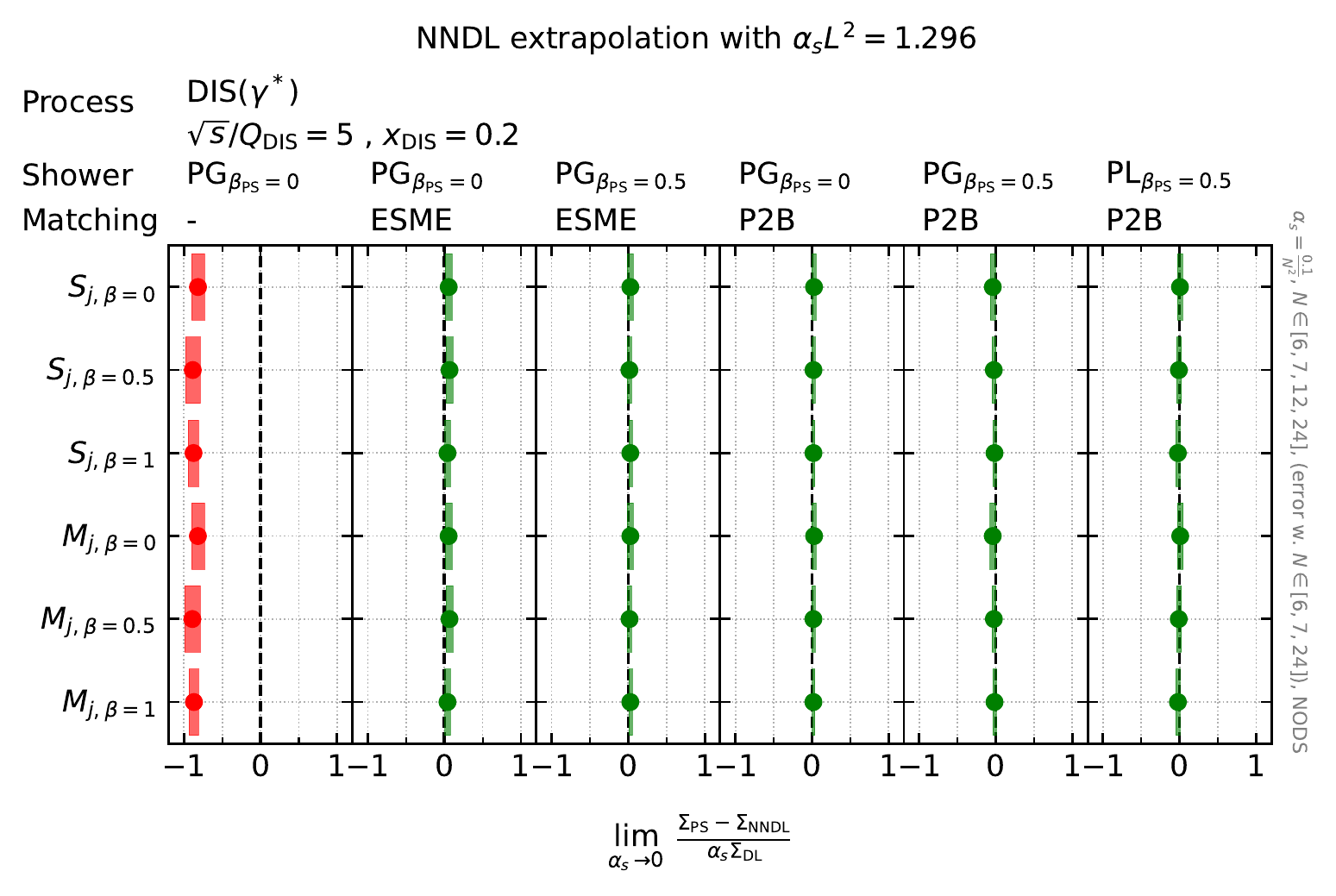}
	\caption{Results of the NNDL accuracy tests at fixed $\xi = \alpha_s L^2$ and fixed Born flavour configuration $\gamma^* d \to d$
  with $\sqrt{s}/Q_\dis=5$ and $x_\dis=0.2$.
  }
	\label{fig:dis-NNDL-tests}
\end{figure}
Fig.~\ref{fig:dis-NNDL-tests} shows corresponding tests for DIS, with
\esme and P2B matching (the latter also with the \PanLocal shower).
The $pp$ and DIS tests represent the first time that NNDL event-shape
accuracy has been demonstrated for a matched parton shower with
incoming hadron beams.

\section{Brief comparison to data and performance studies}
\label{sec:pheno+perf}

\subsection{Comparison to data}
\label{sec:pheno}
Several features are still missing from \PanScales and its \Pythia
interface in order to carry out a full phenomenological comparison to
data with incoming hadrons.
These include QED effects and, in $pp$ collisions, multi-parton interactions.
Therefore in this section we consider only a very first basic
comparison, with the intention of elaborating on the results shown here in
future work.
The showers that we show here are the first to have demonstrated
general NLL accuracy combined with NLO together with NNDL accuracy for
event-shape like observables.
\begin{figure}[tb!]
  \centering
  \includegraphics[width=0.45\textwidth,page=1]{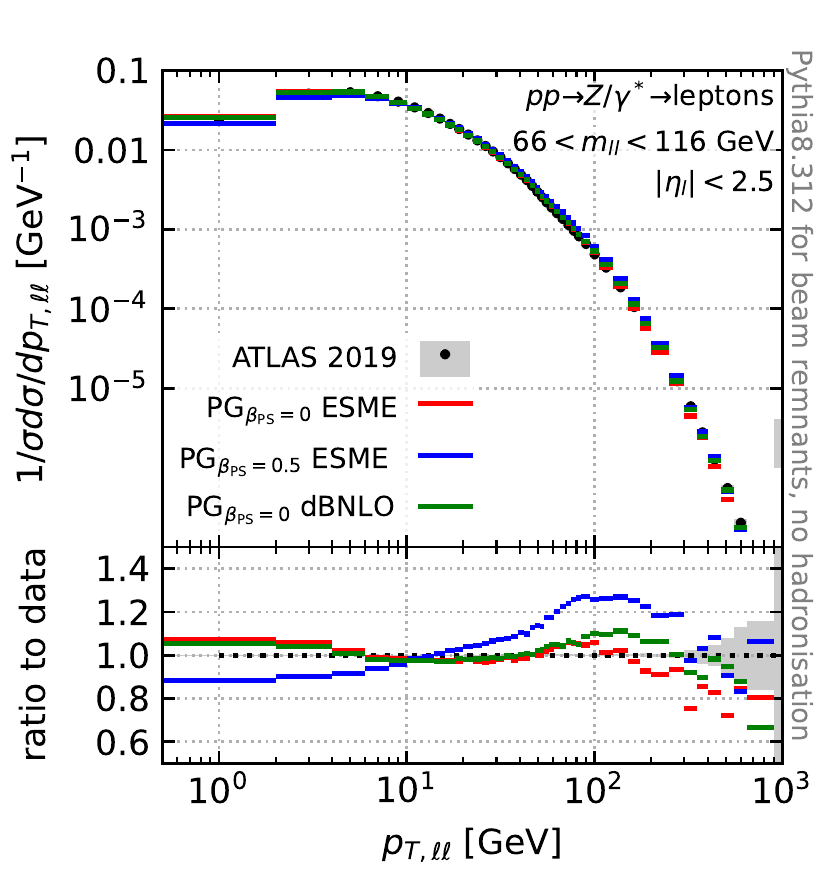}\hfill
  \includegraphics[width=0.45\textwidth,page=2]{data-comparison-runs/rivet/plot-data.pdf}
  \caption{
    \PanScales NLL+NLO matched showers, interfaced with
    \Pythia~\cite{Bierlich:2022pfr}, as compared to $13\TeV$
    QED-Born di-lepton data from the ATLAS collaboration~\cite{ATLAS:2019zci}.
    The left-hand plot is for the di-lepton transverse momentum
    distribution, while the right-hand plot is for the $\phi^*_\eta$
    variable~\cite{Banfi:2010cf}, cf.\ Eq.~(\ref{eq:phi-star-eta}).
    In the Pythia interface, we include Pythia's primordial transverse
    momentum but not hadronisation, QED effects or multi-parton interactions.
  }
  \label{fig:pheno}
\end{figure}
We use a pre-release \PanScales version 0.3 with its
interface~\cite{vanBeekveld:2023ivn} to
Pythia~8.312~\cite{Bierlich:2022pfr} as well as Pythia's
HepMC3~\cite{Buckley:2019xhk} interface to the RIVET
tool~\cite{Bierlich:2019rhm} in order to carry out the ATLAS di-lepton
analysis of Ref.~\cite{ATLAS:2019zci}.
The analysis considers events with two oppositely charged 
leptons (in a ``QED-Born'' definition), each with $p_{t\ell} > 27\GeV$
and $|\eta_\ell| < 2.5$. 
The left-hand plot of Fig.~\ref{fig:pheno} shows the di-lepton $p_t$
distribution, normalised to the total cross section, while the right-hand
plot shows the $\phi^*_\eta$~\cite{Banfi:2010cf} distribution with
\begin{equation}
  \label{eq:phi-star-eta}
  \phi^*_\eta \equiv \tan
  \left(\frac{\pi-\Delta\phi_{\ell\ell}}{2}\right) \sin \theta_\eta^*\,,
  \qquad
  \cos \theta_\eta^* = \tanh \frac{\eta_{\ell^-} - \eta_{\ell^+}}{2}\,.
\end{equation}
Both observables are in the $\beta_\text{obs}=0$ class but they have
substantially different NLL resummation structures.
The figures show curves from the \esme NLO-matching method with the
PanGlobal shower, with $\betaps = 0$ and $0.5$, and the
  \dBNLO method with PanGlobal $\betaps = 0$.
The differences between various shower and matching choices
provide an indication of the size of uncertainties that are associated
with missing higher orders (we defer a more extensive study of
uncertainties to future work). 
Those differences are of the order of $20\%$ and this is consistent
with the size of NNLL corrections observed for the \PanGlobal shower in
Ref.~\cite{vanBeekveld:2024wws} for $e^+e^- \to q\bar q$ event shapes,
though we note that in that case the two $\betaps$ values gave very
similar results even at NLL accuracy.
Within the $20\%$ NLL+NLO uncertainties, as well as the expected size
of higher-order matching uncertainties at large $p_{t,\ell\ell}$ and
$\phi^*_\eta$, there is good agreement with the data.
While we do not have MPI, we find that in a plain Pythia run with the
Monash 2013 tune~\cite{Skands:2014pea}, adding
MPI effects reduces the lowest $p_{t,\ell\ell}$ bin by about $10\%$
and has a negligible effect elsewhere.
We would expect a broadly similar impact with our shower.
\Pythia's primordial transverse momentum \emph{is} included in our
simulation and its impact was a $\mathcal{O}(20\%)$ reduction in the
smallest $p_{t,\ell\ell}$ bin and $\mathcal{O}(5{-}8\%)$ reduction in the few
smallest $\phi_{\eta}^*$ bins.
Note that these effects are likely to depend on the tune and one might
ultimately want to develop updated tunes with the PanScales showers.
\logbook{}{See
  data-comparison-runs/rivet/pp2Zdec-mpi-comparisions/ATLAS_2019_I1768911/d27-x01-y01.pdf
  (pTz) and
  data-comparison-runs/rivet/pp2Zdec-mpi-comparisions/ATLAS_2019_I1768911/d28-x01-y01.pdf
  (phistar) for the plots.}

\subsection{Performance studies}
\label{sec:performance}
Of the matching methods that we have discussed in this article, \dBNLO
is based on a well-established underlying methodology and P2B is an
intrinsically simple and efficient method.
On the other hand the \esme method is qualitatively new and, given that
it offers the prospect of general positive-definite matching, it is
especially important to determine whether its speed performance is at
least comparable to other methods.

Comparisons of speed bring many aspects into play: the efficiency of
the underlying code for matrix-elements, which may be hard-coded or
automatically generated; trade-offs between warm-up time and
event-generation efficiency; the efficiency of unweighting; etc.
There are also delicate technical aspects of speed measurement,
especially as not all codes produce equivalent outputs (e.g.\ full
showered events versus just a first emission).
Some caution is therefore needed in interpreting any timing results.

For our first performance test we considered the $pp \to Z/\gamma^*
\to \ell^+\ell^-$ process at NLO at $\sqrt{s}=13.6$~TeV.
We examined this process with the \powhegboxvt revision
3985~\cite{Alioli:2008gx,Alioli:2010xd}, \herwig version
7.3.0~\cite{Bewick:2023tfi}, \sherpa version
3.0.1~\cite{Sherpa:2024mfk} 
and \mgfive version 3.6.1~\cite{Alwall:2014hca}, starting from
default settings for each.
We focused on the time to produce the NLO event up and to and
including first emission, either explicitly where this was possible,
or deducing it from the difference between NLO+PS and LO+PS runs.
We found the \powhegbox to be both the fastest and the one with the
smallest fraction of negative-weighted events, and so took it as our
baseline.
We used it with the
\texttt{NNPDF40MC\_nlo\_as\_01180}~\cite{Cruz-Martinez:2024cbz} PDF,
which is positive-definite, thus alleviating the one potential extraneous source of
negative-weight events.
We also considered $\gamma^*$-mediated DIS at $\sqrt{s}=317$~GeV and
$Q^2_{\rm DIS}>100$~GeV$^2$.  We used as baseline the
\powhegboxres~\cite{Jezo:2015aia} (revision 4057) generator developed
by some of us~\cite{Banfi:2023mhz},\footnote{For this comparison, we
modified the \powhegboxres to use analytic matrix elements, which are
not enabled by default because they are only valid for the $\gamma^*$,
non-polarised process. This leads to a 20\% speed gain compared to the
default.} and we employed the same \texttt{NNPDF40MC\_nlo\_as\_01180}
PDF set.

\begin{figure}
  \centering
  \includegraphics[width=0.495\textwidth,page=2]{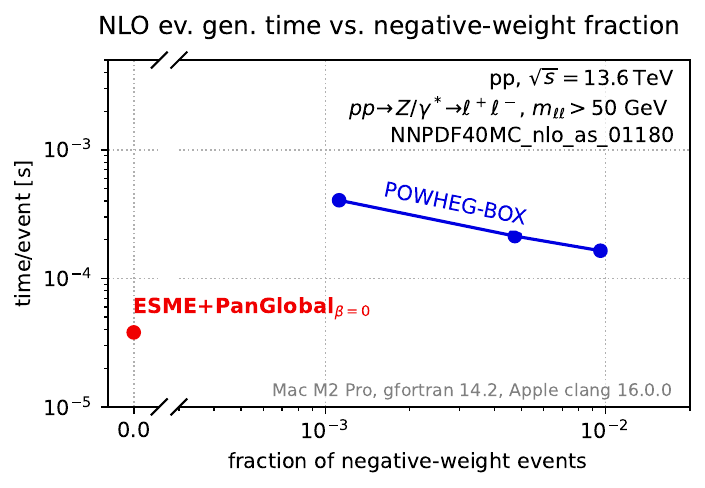}%
  \includegraphics[width=0.495\textwidth,page=2]{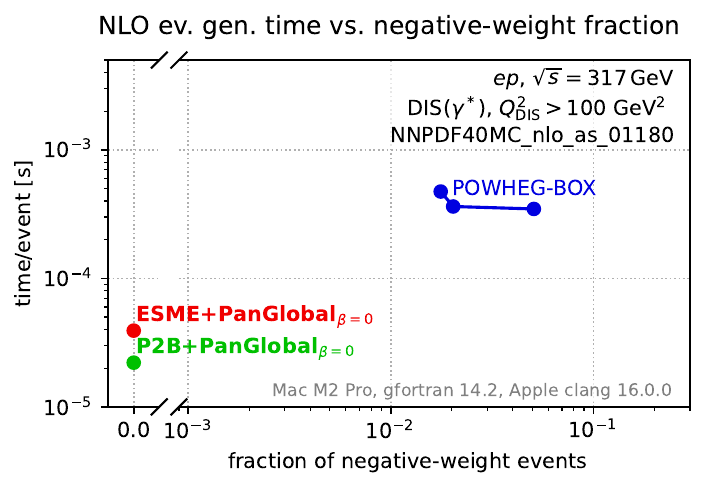}
  \caption{Illustration of the performance of our \ESME implementation
    as compared to \powhegboxvt, for $pp \to Z/\gamma^* \to
    \ell^+\ell^-$ (left panel), and of our \ESME and P2B
    implementations for the $\gamma^*$-mediated DIS as compared to the
    \powhegboxres (right panel).
    The plot shows the time per event versus the fraction of negative
    weights.
    The three \powhegbox points for Drell--Yan production correspond (from right
    to left) to folding choices for $\xi$, $y$ and $\phi$ of 1,1,1,
    2,1,1 and 5,1,1.
    For DIS we instead considered the folding choices: 1,1,1, 1,1,2
    and 1,1,5 (from right to left).
    For all cases, the event generation time is for NLO accuracy with
    just the hardest emission, and is evaluated by running on a single core
    of an Apple M2 Pro processor.
  }
  \label{fig:performance}
\end{figure}

The \powhegbox has so-called \texttt{nfold} parameters for each of the
three phase space variables, $(\xi, y, \phi)$, in the $\Bbar$
integration (cf.\ Section~\ref{sec:dbnlo}).
The left panel of Fig.~\ref{fig:performance} shows the fraction of
negative-weight events in the \powhegbox for Drell--Yan production versus the
timing per event, with the three blue points corresponding to folding
choices of 5,1,1 (left), 2,1,1 (middle) and 1,1,1
(right).\footnote{We also examined 5,5,1 and 2,2,1 and found the same
fractions of negative weights as 5,1,1 and 2,1,1, and slower timings.}
The fraction of negative weights is low for this process, below
$1\%$. For many practical purposes, a user could even set out to
verify that they do not cluster in any specific phase space regions
and then arguably just discard them.
Still this process helps illustrate the trade-off between folding,
event-generation time and negative-weight fraction.
For DIS, we found that the optimal folding is the one
over the $\phi$-variable, and we considered 1, 2 and 5 foldings (from
right to left).
The fraction of negative weight is slightly higher than in Drell--Yan production,
but always below 5\%, and in particular it is equal to 2\% with 2
foldings over the $\phi$-variable, which does not seem to induce an
appreciable speed penalty.
Increasing to 5 foldings leads only to a marginal reduction of
negative-weighted events, but a substantial increase of the run time.
In general, the higher fraction of negative weights compared to the
Drell--Yan case, is due to the lower scale of the process under
consideration.\footnote{Increasing the lower cut on $Q_{\rm DIS}$ to 50 GeV,
and using 2 folds on $\phi$, the fraction of negative weights is
0.2\%, i.e. 10 times smaller, and the time per event is 0.6 ms.}

Fig.~\ref{fig:performance} also shows the timing of the \ESME method
of Section~\ref{sec:ESME-core} (red point) for both Drell--Yan~(left)
and DIS~(right), as well as P2B of Sec.~\ref{sec:p2b} (green point)
for DIS only.  By construction, neither the ESME nor the P2B methods
have any negative-weight events.
ESME is about four times faster than the fastest of the \powhegbox
configurations for Drell--Yan, and ten times for DIS, taking about
$40\,\mu$s per event regardless of the process considered.
It is also interesting to note that the time per \ESME event is only
roughly double the P2B one, despite the more complicated rejection
algorithm involved in \ESME.
This high speed should be put into context: we devoted some effort to
understand the generation of the phase space, which led, e.g.\ to the
lepton-swap technique mentioned in Section~\ref{sec:real-radiation}
and also enabled us to limit the warm-up phase.
We hard-coded our own matrix elements, which allowed the matrix
element for the two lepton-swap configurations to be evaluated in
almost the same time as a single configuration.
We also used a pre-release version of \hoppet
1.3.0~\cite{Salam:2008qg,HoppetInPrep} to evaluate the PDFs at each
$x,\mu_F$ point, which appeared to bring some speed gain relative to
direct LHAPDF~\cite{Buckley:2014ana} evaluation (both optimised with
\texttt{-O3}).
Finally, focusing on Drell--Yan production, one should keep in mind that
parton showering with \PanScales (without the
\Pythia~\cite{Bierlich:2022pfr} interface) would add a further
$40{-}70\,\mu \text{s}$ per event.
However, for Drell--Yan production with today's tools, it would
ultimately be \Pythia's generation of hadronisation and, especially,
multi-parton interactions that would dominate, at about $2\,\text{ms}$
per event.

We also briefly compared Higgs production against the \powhegbox and,
  in that case too, found that \ESME was faster.\footnote{There is one
caveat here, namely that at low $\mu_F \sim 1\GeV$ the gluon
distribution in the \texttt{NNPDF40MC\_nlo\_as\_01180} set approaches
zero at small $x$, but not at moderate $x$.
  This leads to large ratios of
  $g(x_\text{large},\mu_F)/g(x_\text{small},\mu_F)$ in $R/B$ at low
  $\mu_F$.
  The \PanScales code currently uses an $x$-dependent but
  $\mu_F$-independent overhead factor.
  This ends up being set according to the very large overhead required
  at low $\mu_F$, with a corresponding speed impact.
  If we simply freeze the PDF set below $\mu_F = 1.3\GeV$, which has
  limited phenomenological impact, the large overhead is no longer
  required and we find speeds that are only $20\%$ slower than for
  di-lepton production.  }
A final test that we carried out was of our \dBNLO implementation
versus \ESME, for $Z$ production without lepton decays (our
dBNLO implementation with lepton decays relies on the \powhegbox $\tilde B$ and so
is not an independent speed test).
With folding turned off, the \dBNLO negative-event rate was similar to
that of \powhegbox, and the speed was about $30\%$ faster than \ESME. 
It remains to be seen, however, how this observation would generalise
to other processes.

\section{Conclusions}
\label{sec:conclusions}

In this article we have explored a range of methods for NLO matching
such that shower logarithmic accuracy is preserved and even augmented to
NNDL for event-shape like observables.
We focused on colour-singlet production in $pp$ collisions, oriented
$e^+e^- \to 2\,\text{jets}$ and DIS.
From a theoretical point of view, such matching is a milestone in its
own right.
It is also a crucial step towards higher general logarithmic accuracy
in parton showers, specifically NNLL accuracy.
In prior work~\cite{vanBeekveld:2024wws} in an $e^+e^-$ context, NNLL
shower accuracy has been shown to have up to $20\%$ effects, and so is
expected to have broad phenomenological importance, given that many
measurements have accuracies at the few percent level or better.
Thus our work here is an essential step on the way towards bringing
this progress to processes with incoming hadrons.

%

Some of the methods that we used are adaptations of existing
approaches.
That was the case for the \dBNLO method (Section~\ref{sec:dbnlo}),
which adapts the widely used POWHEG method.
It brings a generalisation of the ordering variable, uses a
real-matched shower instead of the FKS map, and a corresponding
$\Delta\Bbar$ added to standard FKS $\Bbar$ to account for the
different relation between the Born and real kinematics.
The projection to Born (P2B) method of Section~\ref{sec:p2b} is most
suited to DIS type processes, where it has been used in the past for NNLO
matching~\cite{Hoche:2018gti}.
One subtlety that we encountered concerns its use with
parity-violating interactions and the correct separation of flavour
and anti-flavour Born channels.

We also explored a qualitatively new approach.
In particular in Section~\ref{sec:Bbar-integer-alg}
(algorithm~\ref{alg:sub-to-int}) we introduced a new core approach ---
Exponentiated Subtraction for Matching Events (ESME) --- that reformulates
the standard Monte Carlo evaluation of $\Bbar$ as a problem of
generating a Sudakov distribution.
It can be straightforwardly made positive definite, while retaining
accuracy up to any specified order in $\as$.
We believe that it has the potential to serve as a foundation that
accommodates many variations and that it should be feasible also for
other codes to adopt and/or adapt it.
The specific variant that we chose was described in
Section~\ref{sec:ESME-core}.
We combined it with a shower-based NLO subtraction method in which we
started from a slicing calculation and used an approximation of the
shower to promote the slicing calculation into a subtraction method.
We envisage that this approach may also have wider applications.

To validate the methods, we carried out tests of both NLO and
event-shape NNDL accuracy.
In particular, given that shower NLO matching typically introduces
terms also beyond NLO, we highlighted the value of studying the
$\as\to0$ limit of the matched result, explicitly extracting the pure
NLO coefficient.
The use of the $\as \to 0$ limit is already widespread in logarithmic
accuracy tests and we applied it also for NNDL validation, verifying
that our matching correctly achieves this milestone, as needed for
future work on high logarithmic accuracy, notably for processes with
incoming hadrons.

A final consideration concerns event-generation efficiency.
Existing methods for reducing the fraction of negative-weight events
bring penalties in speed and/or complexity.
We compared the \esme method with the \POWHEGBOX with various degrees
of folding, cf.\ Fig.~\ref{fig:performance}.
Our implementation of the \esme method is not only positive definite,
but turned out to be
several times faster per event than (unfolded) \POWHEGBOX, which was
the fastest of the public tools we examined and the one with fewest
negative weights.
It also sidestepped the need for a substantial warm-up phase, making
for instance NLO Drell--Yan showered event generation as easy and fast
as LO \Pythia showered event generation.

Taken together, our results represent a key step on the path to higher
logarithmic accuracy in parton showers and also suggest that there may
be significant value in further exploring new matching methods at NLO
and beyond.

The developments presented in this work are available from
\url{https://gitlab.com/panscales/panscales-0.X}, as part of the 0.3.0
release of the PanScales code.

\section*{Acknowledgements}
We are grateful to our \panscales{} collaborators (Mrinal Dasgupta, Basem
El-Menoufi, Keith Hamilton, Pier Monni and Nicolas Schalch) for their work on the
code, the underlying philosophy of the approach and comments on this
manuscript. We also wish to thank Keith Hamilton for help with helicity amplitude calculations. 
We thank Luca Buonocore for sharing his numerical routines to evaluate the matrix elements
presented in Ref.~\cite{Buonocore:2019puv}, which we used during the early stages of this project.
We thank John Campbell for help with MCFM usage.
We are grateful to Fabrizio Caola for discussions about the swap
algorithm, Paolo Nason for helpful comments on the manuscript and both
for discussions on local subtraction schemes,
and to Stefano Forte and Juan Cruz Martinez for sharing a  pre-release of the \texttt{NNPDF40MC\_nlo\_as\_01180} PDF set (and Peter Skands for pointing us to them).
We are also grateful to Stefan H\"oche for discussions related to the
treatment of flavour in Ref.~\cite{Hoche:2018gti}, to Lois Flower for
a comment on the manuscript and to referees for their helpful
suggestions.

This work was supported
by a Royal Society Research Professorship
(RP$\backslash$R1$\backslash$231001) (GPS),
by the European Research Council (ERC) under the European Union’s
Horizon 2020 research and innovation programme (grant agreement No.\
788223, \panscales{}, GPS, SZ), 
by the Science and Technology Facilities Council (STFC) under
grant ST/X000761/1 (GPS, SZ), 
by the Dutch Research Council (NWO) under  project number VI.Veni.232.190 (MvB), 
by the Australian Research Council via Discovery Project DP230103014 (JH),
by the Australian Research Council through a Discovery
Early Career Researcher Award (project number DE230100867) (LS),
and by the Ramón y Cajal program under 
grant RYC2022-037846-I (ASO).
GPS also wishes to thank the Kavli Institute for Theoretical Physics
(KITP) for hospitality and support (grant NSF PHY-2309135) during the
completion of this work.

\appendix

\section{Overview of the \panscales{} showers}
\label{app:summary-panscales}
Here we summarise the kinematic maps for the showers considered in this work
and provide the shower emission probability in the absence of matching. The 
modifications to the latter when considering NLO matching are discussed in App.~\ref{sec:regions}.
A common ingredient entering the kinematic map of the \panscales{} showers and their 
emission kernel is a process-dependent reference four-vector $Q^\mu$, which defines a reference frame for measuring angular distances.
For the processes considered in this paper we set its 
four-momentum ($p_x, p_y, p_z, E$) to:
\begin{itemize}
\item \textbf{$e^+e^-$ collisions:}  for the decay of a colour singlet $X$ 
with momentum 
$p_X^\mu$, we use $Q^\mu=p_X^\mu$, 
i.e., we operate in the rest frame of $X$.
\item \textbf{$pp$ collisions:} when we consider the production of a colour singlet 
$X$ with mass $m_X$ and rapidity $y_X$ in hadron-hadron collisions, we set
\begin{equation}
Q^\mu = m_X \left(0, 0, \sinh y_X, \cosh y_X \right),
\end{equation}
This corresponds to the rest frame of the colour singlet before showering.
\item \textbf{Deep inelastic scattering:} when we consider
  lepton--hadron scattering $\ell(p_1) h(P) \to \ell' (p_2) X$,  
with the standard DIS variables
\begin{equation}
q_{\dis}^\mu = p_1^\mu -p_2^\mu\,, \qquad
x_{\dis} = \frac{-q_{\dis}^2}{2 q_{\dis} \cdot P}\,,
\end{equation}
we define the reference vectors
\begin{equation}
n_{\rm \sss in}^\mu  = x_{\dis} P^\mu, \qquad n_{\rm \sss out}^\mu = q_{\dis}^\mu + n_{\rm \sss in}^\mu,
\label{eq:nrefdis}
\end{equation}
which we employ to define
\begin{equation}
Q^\mu =n_{\rm \sss in}^\mu+n_{\rm \sss out}^\mu = q_{\dis}^\mu  +2 x_{\dis} P^\mu.
\label{eq:Qdis}
\end{equation}
\end{itemize}

\subsection{Shower emission probability}
\label{app:shower-em-prob}
Let us now consider the emission of a parton $k$ from a dipole $i$,\ $j$, with
pre-branching momentum $\tilde{p}_i$,\ $\tilde{p}_j$. We define the invariants
\begin{equation}
\widetilde{s}_{ij} = 2 \tilde{p}_i \cdot \tilde{p}_j\,, \quad \widetilde{s}_{i}
	= 2 \tilde{p}_i \cdot Q\,, \quad \widetilde{s}_{j} = 2 \tilde{p}_j \cdot Q\,,
\end{equation}
where $Q^\mu$ is the process-dependent reference four-vector introduced earlier.
The emission probability is expressed as a function of three variables: $v$, the
ordering variable, which carries the dimension of a transverse momentum, a rapidity-like
variable $\bar{\eta}$ and an azimuthal angle $\phi$.  From such variables we can
build the effective transverse momentum
\begin{equation}
\kappa_t = \kappa_t(v, \bar{\eta}, \widetilde{s}_{ij} , \widetilde{s}_{i}, \widetilde{s}_{j}\,, Q^2; \betaps),  
\label{eq:kappatgen}
\end{equation}
where the specific relation depends on the shower under consideration and is
detailed in App.~\ref{app:kin-maps}, and $0 \leq \betaps < 1$. 
Given $\kappa_t$ and $\bar \eta$,
we can compute two auxiliary variables,
\begin{equation}
\alpha_k = \sqrt{\frac{\kappa_t^2 \tilde{s}_j}{{s}_{ij}\tilde{s}_i}}
	e^{+\bar{\eta}}, \qquad \beta_k = \sqrt{\frac{\kappa_t^2
	\tilde{s}_i}{{s}_{ij}\tilde{s}_j}} e^{-\bar{\eta}},
\label{eq:alphakbetak}
\end{equation}
which we use to build the momentum fractions $z_i$ and $z_j$ that enter
into the emission probability. In particular we have for final-state 
branchings
\begin{equation}
z_i = \alpha_k, \quad z_j = \beta_k, \quad \text{(final-state)}\,,
\end{equation}
while for initial-state branchings,
\begin{equation}
z_i= \frac{\alpha_k}{1+\alpha_k}, \quad z_j = \frac{\beta_k}{1+\beta_k} \quad \text{(initial-state)}\, .
\end{equation}
The emission probability then reads
\begin{equation}
d\mathcal{P}_{\itilde\jtilde \to ijk}  = \sum_{\ell =i,j} \frac{\alpha_s^{\rm \sss eff}(\mu_R)}{\pi} \mathd\ln v
	\, \mathd\bar \eta \frac{\mathd\phi}{2\pi} \frac{\partial \ln
	\kappa_t}{\partial \ln v} \mathcal{L}_{\ell
	\tilde{\ell}}(\tilde{x}_{\ell}, z_{\ell}, \mu_F) z_\ell
	P_{\ell k}^{\rm \sss IS/FS}(z_{\ell}) f(\bar{\eta}_{\ell}), 
\label{eq:emsnprob}
\end{equation}
where 
\begin{itemize}
\item $\alpha_s^{\rm \sss eff}(\mu_R)$ is the QCD coupling constant evaluated at the
	scale $\mu_R=\kappa_t$ according to the Catani-Marchesini-Webber
		prescription~\cite{Catani:1990rr};
\item The luminosity factor is given by
	\begin{equation}
	\mathcal{L}_{\ell \tilde{\ell}}(\tilde{x}_{\ell}, z_{\ell}, \mu_F) =
	\begin{cases}
	\frac{f_{\ell}\left(\frac{\tilde{x}_\ell}{1-z_\ell}, \mu_F
	\right)}{(1-z_\ell) f_{\tilde{\ell}}(\tilde{x}_\ell, \mu_F)} &\mbox{ for initial state (IS)} \, ,\\
	1 &\mbox{for final-state (FS)}\, , 
	\end{cases}
	\end{equation}
	where $f_{a}(x,\mu_F)$ is the parton distribution function~(PDF) for a parton
	with flavour $a$ and longitudinal momentum fraction $x$, evaluated at the
	factorisation scale $\mu_F$, which we take to be equal to 
	\begin{equation}
	\mu_F = v \left( \frac{Q}{v}\right)^{\frac{\betaps}{1+\betaps}}.
	\label{eq:muF}
	\end{equation}
\item The splitting functions ${P}_{\ell k}^{\rm \sss IS/FS}(z_{\ell})$ 
are given by
 \begin{equation}
{P}_{\tilde{\ell} \ell}^{\rm \sss FS}(z_{\ell}) = {P}_{\tilde{\ell}
	\to \ell, k}^{\rm \sss FS}(z_{\ell}), \qquad {P}_{\tilde{\ell}
	\ell}^{\rm \sss IS}(z_{\ell}) = (1-z_{\ell}){P}_{\ell \to \tilde{\ell},
	k}^{\rm \sss IS}(z_{\ell}),
\end{equation}
where ${P}_{\ell \to \tilde{\ell},
	k}^{\rm \sss IS}(z_{\ell})$ are opportunely symmetrised DGLAP splitting functions, and are reported in appendix~A of Ref.~\cite{vanBeekveld:2022zhl}.

\item $f(\bar{\eta}_{\ell})$ is a function used to partition the emission
	probability between the leg $\itilde $ and $\jtilde$.
	For an antenna shower,  if $\itilde $ carries a colour index, and $\jtilde $
	carries the anti-colour one, we use $\bar{\eta}_i = +\bar \eta$,
	$\bar{\eta}_j = -\bar \eta$ and
	\begin{equation}
	f(\bar{\eta}) = f_{\rm \sss ant}(\bar{\eta}) =\frac{1}{1+e^{-2\bar\eta}}.
	\label{eq:fant}
	\end{equation}
	For a dipole shower, the two contributions in Eq.~\eqref{eq:emsnprob} are
	handled separately, so that we can imagine $\itilde$ always
	being the emitter and $\bar{\eta}_{\ell}=\bar{\eta}_i =\bar
	\eta$, and
	\begin{equation}
	f(\bar{\eta}) = g_{\rm \sss dip}(\bar{\eta}) = \begin{cases}
	0 & \mbox{if } \bar{\eta} <-1 \\
	\frac{15}{16}\left(\frac{\bar{\eta}^5}{5}-\frac{2\bar{\eta}^3}{3}+\bar{\eta}
	+\frac{8}{15} \right) & \mbox{if } -1\leq \bar{\eta} \leq 1 \\
	1 & \mbox{if } \bar\eta>1
	\end{cases}.
	\label{eq:fdip}
	\end{equation}
\end{itemize}
In the following we specify how to build $\kappa_t$, and the new momenta given
the shower variables $v$, $\bar{\eta}$ and $\phi$ for two variants of the
\panscales{} showers, namely \panglobal and \panlocal.

\subsection{Kinematic maps}
\label{app:kin-maps}
We now the discuss the emission generation 
for two variants of the \panscales{} showers, 
namely \panglobal{}~\ref{app:panglobal} 
and \panlocal{}~\ref{app:panlocal}.
For the latter, we introduce a new interpretation of 
the relation between the ordering variable $\ln v$ 
and the actual Sudakov decomposition used to write 
the new momenta, which enables better phase space coverage 
in the presence of initial-state radiation, 
as compared to the maps used in Refs.~\cite{vanBeekveld:2022zhl,vanBeekveld:2022ukn,vanBeekveld:2023chs}.
 
\subsubsection{The \panglobal{} shower}
\label{app:panglobal}
The \panglobal{} shower is an antenna shower with 
local longitudinal momentum conservation, 
but global transverse momentum conservation.
The first step is the calculation of the 
effective transverse momentum $\kappa_t$ 
of Eq.~\eqref{eq:kappatgen}, which is given by
\begin{equation}
\kappa_t  \equiv \left(\frac{\tilde{s}_i \tilde{s}_j}{\tilde{s}_{ij}Q^2} \right)^{\betaps} v e^{\betaps |\bar\eta|}, 
\label{eq:kappatdef}
\end{equation}
with $0\leq \betaps <1$. 
The Jacobian $\partial \ln \kappa_t /\partial \ln v$ appearing in eq.~\eqref{eq:emsnprob} is thus always 1.
One then builds the variables $\alpha_k$ and $\beta_k$ 
from Eq.~\eqref{eq:alphakbetak} that are used to define 
the intermediate momenta
\begin{subequations}
\begin{align}
\label{eq:panglobal-map}
\bar p_k^{\mu} =&\rl( \alpha_k \tilde{p}_i^{\mu}  + \beta_k \tilde{p}_j^{\mu}  + k_\perp^{\mu} ), \\
\bar p_i^{\mu}  =&\rl(1\mp \alpha_k)\tilde{p}_i ^{\mu} , \\
\bar p_j ^{\mu} =&\rl(1\mp \beta_k)\tilde{p}_j^{\mu},
\end{align}
\end{subequations}
where $k_\perp^{\mu} $ is a four-vector orthogonal 
to $\tilde{p}_i^{\mu} $ and $\tilde{p}_j^{\mu} $ 
with $k_\perp^2 = -\alpha_k \beta_k \tilde{s}_{ij}$, and
the sign $\mp$ takes into account if the leg is a 
final-state one ($-$) or an initial-state ($+$) one. 
The factor $\rl$ is equal to 1 for initial-initial (II) 
and initial-final (IF) dipoles, while it is different from 
1 for final-final (FF) dipoles. In particular, when considering
the decay of a colour singlet in $e^+e^-$ we define~\cite{FerrarioRavasio:2023kyg}
\begin{equation}
\rl = \frac{-\tilde{p}_m \cdot \bar{p}_{ijk}+\sqrt{(\tilde{p}_m \cdot \bar{p}_{ijk})^2+\bar{p}_{ijk}^2(Q^2-\tilde{p}_m^2)}}{ \bar{p}_{ijk}}, \quad (e^+e^-)
\label{eq:rLee}
\end{equation}
where $\tilde{p}_m = Q-\tilde{p}_i-\tilde{p}_j$ and $\bar{p}_{ijk}=\bar{p}_i+\bar{p}_j+\bar{p}_k$. 
For processes involving at least one initial-state hadron 
we use~\cite{vanBeekveld:2023chs}
\begin{equation}
\rl = \frac{\tilde{s}_i + \tilde{s}_j}{\tilde{s}_i + \tilde{s}_j+2k_\perp \cdot Q}  \quad (\text{$pp$ and DIS}).
\label{eq:rLhadr}
\end{equation}
The choice in Eq.~\eqref{eq:rLee} ensures that 
the mass of the colour-singlet is left unchanged, 
while the choice in Eq.~\eqref{eq:rLhadr} ensures that
the energy of the dipole in the rest frame of $Q^\mu$ does not change.
This factor was absent in the original formulation of the \panglobal{}
showers~\cite{Dasgupta:2020fwr,vanBeekveld:2022zhl} and it is
necessary~\cite{FerrarioRavasio:2023kyg} to avoid an issue of
long-distance correlations that otherwise arises with triple-collinear
configurations.

The relation between the final momenta $p_l^{\mu} $ and
$\bar{p}_l^{\mu}$ is process dependent.
\begin{itemize}
\item \textbf{$e^+e^-$:} for the decay of the colour singlet, one 
defines $\bar{Q}^{\mu} = \bar{p}_i^{\mu}  + \bar{p}_j ^{\mu} +\bar{p}_k^{\mu}  +\tilde{p}_m^{\mu}$, 
that corresponds to the new total final-state momentum, and applies 
the following boost to all final-state particles: 
\begin{equation}
\Lambda^{\mu \nu} = g^{\mu \nu} + \frac{2 Q^\mu \bar{Q}^\nu}{Q^2} -  \frac{2 (Q+\bar{Q})^\mu(Q+ \bar{Q})^\nu}{(Q+\bar{Q})^2}.
\end{equation}
\item  \textbf{$pp$:} when considering the production of a colour singlet 
$X$ in hadron collisions,%
\footnote{A discussion on the generalisation of our map 
for generic hadron-hadron collider processes can be found 
in Ref.~\cite{vanBeekveld:2022zhl}.} we use the colour singlet 
to absorb the transverse-momentum imbalance, and then rescale 
the beams to ensure longitudinal momentum conservation.
In practice, for each final-state particle, 
excluding the colour singlet, we define
\begin{equation}
p_l^{\mu}  = \bar{p}_l^{\mu} \; \mbox{ if }l \in i,j,k, \quad p_l^{\mu} =\tilde{p}_l^{\mu}\, \mbox{otherwise}.
\end{equation}
We then calculate ${p}_m$, i.e., the momentum of 
all final-state particles excluding $X$, and we 
decompose it along the directions of the hadron beams 
$P_a^{\mu} $ and $P_b^{\mu} $, to get
\begin{equation}
\bar{p}_m^{\mu}  = a_m P_a^{\mu} + b_m P_b^{\mu}  + q_\perp^{\mu} ,
\end{equation}
where $q_\perp$ is the transverse momentum component. 
The momentum of the colour-singlet is modified to be
\begin{equation}
p_X^\mu = \sqrt{\frac{|q_\perp^2| + m_X^2}{\sqrt{S}}}\left( e^{y_X} P_a^\mu  + e^{-y_X} P_b^\mu\right) - q_\perp^\mu,
\end{equation}
with $S=(P_a+P_b)^2$, and we reset the momenta of the 
incoming partons to be
\begin{equation}
\!\!p_a^\mu = \left( a_m + e^{y_X}\sqrt{\frac{|q_\perp^2| + m_X^2}{\sqrt{S}}}\right) P_a^\mu , \;\;\;\;
p_b^\mu = \left( b_m + e^{-y_X}\sqrt{\frac{|q_\perp^2| + m_X^2}{\sqrt{S}}}\right) P_b^\mu. \!
\end{equation}
\item \textbf{DIS:} for lepton-hadron collisions, we first calculate the total 
final-state momentum, excluding the final state lepton, $\bar{p}_X$, 
and we decompose it along the directions $n_{\rm \sss in}^{\mu}$ and 
$n_{\rm \sss out}^{\mu}$ introduced in Eq.~\eqref{eq:nrefdis}
\begin{equation}
\bar{p}_m^\mu = \frac{|q_\perp^2| + \bar{p}_m^2}{b_m} n_{\rm \sss in}^\mu + b_m \, n_{\rm \sss out}^\mu + q_\perp^{\mu} .
\label{eq:disFSpreboost}
\end{equation}
The incoming parton momentum is reset to (by construction $Q^2=-q_{\dis}^2$, 
see Eq.~\eqref{eq:Qdis})
\begin{equation}
p_a^\mu = \frac{\bar{p}_m^2 + Q^2}{Q^2} n_{\rm \sss in}^\mu,
\label{eq:beamindis}
\end{equation}
while all the final state partons are boosted according to
\begin{equation}
\Lambda^{\mu \nu} = g^{\mu \nu} + \frac{2 n_{\rm \sss in}^{\mu}}{Q^2}\left[(b_m-1) n_{\rm \sss out}^\nu + \frac{|q_\perp^2|}{b_m Q^2} n_{\rm \sss in}^\nu + q_\perp^\nu \right] + \frac{2 n_{\rm \sss out}^{\mu} n_{\rm \sss in}^{\nu}}{Q^2} \frac{1-b_m}{b_m} - \frac{2 q_\perp^{\mu} n_{\rm \sss in}^{\nu}}{b_m Q^2}.
\label{eq:disboost}
\end{equation}
It is easy to see that
\begin{equation}
p_m \equiv \Lambda^{\mu}_{\phantom{\mu} \nu} \bar{p}_m^{\mu} =  \frac{\bar{p}_m^2 }{Q^2} n_{\rm \sss in}^\mu + n_{\rm \sss out}^\mu,
\end{equation}
so that $p_m^{\mu} - p_a ^{\mu}= q_{\dis}^{\mu}$, 
as required for momentum conservation.
Notice that the boost in Eq.~\eqref{eq:disboost} was specifically 
designed to avoid assigning a large transverse-momentum component 
to partons aligned along the incoming-beam direction 
$n_{\rm \sss in}^{\mu}$~\cite{vanBeekveld:2023chs}.
\end{itemize}

\subsubsection{The (new) \panlocal{} shower}
\label{app:panlocal}
We now present a refined version for handling 
initial-state radiation in the \panlocal{} shower, as compared to that presented in 
Refs.~\cite{vanBeekveld:2022zhl,vanBeekveld:2023chs}, 
that is well-suited for generic processes and has an 
improved phase-space coverage.
Again, the first step consists in calculating 
the effective transverse momentum $\kappa_t$
\begin{equation}
\kappa_t  \equiv \min \left(\tilde{\kappa}_t, \mu_F \right), 
\end{equation}
where $\tilde{\kappa}_t$ coincides with the \panglobal{} 
definition in Eq.~\eqref{eq:kappatdef}, while $\mu_F$ is 
defined in Eq.~\eqref{eq:muF}. This ensures we switch to 
transverse-momentum ordering when, in the rest frame of 
$Q^{\mu}$, the emission is hard-collinear with energy larger 
than $\sqrt{Q^2}$.
This is necessary to avoid long-distance correlations 
in the presence of very energetic collinear emissions, 
as pointed out in Ref.~\cite{vanBeekveld:2022zhl}.
Notice that now the Jacobian in Eq.~\eqref{eq:emsnprob} will be
\begin{equation}
\frac{\partial \ln \kappa_t}{\partial \ln v}
=
\begin{cases}
1 & \mbox{ for } \kappa_t = \tilde{\kappa}_t \\
\frac{1}{1+\betaps} & \mbox{ for } \kappa_t = \mu_F.
\end{cases}
\end{equation}
Given $\kappa_t$ and $\bar{\eta}$, we then build the auxiliary 
variables $\alpha_k$ and $\beta_k$ as in Eq.~\eqref{eq:alphakbetak}.
These variables are then employed to build the coefficients 
$a_k$ and $b_k$, which are used to construct the momenta of 
the new emission $\bar p_k$
\begin{equation}
\bar p_k^{\mu} = a_k \tilde{p}_i ^{\mu} + b_k \tilde{p}_j ^{\mu} + k_\perp^{\mu} ,
\end{equation}
with $k_\perp^2 = - a_k b_k \tilde{s}_{ij}$.
How $a_k$ and $b_k$ are related to the auxiliary variables 
$\alpha_k$ and $\beta_k$, and how the recoiled momenta 
$\bar p_{i,j}$ are defined, depends on the type of the dipole.
In the following we consider only the dipole variant, 
and not the antenna one, so we need to distinguish between 
IF (initial-final) and FI (final-initial) dipoles, where the 
first label is used to denote the emitter.
The transverse momentum is conserved locally within the 
dipole, and in particular it is absorbed only by the emitter.
\begin{itemize}
\item For FF and FI dipoles, we choose
\begin{equation}
a_k = \alpha_k, \qquad b_k = \beta_k 
\end{equation}
and
\begin{subequations}
\begin{align}
\bar p_i =& (1-a_k) \tilde{p}_i + \frac{a_k b_k}{1-a_k}\tilde{p}_j - k_\perp\, , \\
\bar p_j =& \frac{1-a_k \mp b_k}{1-a_k} \tilde{p}_j,
\end{align}
\end{subequations}
where the sign $-$ is used when $j$ is a final-state spectator, 
and the sign $+$ is used for a FF dipole.
\item For an II map, we choose
\begin{equation}
a_k = \frac{\alpha_k}{\sqrt{1+\alpha_k \beta_k}}, \qquad b_k = \beta_k \frac{\left(\alpha_k +\sqrt{1+\alpha_k \beta_k}\right)^2 }{\sqrt{1+\alpha_k \beta_k}},
\label{eq:akbkIIPL}
\end{equation}
and the recoiled momenta read
\begin{subequations}
\begin{align}
\bar p_i =& (1+a_k) \tilde{p}_i + \frac{a_k b_k}{1+a_k}\tilde{p}_j + k_\perp\, , \\
\bar p_j =& \frac{1+a_k + b_k}{1+a_k} \tilde{p}_j\, , \\
\bar p_m = & \tilde{p}_i +\tilde{p}_j,
\end{align}
\end{subequations}
where $\bar p_m$ denotes the collection of all  
other final state particles, except $k$.
This choice ensures that the transverse momentum 
of $\bar{p}_k$ with respect to the new beams 
$\bar{p}_i$ and $\bar{p}_j$ is equal to $\kappa_t^2$, 
and that the difference in rapidity between $\bar{p}_k$ 
and $\bar{p}_m$ is equal to
\begin{equation}
y_k -y_m = \frac{1}{2} \ln \frac{(\bar p_k \cdot \bar{p}_j)(\bar p_m \cdot \bar{p}_i)}{(\bar p_k \cdot \bar{p}_i)(\bar p_m \cdot \bar{p}_j)}=\bar \eta +\frac{1}{2}\ln \frac{\tilde{s}_i}{\tilde{s}_j}.
\end{equation}
\item For an IF map we instead adopt
\begin{equation}
a_k = \frac{\alpha_k}{(1-\beta_k)^2-\alpha_k\beta_k}, \qquad
b_k = \frac{\beta_k(1+\alpha_k-\beta_k)^2}{(1-\beta_k)^2-\alpha_k\beta_k},\label{eq:akbkIFstandard}
\end{equation}
and
\begin{subequations}
\begin{align}
\bar p_i =& (1+a_k) \tilde{p}_i + \frac{a_k b_k}{1+a_k}\tilde{p}_j + k_\perp\, , \\
\bar p_j =& \frac{1+a_k - b_k}{1+a_k} \tilde{p}_j. 
\end{align}
\end{subequations}
The expressions in Eq.~\eqref{eq:akbkIFstandard} 
are derived such that the invariants obtained with the IF map 
exactly match those of the corresponding FI map.
From Eq.~\eqref{eq:akbkIFstandard} it is also easy to 
notice that there is a non-singular region corresponding 
to large $\beta_k$, where $(1-\beta_k)^2 < \alpha_k\beta_k$, 
such that $a_k$ and $b_k$ can become negative.
If we look at the corresponding FI map in the dipole frame, 
this would correspond to a case where the original final-state 
leg develops a longitudinal component along the incoming 
parton $\tilde{p}_i$ which is larger than the residual 
component along $\tilde{p}_j$ (i.e., if we think about the 
first emission in DIS, the original final-state leg 
recoils in the remnant hemisphere).
Thus, to populate this region, when 
$(1-\beta_k)^2 < \alpha_k\beta_k$ we instead use
\begin{equation}
b_k = -\frac{\alpha_k}{(1-\beta_k)^2-\alpha_k\beta_k}, \qquad
a_k = -\frac{\beta_k(1+\alpha_k-\beta_k)^2}{(1-\beta_k)^2-\alpha_k\beta_k},\label{eq:akbkIFflip}
\end{equation}
and
\begin{subequations}
\begin{align}
\bar p_i =& \frac{a_k b_k}{b_k-1} \tilde{p}_i + (b_k-1)\tilde{p}_j + k_\perp\,, \\
\bar p_j =& \frac{1+a_k - b_k}{b_k -1} \tilde{p}_i. 
\end{align}
\end{subequations}
\end{itemize}
Notice that in the soft and in the collinear limit 
($\beta_k \ll 1$), both for the IF and II maps we 
obtain $a_k \sim \alpha_k$ and $b_k \sim \beta_k(1+\alpha_k)^2$, 
as originally implemented in the old map of Ref.~\cite{vanBeekveld:2022zhl}.
All the FF, FI, II and IF maps are fully local, 
in the sense $\pm \tilde{p}_i \pm \tilde{p}_j = \pm p_i \pm p_j+p_k$ 
($+$ for outgoing partons, $-$ for incoming ones).
For FF and FI, we can use $p_l =\bar{p}_l$, where 
$p_l$ are the final momenta used to update the event record.
For IF and II maps, the incoming momentum $\bar p_i$ 
is no longer aligned with the beam, so we need to apply 
a Lorentz transformation to the momenta in the event 
to realign the beams.
The Lorentz transformation depends upon the process 
under consideration.
\begin{itemize}
\item \textbf{pp:} we apply a boost 
and a rotation so that $\bar p_i$ (the emitting initial-state parton) 
and $\bar p_b$ (the other initial-state parton, which coincided 
with $\bar p_j$ for an II map) are back-to-back and aligned 
along the $z$ axis.
For the case of colour-singlet production, we then apply 
a longitudinal boost so that the rapidity of the colour 
singlet is preserved.\footnote{An alternative option, suited for generic processes, 
is to apply a longitudinal boost so that $p_b = \tilde{p}_b$, 
i.e. the other initial-state beam is preserved.}
\item \textbf{DIS:} we first apply a rotation 
to the initial-state and to the final-state partons (i.e. all the 
particles but the leptons) so that $\bar p_i$ is aligned along 
the direction of the incoming proton.
Then we decompose the total partonic final-state 
momentum as in Eq.~\eqref{eq:disFSpreboost}, and we apply the boost in 
Eq.~\eqref{eq:disboost} to all the partons (including $\bar{p}_i$).
This boost simply acts as a rescaling for everything 
parallel to $n_{\rm \sss in}^\mu$, so that the final expression 
for $p_i^\mu$ corresponds to the one in Eq.~\eqref{eq:beamindis}.
\end{itemize}


\section{Hardest emission matrix elements in the \panscales{} showers}
\label{sec:regions}
When considering NLO matching, the effective parton-shower 
matrix element (see Eq.~\eqref{eq:emsnprob}) is replaced with 
\begin{equation}
\label{eq:app-shower-nlo-weight}
\sum_{\ell =i,j} 
\mathcal{L}_{\ell \tilde{\ell}}(\tilde{x}_{\ell}, z_{\ell}, \mu_F) z_\ell
	P_{\ell k}^{\rm \sss IS/FS}(z_{\ell}) f(\bar{\eta}_{\ell}) \to
	\frac{d\Phi}{d\PhiB d\ln \kappa_t d \bar \eta d\phi}\frac{R_{ijk}(\mu_R,\mu_F;
	\Phi)}{{B}_{\itilde\jtilde}(\mu_R,\mu_F;\PhiB)},
\end{equation} 
for the first emission only. There are three different
ingredients in the NLO shower weight given by Eq.~\eqref{eq:app-shower-nlo-weight}. 
First, we have introduced
a process-dependent Jacobian associated with the transformation from 
the radiative phase-space to the shower variables.
More concretely, the phase-space factor is equal to   
\begin{equation}
  \frac{d\Phi}{d\PhiB d\ln \kappa_t d \bar \eta d\phi} =
    \frac{k_t^2}{16\pi^2} \frac{d\phi}{2\pi} \times
  \begin{cases}
  \displaystyle
   1 & \mbox{ for colour-singlet production, } 
   \\
  \displaystyle \frac{s_{ij} + s_{ik}}{s_{ij}} & \mbox{ for DIS (with } i \mbox{
    IS parton, }k \mbox{ the emitted one).} 
  \end{cases}
  \end{equation}
where $k_t$ is the physical transverse momentum of the emission.
The corresponding phase-space factor for $e^+e^-$
is given in Appendix C.1 of Ref.~\cite{Hamilton:2023dwb}.
Another ingredient in Eq.~\eqref{eq:app-shower-nlo-weight} is the 
Born squared matrix-element
\begin{equation}
\label{eq:app-born-me}
{B}_{\itilde\jtilde}(\mu_R,\mu_F; \PhiB) = f_{\itilde}(\tilde{x}_i,
	\mu_F)  f_{\jtilde}(\tilde{x}_j, \mu_F)
	\frac{|M^{(0)}(\PhiB)|^2}{2\tilde{s}},
\end{equation}
evaluated at the underlying Born phase-space point $\PhiB$. This includes 
the tree-level Born matrix element divided by the flux factor, $|M^{(0)}(\PhiB)|^2/(2\tilde{s})$,
and the product of PDFs evaluated at scale $\mu_F$ as given in Eq.~\eqref{eq:muF}. 
Similarly, $R_{ijk}(\mu_R,\mu_F; \Phi)$ is the real matrix
element, stripped of a factor $\alpha_s/\pi$ 
\begin{equation}
R_{ijk}(\mu_R,\mu_F;\Phi) = \frac{\pi}{\alpha_s(\mu_R)} f_{i}(x_i,
	\mu_F)  f_{j}(x_j, \mu_F) \frac{|M^{(0)}(\Phi)|^2}{2s}.
\end{equation}
As discussed in the main text, the real-matrix element needs to be  
partitioned between all possible emitting dipoles. Denoting with $R(\Phi)$ 
the total real matrix element (including PDF and $\pi/\as$ factors) we write
\begin{equation}
  R_{ijk}(\mu_R,\mu_F;\Phi) = \sum_{p} R_p(\Phi)=\sum_{p} R(\Phi) w_{p}(\Phi),
\end{equation}
where the $w_{p}(\Phi)$ are the partitioning functions (satisfying
$0 \leq w_{p}(\Phi) \leq 1$) that we use to separate our total real
cross section into several blocks, each of them to be interpreted as a
dipole emission probability.
The $e^+e^-$ case 
was worked out in Appendix C of Ref.~\cite{Hamilton:2023dwb}. Here, 
we discuss how to build the partitioning for all other processes. 
\begin{itemize}
\item \textbf{Drell Yan}. Let us consider the LO process $ q \bar{q} \to Z$, which is characterised by the following oriented dipoles:
  $(q,\bar{q})$ and $(\bar{q}, q)$, where the first element denotes the emitter. 
  If we assume the quark moves along the positive $z$ direction, we have
  \begin{enumerate}
  \item gluon emission from the dipole $(q,\bar{q})$
    \begin{equation}
      R_1(\Phi) = R_{q\bar q \to Z g}(\Phi) \times        
      \begin{cases}
        g_{\rm \sss dip}(y-y_Z) &\mbox{ for \PanLocal} \\
        f_{\rm \sss ant}(y-y_Z) &\mbox{ for \PanGlobal}, 
      \end{cases}
    \end{equation}
    where $y$ is the rapidity of the emission, and $y_Z$ is the rapidity of the $Z$ boson;
  \item gluon emission from the dipole $(\bar{q},q)$
    \begin{equation}
      R_2(\Phi) = R_{q\bar q \to Z g}(\Phi)-R_1(\Phi);
    \end{equation}
  \item anti-quark emission from the dipole  $(q,\bar{q})$
   \begin{equation}
      R_3(\Phi) = R_{g\bar q \to Z \bar q}(\Phi) ;       
   \end{equation}
  \item quark emission from the dipole $(\bar{q},q)$
   \begin{equation}
      R_4(\Phi) = R_{q g \to Z q}(\Phi).       
   \end{equation}   
  \end{enumerate}
\item \textbf{Gluon fusion}. Let us consider the LO process $ g_1 g_2 \to H$, which is characterised by the following oriented dipoles:
  $(g_1^{C},g_2^{A})$, $(g_1^{A},g_2^{C})$, $(g_2^C,g_1^A)$
  and $(g_2^A,g_1^C)$, where again the first element is the emitter. The prefix $C$ and $A$ denote if the gluon carries a colour or an anti-colour index.  
  If we assume $g_1$ moves along the positive $z$ direction, we have
  \begin{enumerate}
  \item gluon emission from the dipole $(g_1^C, g_2^A)$
    \begin{equation}
      R_1(\Phi) = \frac{1}{2}R_{g\bar g \to H g}(\Phi) \times        
      \begin{cases}
        g_{\rm \sss dip}(y-y_H) &\mbox{ for \PanLocal} \\
        f_{\rm \sss ant}(y-y_H) &\mbox{ for \PanGlobal}, 
      \end{cases}
    \end{equation}
    where $y$ is the rapidity of the emission, and $y_H$ is the rapidity of the $H$ boson;
  \item gluon emission from the dipole $(g_1^A, g_2^C)$
 \begin{equation}
      R_2(\Phi) = R_1(\Phi);      
 \end{equation}
\item gluon emission from the dipole $(g_2^C, g_1^A)$
    \begin{equation}
      R_3(\Phi) = \frac{1}{2} R_{g\bar g \to H g}(\Phi) -  R_2(\Phi)
    \end{equation}
 \item gluon emission from the dipole $(g_2^A, g_1^C)$
 \begin{equation}
       R_4(\Phi) = R_3(\Phi);      
  \end{equation}
 \item anti-quark emission from the dipole $(g_1^C, g_2^A)$
 \begin{equation}
   R_5(\Phi) = R_{\bar{q} g \to H\bar{q}} (\Phi);
 \end{equation}
\item anti-quark emission from the dipole $(g_2^C, g_1^A)$
\begin{equation}
  R_6(\Phi) = R_{g\bar{q} \to H\bar{q}} (\Phi);
\end{equation}    
\item quark emission from the dipole $(g_1^A, g_2^C)$
\begin{equation}
  R_7(\Phi) = R_{q g \to Hq} (\Phi)
\end{equation}
\item quark emission from the dipole $(g_2^A, g_1^C)$
\begin{equation}
  R_8(\Phi) = R_{gq \to Hq} (\Phi).
\end{equation}   
  \end{enumerate}
\item \textbf{Deep inelastic scattering}. Let us consider the LO process $\ell q_I \to \ell q_F$, and assume the momenta are in the Breit frame and $q_F$ moves along the positive $z$ direction.
We have three regions:
\begin{enumerate}
  \item gluon emission from the dipole $(q_F,q_I)$
    \begin{equation}
      R_1(\Phi) = R_{\ell q \to \ell q g }(\Phi) \times        
      \begin{cases}
        g_{\rm \sss dip}(y) &\mbox{ for \PanLocal} \\
        f_{\rm \sss ant}(y) &\mbox{ for \PanGlobal}, 
      \end{cases}
    \end{equation}
    where $y$ is the rapidity of the emission in the Breit frame;
  \item gluon emission from the dipole $(q_I,q_F)$
    \begin{equation}
      R_2(\Phi) = R_{\ell q \to \ell q g }(\Phi) -R_1(\Phi);
    \end{equation}
  \item anti-quark emission from the dipole $(q_I,q_F)$
  \begin{equation}
    R_3(\Phi) = R_{\ell g \to \ell q \bar q }(\Phi) \Theta(p_{\bar q} \cdot p_g < p_q \cdot p_g  ),
    \label{eq:DISisrq}
  \end{equation}
  where, as discussed before the $\Theta$ prevents double counting
  this contribution since it can also be reached starting from the LO process $\ell \bar{q}_I \to \ell \bar{q}_F$.
\end{enumerate}
\end{itemize}

\section{Slice to subtraction expressions for processes with two coloured legs}
\label{sec:S2S-incoming-hadrons}

In this Appendix we give the expressions for the $\bar{B}_C(\Phi_B)$
weights, and the corresponding differential counterterms, for the
PanGlobal $pp$, DIS and $e^+ e^-$ shower variants for all processes
considered in this paper. 
\subsection{$e^+e^- \to q\bar{q}$ and $H\to gg$}
For the $e^+ e^-$ \PanGlobal shower we use a counterterm real-radiation
probability given by (cf.\ Section \ref{sec:counterterm-from-slicing})
\begin{align}
  \label{eq:ee-diff-counterterm}
  \frac{C(\Phi)}{B_0(\PhiB)} d\Phirad \to \frac{dv}{v} d\bar \eta
  \frac{d\phi}{2\pi} 
  \frac{\as}{\pi}  z P_{ij}(z),
  \quad
  \ln z = \bar\eta - \bar\eta_{\max},
  \quad
  0 < \bar\eta < \bar\eta_{\max} = \ln Q/v,
\end{align}
where the splitting functions that enter into the $e^+e^- \to
q\bar{q}$ and $H\to gg$ processes are given by
\begin{subequations}
  \begin{align}
    P_{gq}(z)&=C_F\frac{1+(1-z)^2}{z}, \\
    P_{gg}(z)&=C_A\left[\frac{1 + (1-z)^3}{z} + w_{gg}(1-2z)\right], \\
    P_{qg}(z)&=2T_R\left[(1-z)^2-w_{qq}(1/2-z)\right]\,. 
  \end{align}
\end{subequations}
The factors $w_{qq}$ and $w_{gg}$ govern the partitioning of the
$P_{qg}$ and $P_{gg}$ splitting functions as they enter in the
PanGlobal shower~\cite{Dasgupta:2020fwr}, and are by default set to
$0$. These counterterms are simple enough to be integrated
analytically, and following the procedure of
Section~\ref{sec:counterterm-from-slicing} we arrive at the following
weight for $e^+e^- \to q\bar{q}$
\begin{align}
\Bar{B}_C^{\text{PG},q\bar{q}}(\Phi_B) &= B^{e^+ e^- \to q\bar{q}}_0(\PhiB) \left( 1
+ \frac{\alpha_s C_F}{2\pi}\left[5 - \frac{\pi^2}{3}\right]\right)\,.
\end{align}
Similarly, we can derive the effective NLO weight for $H\to gg$, which
enters into the decay width
\begin{align}
\Bar{B}_C^{\text{PG},gg}(\Phi_B) &=  B^{H \to gg}_0(\PhiB) \left[1 + \frac{\alpha_s(\mu_R)}{2\pi}\left(C_A \left(\frac{167}{9} - \frac{\pi^2}{3}\right) - T_R n_f  \frac{46}{9} + 8 \pi b_0 \ln \frac{\mu_R}{Q}\right)\right]\,.
\end{align}

\subsection{Worked  example for Drell--Yan production}
\label{app:slice-pp2Z}
Before giving the equations that enter into processes with
initial-state radiation, it is instructive to look at a detailed
example. For this purpose we will look at $pp \to V$. The results
presented here are independent of whether or not the vector boson is
allowed to decay.
The first step is to formulate a shower counterterm where the ordering variable
resembles that of the actual shower in the infrared limits, and the
emission probability reproduces the correct singularity structure of
the full matrix element. This ``approximate shower'' has to be simple
enough that it can be integrated at order $\as$ above some slicing
cutoff.
It will have approximate phase-space bounds and will not need an
explicit kinematic mapping other than in the IR.
For our $pp$ showers, we will define a shower that has a single
(unregularised) leading-order splitting function $p_{ij}(z)$ for each
side of the event, i.e.\ $p_{ij}(z_1)\Theta(\bar{\eta}>0)$ and
$p_{ij}(z_2)\Theta(\bar{\eta}<0)$, where $z_1$ and $z_2$ are momentum
fractions for the forward and backward going beams respectively.
We consider the $\betaps = 0$ \PanGlobal shower in the $\bar{\eta}>0$
hemisphere, and define (see
App.~\ref{app:summary-panscales})
\begin{align}
\label{eq:z-defn}
\bar{z} \equiv 1-z =\frac{1}{1+\alpha_k}\,, \quad \alpha_k = \frac{\kappa_t}{{Q}} e^{\bar\eta}\,,
\end{align}
where $Q$ is the invariant mass of the colour singlet.
The requirement $\bar\eta >0$ translates to 
\begin{align}
\label{eq:z-limit-approx}
\bar{z} < \frac{{Q}}{{Q}+\kappa_t}\,.
\end{align}
The counterterm real-radiation probability is then given by
\begin{align}
  \label{eq:CTDY}
  \frac{C(\Phi)}{B_0(\PhiB)} d\Phirad \to \frac{d\kappa_t}{\kappa_t} d\bar \eta
  \frac{d\phi}{2\pi} 
  \frac{\as}{\pi}  (1-\bar{z}) p_{qk}(\bar{z}) \frac{f_{k}(x/\bar{z})}{f_q(x)} \Theta\left(x<\bar{z}<\frac{{Q}}{{Q}+\kappa_t}\right)\,,
\end{align}
where $x$ is the momentum fraction of the incoming parton. For
$q\bar{q} \to Z$ there are two channels to consider: one where the
flavour is preserved, and one where the (anti-)quark backwards evolves
into a gluon.
The corresponding splitting functions are given by
\begin{subequations}
  \begin{align}
    p_{qq}(\bar{z}) &= C_F\left[\frac{1+\bar{z}^2}{1-\bar{z}}\right]\,, \\
    p_{qg}(\bar{z}) &= T_R [\bar{z}^2 + (1-\bar{z})^2]\,.
  \end{align}
\end{subequations}
Above a slicing cutoff $\kappa_t = {Q} e^{L}$, the approximate shower will
give a cross section of
\begin{align}
\label{eq:shower-approx-pp-1}
\frac{C_{\text{int}}(v>{Q}e^L)}{B_0(\PhiB)} =
  2\frac{\as}{2\pi}\sum_{i=1,2}\sum_{k\in{q,g}}
  \int \frac{d\kappa_t}{\kappa_t} \frac{d\bar{z}}{\bar{z}}
  p_{qk}(\bar{z}) \frac{f_{k}(x_i/\bar{z})}{f_q(x_i)}
  \Theta\left(x_i<\bar{z}<\frac{{Q}}{{Q}+\kappa_t}\right)
\Theta({Q}e^L < \kappa_t < {Q})\,,
\end{align}
where $f_i(x) \equiv f_i(x,\mu_F=\kappa_t)$ denote the PDFs for flavour $i$ and
momentum fraction $x$. 
This may be written in a form that separates the PDF dependence as
\begin{align}
\label{eq:shower-approx-pp-2}
  \frac{C_{\text{int}}(v>{Q}e^L)}{B_0(\PhiB)}
  &= 2\frac{\as}{2\pi}\int_{{Qe}^L}^{{Q}} \frac{d\kappa_t}{\kappa_t} \Bigg[ 2 \int_0^{{Q}/({Q}+\kappa_t)} d\bar{z}\, p_{qq}(\bar{z}) \nonumber \\
  & \qquad \qquad
	+ \sum_{i=1,2} \int_{x_i}^1
    \frac{d\bar{z}}{\bar{z}}\left(p_{qq}(\bar{z})\Theta\left(\bar{z} <
    \frac{{Q}}{{Q}+\kappa_t}\right)\right)_+ \frac{f_q(x_i/\bar{z})}{f_q(x_i)}
    \nonumber
  \\
  & \qquad \qquad+ 
	 \sum_{i=1,2} \int_{x_i}^{{Q}/({Q}+\kappa_t)} \frac{d\bar{z}}{\bar{z}} p_{qg}(\bar{z}) \frac{f_g(x_i/\bar{z})}{f_q(x_i)} \Bigg]. 
\end{align}
In the limit of large negative $L$, the first line gives us
\begin{align}
\label{eq:no-pdf-approx}
4\int_{{Qe}^L}^{{Q}} \frac{d\kappa_t}{\kappa_t}  \int_0^{{Q}/({Q}+\kappa_t)} d\bar{z}\, p_{qq}(\bar{z})  = 4 C_F L^2 + 6C_F L +  \bar{H}^{(1)}_{qq} + \mathcal{O}(e^L)\,, 
\end{align}
with
\begin{align}
\label{eq:H1bar-DY}
\bar{H}^{(1)}_{qq} = C_F \left(1 +6 \ln2 + \frac{2\pi^2}{3} \right)\,.
\end{align}
For the last two lines of Eq.~\eqref{eq:shower-approx-pp-2} we replace
$\Theta\left(\bar{z} < \frac{{Q}}{{Q}+\kappa_t}\right)$ with
$1-\Theta\left(\bar{z} > \frac{{Q}}{{Q}+\kappa_t}\right)$ and interchange the
$\kappa_t$ and $\bar{z}$ integration orders.
We get for these last two lines
\begin{align}
	\label{eq:pdf-approx}
	= -2L\frac{(P_{qk} \otimes f_{k})(x_i)}{f_q(x_i)} + \frac{(\bar{C}^{(1)}_{qk}\otimes f_{k})(x_i)}{f_q(x_i)} + i\leftrightarrow j\,,
\end{align}
where $P_{qk}$ is now the regularised splitting function and 
\begin{subequations}
    \label{eq:approx-CcoeffDY}
  \begin{align}
    \bar{C}^{(1)}_{qq}(\bar{z}) &= -2\left(p_{qq}(\bar{z})\ln\frac{\bar{z}}{1-\bar{z}}\Theta(\bar{z}>1/2) \right)_+ \!, \\
    \bar{C}^{(1)}_{qg}(\bar{z}) &= -2 p_{qg}(\bar{z})\ln\frac{\bar{z}}{1-\bar{z}}\Theta(\bar{z}>1/2)\,.
  \end{align}
\end{subequations}
The complete result for the counterterm thus becomes 
\begin{align}
\label{eq:shower-approx-pp-final}
\frac{C_{\text{int}}(v>{Q}e^L)}{B_0(\PhiB)} =
	\frac{\alpha_s}{2\pi}\left[4 C_F L^2 + 6C_F L +  \bar{H}^{(1)}_{qq} +
	\frac{\left((-2LP_{qk} + \bar{C}^{(1)}_{qk})\otimes
	f_{k}\right)(x_i)}{f_q(x_i)} + i\leftrightarrow j \right]\,.
\end{align}
We tabulate and then interpolate the results of the convolutions with
\hoppet, which we adapted in order to accurately handle the
$\Theta(\bar z > 1/2)$ in Eq.~(\ref{eq:approx-CcoeffDY}).

To work out $\bar B_C$ we also need a slicing calculation in the
shower variable, $\Sigma^{\text{NLO}}_{\text{PG}}(v<{Q}e^L)$. The first
emission of the \PanGlobal shower coincides with the transverse
momentum of the leading jet.
This means we may use the results of Ref.~\cite{Banfi:2012jm}
(cf.\ the supplemental material therein) directly to obtain
$\Sigma^{\text{NLO}}_{\text{PG}}(v<{Q}e^L)$.
At $\mathcal{O}(\alpha_s)$, at fixed Born flavours with momentum fractions $x_i$
and $x_j$, we can write those results as
\begin{align}
\label{eq:resum-nlo-result}
\frac{\Sigma(v < {Q}e^L) }{B_0(\PhiB)} = 1 +
	\frac{\alpha_s}{2\pi}\left[-2A_q^{(1)}L^2 + 2B_q^{(1)}L +
	H^{(1)}_{q\bar{q}} + \frac{\left((2L P_{ik} +
	C^{(1)}_{ik})\otimes f_{k}\right)(x_i)}{f_i(x_i)} + i\leftrightarrow
	j\right]\,,
\end{align}
with 
\begin{subequations}
  \begin{align}
&A_q^{(1)} = 2C_F\,, \quad B^{(1)}_q = -3C_F\,, \quad H^{(1)}_{q\bar{q}} = C_F \left(-8 +
	\frac{7\pi^2}{6}\right)\,, \\ &\label{eq:Cij}C_{ij}^{(1)}(z) =
	-P_{ij}^{(0),\epsilon}(z) - \delta_{ij}\delta(1-z) C_F \frac{\pi^2}{12}
	+ 2P_{ij}(z) \ln \frac{Q}{\mu_F}\,.
\end{align}
\end{subequations}
Here $P^{(0),\epsilon}_{ij}$ denote the $\epsilon$-dependent part of the
leading-order splitting functions in $D=4-2\epsilon$ dimensions
\begin{subequations}
  \begin{align}
    P_{qq}^{(0),\epsilon}(z) &= -C_F(1-z)\,,\\
    P_{gq}^{(0),\epsilon}(z) &= -C_F z\,,\\
    P_{qg}^{(0),\epsilon}(z) &= -2 T_R z(1-z)\,,\\
    P_{gg}^{(0),\epsilon}(z) &= 0.
  \end{align}
\end{subequations}
where $P_{qg}^{(0),\epsilon}(z)$ is for a single flavour or
anti-flavour of quark (not the sum of flavour and anti-flavour).
Adding together Eqs.~\eqref{eq:shower-approx-pp-final} and
\eqref{eq:resum-nlo-result} gives us
\begin{align}
  \label{eq:BbarDY}
  \bar{B}_{C}^{\text{PG},pp\to V} = B^{pp\to V}_0(\PhiB) \left[1 + \frac{\alpha_s(\mu_R)}{2\pi}\left(H^{(1)}_{q\bar{q}} +
    \bar{H}^{(1)}_{q\bar{q}} +\frac{\left((C^{(1)}_{ik} +
      \bar{C}^{(1)}_{ik})\otimes f_{k}\right)(x_i, \mu_F)}{f_i(x_i, \mu_F)} +
    i\leftrightarrow j \right)\right]\,.
\end{align}
\subsection{Results for gluon fusion Higgs production and DIS}
In order to state an equivalent result to Eq.~\eqref{eq:BbarDY} for
gluon fusion Higgs production, we need first to give the splitting
functions that enter into Eqs.~\eqref{eq:CTDY} and
\eqref{eq:shower-approx-pp-1}
\begin{subequations}
  \begin{align}
    p_{gq}(\bar{z}) &= C_F\left[\frac{1+(1-\bar{z})^2}{\bar{z}}\right]\,, \\
    \tilde{p}_{gg}(\bar{z}) &= 2C_A \left[\frac{\bar{z}}{1-\bar{z}}+\frac{\bar{z}(1-\bar{z})}{2}\right]\,, \\
    p_{gg}(\bar{z}) &= \tilde{p}_{gg}(\bar{z}) + \tilde{p}_{gg}(1-\bar{z}) \,.
  \end{align}
\end{subequations}
It is worth noting that the regularised $P_{gg}$ splitting function
can be written as
\begin{equation}
  P_{gg} = (\tilde{p}_{gg}(\bar{z}))_+ + \tilde{p}_{gg}(1-\bar{z}) - \frac{4 n_f T_R}{6}\delta(1-\bar{z})\,.
\end{equation}
For Higgs production in the limit of a large top-quark mass and
vanishing bottom-quark mass, 
the counterterm real-radiation probability is then given by
\begin{align}
  \label{eq:CTggF}
  \frac{C(\Phi)}{B_0(\PhiB)} d\Phirad \to \frac{d\kappa_t}{\kappa_t} d\bar \eta
  \frac{d\phi}{2\pi} 
  \frac{\as}{\pi}  (1-\bar{z}) p_{gk}(\bar{z}) \frac{f_{k}(x/\bar{z})}{f_g(x)} \Theta\left(x<\bar{z}<\frac{{Q}}{{Q}+\kappa_t}\right)\,,
\end{align}
while the NLO weight is 
\begin{align}
  \label{eq:BbarggH}
  \bar{B}_{C}^{\text{PG},pp\to H} = B^{pp\to H}_0(\PhiB) \left[1 + \frac{\alpha_s(\mu_R)}{2\pi}\left(H^{(1)}_{gg} +
    \bar{H}^{(1)}_{gg} +\frac{\left((C^{(1)}_{ik} +
      \bar{C}^{(1)}_{ik})\otimes f_{k}\right)(x_i, \mu_F)}{f_i(x_i, \mu_F)} +
    i\leftrightarrow j \right)\right]\,,
\end{align}
where
\begin{align}
  H^{(1)}_{gg} = C_A \left(5 + \frac{7}{6} \pi^2\right) - 3 C_F+ 8 \pi b_0 \ln \frac{\mu_R}{Q}\,, \quad
  \bar H_{gg}^{(1)} = C_A \left(\frac{1}{6}+\frac{2\pi^2}{3} + \frac{22}{3} \ln 2\right)\,,
\end{align}
and
\begin{subequations}
  \begin{align}
    \bar C^{(1)}_{gq}
    &=
      -2p_{gq}(\bar{z}) \ln\frac{\bar{z}}{1-\bar{z}} \Theta(\bar{z}>1/2)\,, \\
    \bar C^{(1)}_{gg} & = -2 \left[\tilde p_{gg}(\bar{z}) \ln\frac{\bar{z}}{1-\bar{z}} \Theta(\bar{z}>1/2)\right]_+
                        -2 \tilde p_{gg}(1-\bar{z}) \ln\frac{\bar{z}}{1-\bar{z}} \Theta(\bar{z}>1/2)\,.
  \end{align}
\end{subequations}
The $C^{(1)}_{ik}$ is given as in eq.~\eqref{eq:Cij} with the
replacement $C_F \to C_A$.
Finally, for DIS we need to combine elements of the $e^+ e^-$ and $pp$
analyses above.
Specifically, the differential counterterms are given by
Eq.~(\ref{eq:ee-diff-counterterm}) in the current hemisphere and
Eq.~(\ref{eq:CTDY}) in the remnant hemisphere, and the sum of the
integrated counterterm and virtual correction is
\begin{align}
  \Bar{B}_C^{\text{PG},\text{DIS}}(\Phi_B) &= B^{\text{DIS}}_0(\PhiB)\left[1 + \frac{\alpha_s(\mu_R)}{2\pi}\left(H_{\rm DIS}^{(1)} + \bar{H}_{\rm DIS}^{(1)} + \frac{\left((C^{(1)}_{ik} +
      \bar{C}^{(1)}_{ik})\otimes f_{k}\right)(x_i, \mu_F)}{f_i(x_i, \mu_F)} \right)\right]\,,
	\end{align}
with 
\begin{align}
  H^{(1)}_{\text{DIS}} = - 8C_F, \quad
  \bar H_{\text{DIS}}^{(1)} = C_F \left(7-\frac{\pi^2}{4} + 3 \ln 2\right)\,.
\end{align}

\section{Lepton-swap algorithm for the Drell--Yan process}
\label{app:swap}
In this section we describe the swap algorithm that we have
implemented to efficiently generate the hardest emission in Drell--Yan
processes, where the lepton-anti lepton asymmetry can lead to a
nearly (exactly) vanishing Born squared matrix element $B_0(\PhiB)$ in
neutral-current (charged-current) production, causing the ratio
$R(\PhiB, \Phi_{\rm rad})/B_0(\PhiB)$ to become very large away from a
singular configuration.

\paragraph{Hardest radiation generation.} We aim to generate an
emission given (a) the underlying Born phase space $\PhiB$ and (b)
radiation variables $\Phi_{\rm rad}$.
For standard multiplicative matching, we generate the hardest emission
with the following branching probability,
\begin{equation}
\label{eq:app-shower-nlo-weight-simplified}
\frac{d P}{d\Phi_\text{rad}} = \frac{R(\PhiB,\Phi_\text{rad})}{B_0(\PhiB)}.
\end{equation}
cf.\ Eq.~(\ref{eq:2}).
When we activate the swap algorithm we instead use
\begin{equation}
  \frac{dP^{\rm (swap)}(\PhiB,\Phi_\text{rad})}{d\Phi_\text{rad}}
  = \frac{R(\PhiB,\Phi_\text{rad})+R(\PhiB',\Phi_\text{rad})}{B_0(\Phi)+B_0(\PhiB')},
\label{eq:swapemission}
\end{equation} 
where $\PhiB'$ is the underlying Born phase space in which the lepton momenta have been swapped.
It is important to notice that Eq.~\eqref{eq:app-shower-nlo-weight-simplified} and Eq.~\eqref{eq:swapemission} become identical in the limit where the emission is either soft or collinear.

If an emission is accepted, one computes
\begin{equation}
f_{\rm B} = \frac{B_0(\PhiB)}{B_0(\PhiB)+B_0(\PhiB')}, \qquad f_{\rm R} = \frac{R(\PhiB,\Phi_\text{rad})}{R(\PhiB,\Phi_\text{rad})+R(\PhiB',\Phi_\text{rad})},
\end{equation}
and swaps the lepton kinematics with probability
\begin{equation}
p^\text{(swap)} = \max \left(0, 1-\frac{f_{\rm R}}{f_{\rm B}}\right).
\end{equation}
The swap ensures that the lepton kinematics in the presence of a
resolved real emission is distributed according to $R(\PhiB,
\Phi_\text{rad}$).
This effectively corrects for Eq.~(\ref{eq:swapemission}), bringing
the real generation probability back to
Eq.~(\ref{eq:app-shower-nlo-weight-simplified}).

\paragraph{Virtual corrections in \dBNLO.}
Any change in the generation of real radiation implies also a
corresponding change in the $\Bbar$ NLO normalisation factor for the
Born configuration, which now reads
\begin{equation}
  \label{eq:Bbar-swap-with-counterterm}
  \Bbar^\text{(swap)}(\PhiB) = B_0(\PhiB) + 
    V(\PhiB) +    
    C_\text{int}(\PhiB)
    + 
    \int \left[B_0(\PhiB)
      \frac{dP^{\rm (swap)}(\PhiB,\Phi_\text{rad})}{d\Phi_\text{rad}}
      - C(\Phi) \right] \, d\Phi_\text{rad},
\end{equation}
where we rely on the property that $V(\Phi_B)/V(\Phi_B') =
B_0(\Phi_B)/B_0(\Phi_B')$ and similarly for the counterterm, both
differential and integrated.

\paragraph{Virtual corrections in \ESME.}
For the \ESME algorithm, the normalisation of Born-like events is
given by $\bar{B}_C$ of Eq.~\eqref{eq:Bbar-C}, which is a function of
the Born amplitude squared $B_0$, the virtual one $V$ and the
integrated real counterterm $C_{\rm int}$.
Recall the ratios $V(\PhiB)/ B_0(\PhiB)$ and
$C_{\rm int}(\PhiB)/B_0(\PhiB)$ are independent from the decay angles
of the vector boson.
Thus $\bar{B}_C$ stays the same for the lepton-swapped configuration.
The real radiation part of \esme uses
$dP^{\rm (swap)}(\PhiB,\Phi_\text{rad})$ of
Eq.~\eqref{eq:swapemission}.
This ensures that Eq.~(\ref{eq:Bbar-swap-with-counterterm}) is
automatically reproduced, without the need to modify the Born event
normalisation.

\bibliographystyle{JHEP}
\bibliography{MC}

\end{document}